\documentclass[twocolumn]{aastex631}
\definecolor{myblue}{RGB}{0, 0, 180}

\def\Mpc{{\rm Mpc}}
\def\kpc{{\rm kpc}}

\def\kms{{\rm km}\,{\rm s}^{-1}}

\def\SEDz*{\emph{SEDz*}}

\gdef\deg{$^{\circ}$}
\gdef\kms{km s$^{-1}$} 
\gdef\gappr{\hbox{$_>\atop{^\sim}$}}
\gdef\lappr{\hbox{$_<\atop{^\sim}$}}

\def\etal{\hbox{et al.}}

\gdef\eg{{\it e.g.,\ }}

\gdef\ltsima{$\scriptscriptstyle \; \buildrel < \over \sim \;$}
\gdef\simlt{\lower.3ex\hbox{\ltsima}}

\gdef\gtsima{$\scriptscriptstyle \; \buildrel > \over \sim \;$}
\gdef\simgt{\lower.3ex\hbox{\gtsima}}

\gdef\about{\raise.3ex\hbox{$\scriptscriptstyle \sim $}}

\def\gs{\mathrel{\raise0.35ex\hbox{$\scriptstyle >$}\kern-0.6em 
\lower0.40ex\hbox{{$\scriptstyle \sim$}}}}
\def\ls{\mathrel{\raise0.35ex\hbox{$\scriptstyle <$}\kern-0.6em 
\lower0.40ex\hbox{{$\scriptstyle \sim$}}}}

\def\etal{\hbox{et al.}}

\def\kms{\rm{\hbox{\,km s$^{-1}$}}}
\def\24m{\hbox{24\,$\micron$}$\,$}
\def\10-18{\hbox{$\times~10^{-18}$}}
\def\deg{\hbox{$^{\circ}$}}

\def\T0{{$t_0$}}
\def\SBF{\hbox{\emph{SBF}}}
\def\FourStar{\hbox{\emph{FourStar}}}
\def\2MASS{\hbox{\emph{2MASS}}}

\accepted{for Publication in The Astrophysical Journal}
\shorttitle{Return to the Great Attractor}
\shortauthors{Dressler \& Monson }
\graphicspath{{./}{figures/}}

\begin{document}

\title{Return to the Great Attractor}

\author[0000-0002-6317-0037]{Alan Dressler}
\affiliation{Carnegie Science Observatories, 813 Santa Barbara St., Pasadena, CA 91101, USA}

\correspondingauthor{Alan Dressler}
\email{dressler@carnegiescience.edu}

\author[0000-0002-0048-2586]{Andrew Monson}
\affiliation{Steward Observatory, The University of Arizona, 933 N. Cherry Ave., Tucson, AZ 85721, USA}




\begin{abstract}

We used the FourStar near-IR camera on Magellan-Baade to obtain high S/N H-Band imaging of 66 galaxies with radial velocities of $2000 \leq V\leq 5000$ \kms. Our goal was to use the superior distance measurements of surface-brightness-fluctuations (\emph{SBF}) to derive ``peculiar velocities'' to test earlier claims that the \emph{CMB dipole anisotropy}, equivalent to $\approx$600\kms\ with respect to the Local Group, arises from a `local' overdensity in the galaxy/dark-matter distribution --- the \emph{Great Attractor}. \emph{SBF’s} ability to measure distances with $\sim$5\% accuracy confirms a strong flow over a steradian of the sky peaking at $V_{pec}$ $\sim 1000$ \kms\ and converging to zero at D~$\approx$~70~\Mpc\ from the Local Group. The modest spatial extent of this flow $R_{V}\approx5000$ \kms\  is consistent with the original Great Attractor model (a diameter $D\lappr140$~Mpc), as well as the magnitude and direction of the CMB dipole anisotropy, and the power spectrum of CMB fluctuations --- the latter two arguably the most secure measurements in astrophysics. In contrast, our results are at-odds with reports of comparable amplitude `bulk flows’ on scales of hundreds of Mpc that themselves may be inconsistent with the expected fluctuations in the CMB for a $\Lambda$CDM universe.  We  contend that only distance-estimators as accurate as \SBF\ are able settle the question of whether the CMB dipole arises from the gravitational influence of large-scale structure \emph{within, or without} $\sim$100 Mpc of the Local Group.  

\end{abstract}


\title{Return to the Great Attractor: Strong Evidence for a Steradian-sized Flow Converging at $\sim$70~Mpc within the GA Supercluster and Aligned with the CMB Dipole}



\section{Introduction: The Issue}
\label{sec:Introduction}

The discovery of the cosmic microwave background (CMB) \citep{1965ApJ...142..419P,
1965ApJ...142..414D} legitimized cosmology as an observational science. In subsequent decades, challenging `microwave telescopes,' most notably \emph{COBE} \citep{1990ApJ...354L..37M}
\emph{WMAP} \citep{2003ApJS..148....1B, 2003ApJS..148..175S}, 
and \emph{Planck} \citep{2009arXiv0911.3101B} opened a treasure trove of secrets of how the universe began, a direct connection between the properties of the one we inhabit to its astonishingly different beginnings --- among the most remarkable achievements in science. The discovery of the CMB dipole anisotropy \citep{1995ApJL..32..173L, 1997mba..conf...69L} was surprising but welcomed, because it drove our exploration of the large-scale structure at the present epoch, the `main feature' --- structure growth from a scant 400,000 years after the Big Bang to today.  This area is rich enough in basic physics
\citep{1997astro.ph..5101S} that the pursuit of ever-more-detailed data continues in the decades since.  The CMB dipole anisotropy seemed like  a `short subject' --- soon tagged as evidence for substantial departures from smooth Hubble flow that would produce a Doppler-like signal. But the simplicity of this explanation has yet to be definitively demonstrated, with dozens of alternative explanations due to properties of the CMB itself that range from the exotic to the bizarre (\eg \citet{1991PhRvD..44.3737T}, \citet{2006ASPC..351..112B}, \citet{2013JCAP...10..049M}, \citet{2013PhRvD..87b3524B}, \citet{2017arXiv171209917Y}
\citet{2023ApJ...947...47K}...)

The 1980's brought the first attempts to understand the source of the CMB dipole anisotropy, amounting to $\approx$600\kms\ when attributed to the \emph{peculiar motion} of the `Local Group' of galaxies in the direction $L,B$(Galactic) $\approx$ 270\deg,~+30\deg. Departures from a monotonic and smooth Hubble expansion are expected if the local gravity field is perturbed by overdensities or voids in the galaxy distribution.\footnote{Presumably baryons and galaxies both trace the distribution of dark matter: any departure of would be extremely important for understanding how galaxies are `made.'}  Large-scale maps of redshift distances to thousands of galaxies \citep{1987IAUS..124..301G, 1997AJ....114.2205G, 1993ASPC...51...81O} demonstrated the abundance and universality of large-scale structures. The \emph{Local Group}, a gravitationally bound system of the Milky Way, Andromeda (M31), and Triangulum (M33) (with its collection of smaller companions and satellites) resides at the edge of one such overdensity --- the Virgo supercluster. Centered on the rich Virgo Cluster of galaxies, it is $\sim$30 Mpc in diameter and consists of hundreds of major, massive galaxies. 

\begin{figure*}
\centering

\includegraphics[scale=0.6]{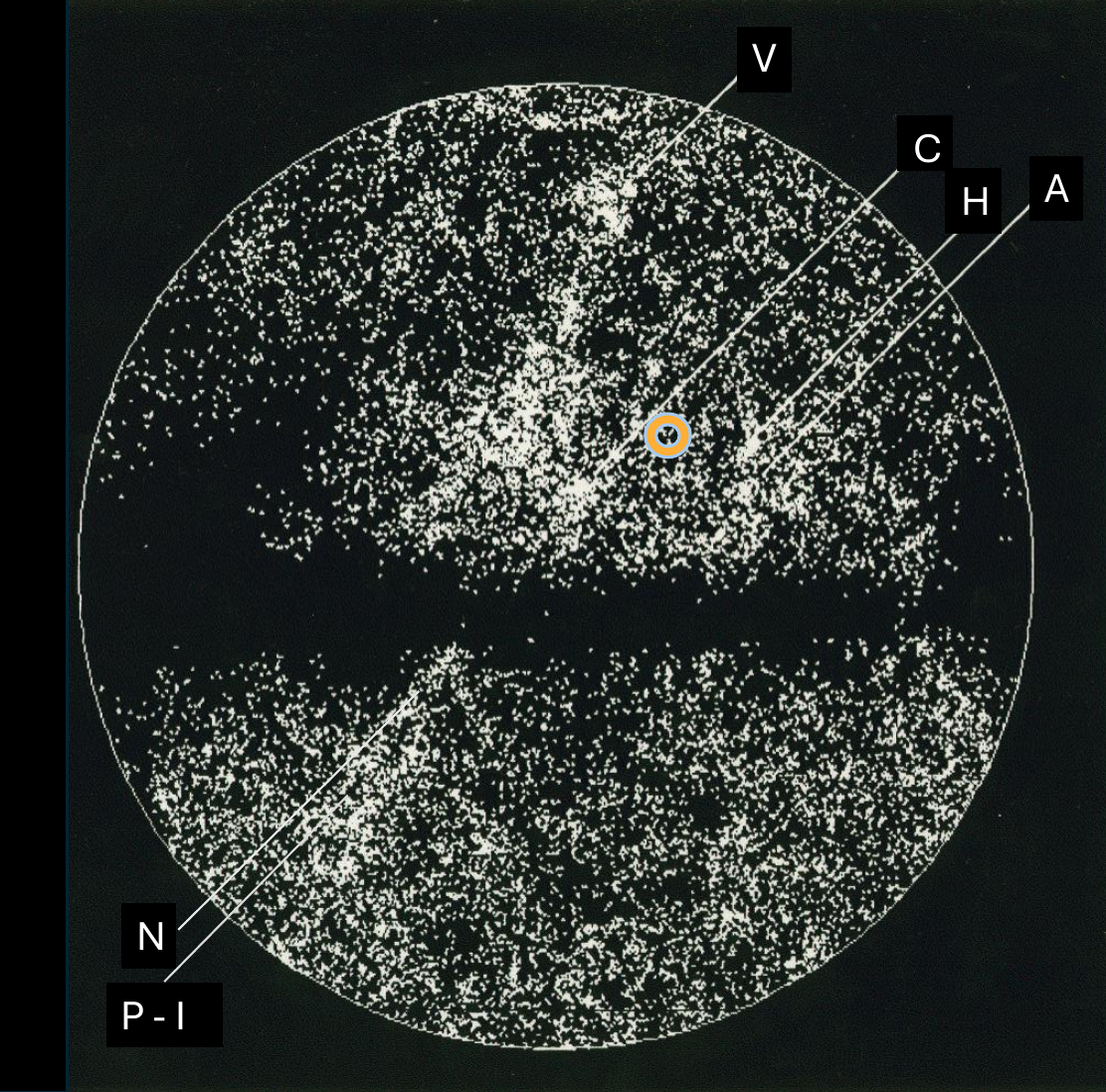}

\caption{Galaxy counts over one hemisphere of the sky from \citet{1987MNRAS.225..213L}.   Linked to familiar clusters of galaxies and their superclusters --- Virgo, Centaurus, Hydra, Antila, Pavo-Indus, Norma, the enormous `Great Attractor' outshines them all --- even where hidden by Milky Way dust. The orange doughnut is the direction of the CMB dipole, about 20\deg\ off the GA center (upper-left from Centaurus and straight-left of the CMB dipole direction.  Lacking the presence of a huge cluster, or peculiar motions of a much larger sample  of its members, The GA center remains approximate.}
\label{fig:Lahav-sky}
\end{figure*}

\begin{figure*}
\centering

\includegraphics[scale=0.7]
 {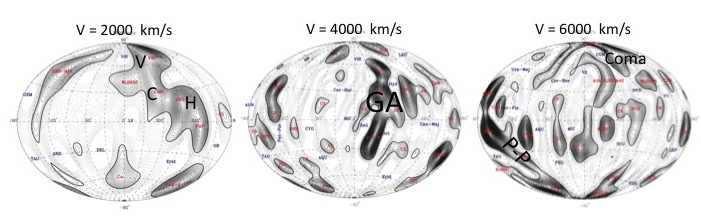}

\caption{\normalsize All-sky shells of galaxy density from the \2MASS survey \citep{2006MNRAS.373...45E}. Within $\emph{V} < 6000 \kms)$, the most prominent mass concentration is the superposition of the Virgo, Hydra, and Centaurus clusters and their superclusters $\emph{V} \sim2000 \kms$ and the GA supercluster $\emph{V} \sim 4000 \kms$ that are the subject of this study.}

\label{fig:Erdogu_shells}
\end{figure*}

Detecting departures from pure Hubble flow that build clusters and superclusters\footnote{This is commonly called ``infall'' for the overdensities and ``outflow" for the voids, but is, of course, just the retarding or speeding of the local Hubble expansion.} involves more than a galaxy's `Hubble distance' (Hubble's law + a \emph{redshift}) but also a distance measurement that is \emph{independent} of its redshift. \citet{1977A&A....54..661T} found a good correlation of the speed of the galaxy's rotation with its luminosity (\eg{mass}) that provided such a redshift-independent distance.  Comparing a galaxy's \emph{observed} velocity to this `\emph{TF-predicted-distance} $\times$ Hubble constant' distance yields its \emph{peculiar velocity}.  Measurements like these were developed and calibrated by \citet{1980ApJ...237..655A} and \citet{1980ApJ...238..458M} and applied to the Virgo Supercluster by \citet{Aaronson1982}, who measured the Local Group's peculiar velocity as $V_{pec}\sim$~250\kms.  This amounted to a significant though still minority component of the CMB dipole, in a direction $\sim$40\deg\ from the dipole, leaving the primary component(s) unexplained, presumably the effect of other large-scale structures.

Coincident with these studies, a group of 7 (including AD),  dubbed the ``Seven Samurai" (hereafter, 7S)\footnote{D. Lynden-Bell, S.M. Faber, D. Burstein, R.L. Davies, A. Dressler, R.J. Terlevich, G. Wegner} began an all-sky survey of elliptical galaxies \citep{1988IAUS..130..169F}. Its goal was to compare their basic properties --- light, mass, kinematics, and metal-abundance.  The 7S team was encouraged by a then-common belief that ellipticals could be key in understanding galaxy formation and growth, because they were the simplest galaxies --- a single structure and an old stellar population. This turned out to be a mirage, but a different reward came from the expedition. 

The 7S team collected photometric and spectroscopic data from observatories around the world to assemble an all-sky sample of 469 elliptical galaxies and their cousin `S0' galaxies (together, \emph{`early-type')}, with emphasis on carefully calibrating observations from different telescopes/instruments covering the full sky. Interesting relationships were found, but an especially useful one --- the \emph{fundamental plane} (FP, \cite{Dressler1987a}, \cite{1987ApJ...313...59D}) --- showed a good relationship of metric-diameter, velocity dispersion, and surface brightness for early-type galaxies in the Coma and Virgo clusters. The FP's utility as a distance indicator for elliptical and S0 galaxies --- accurate to $\sim$25\% --- is analogous to that of the Tully-Fisher relation for spiral galaxies.  When applied to the full sample, 7S found that --- compared to the observations of galaxies in clusters that defined the FP --- there were \emph{significantly larger residuals.}    Specifically, comparing FP distances to Hubble distances revealed substantial positive \emph{peculiar motions} for galaxies over a vast part of the southern sky.  Reinforcing this pattern was a visual map of that extensive area, shown in Figure~\ref{fig:Lahav-sky}: a huge swath of light (each galaxy a white dot) from a vast supercluster, previously unidentified as such. Large peculiar motions of $V_{pec}\sim$~500-1500\kms\ pointed in the direction of the supercluster --- close to the direction of the CMB dipole.

An analysis of the \2MASS galaxy survey \citep{2006MNRAS.373...45E} sliced the local volume into shells of radial velocity 0 $<$~V~$<$~2000\kms, 2000 $<$~V~$<$~4000, and 4000 $<$~V~$<$~6000. In accord with this projection of galaxies on the sky, the highest density of galaxies is found in the middle shell (Figure~\ref{fig:Erdogu_shells}) --- at the proper distance, and in the right direction to be the `overdensity' principally responsible for the CMB dipole anisotropy.

\citet{Dressler1987a} described the result as a ``large-scale streaming" but were soon using the term ``bulk flow," which became the favored description when there is no clear convergence of a flow (which is often the case --- see Section~\ref{sec:Mapping_Further}).  However, the map of peculiar motions in that paper suggested a \emph{converging} flow at V~$\sim$~5000\kms\ toward what the 7S team was then calling the ``Great Attractor" (GA), and \citet{1988ApJ...326...19L} made a quantitative GA model that included maps of the sky showing `galaxy peculiar motions' as vectors. The amplitude and direction of the flow toward the GA center, when added to the gravitational pull of the Virgo supercluster, matched --- within errors --- the CMB dipole direction and its amplitude.

Still, in a comparison to four other similar studies, \citet{Dressler1987a} found marginal and in some cases inconsistent results.  In hindsight, it seems that distance estimators of $\sim$25\%\ accuracy, while adequate for relatively nearby structures like the Virgo supercluster, were marginal at best at V$\gappr{4000}$\kms, because, with a \emph{S/N} $\lappr$ 1 for each galaxy, \emph{systematic errors} can (and do) dominate.  This was the primary motivation for the present study, which was not worth attempting until much better distance estimators became available. Section \ref{sec:Mapping_Cosmic_Flows} further explores this issue.

A final issue, much discussed at the time of \citet{Dressler1987a} and \citet{1988ApJ...326...19L}, was whether galaxy flows on the scale of the Great Attractor, $\sim$100\ Mpc, were at-odds with \emph{cold-dark-matter} cosmology, as exemplified by \citep{1989ApJ...344....1G}.  In the (then) much-favored $\Omega_m$ = 1 (``critical density") model, coherent structures as large as the Great Attractor should be a $\sim$1\% occurrence, thus violating the \emph{anthropic cosmological principle} --- our place in the universe must not be `special.'  Amos Yahil, a co-author of \citet{1989ApJ...344....1G}, rose up at a famous conference and announced that these ``Seven Samurai" (hence, the name) --- with their slashing away at accepted, \emph{known} physics --- just had to be wrong! However, a soon-to-be change-of-heart in the astrophysics community --- giving up the theoretical ``closed universe" for an observationally persuasive $\Omega_m\sim$~0.25 --- helped the Great Attractor model live another day.  Our location in the universe returned to ``nothing special"...although, thanks to \emph{us}, nothing could be further from the truth.

As discussed in Section~\ref{sec:NOT_GA}, the objections of those theorists, though long-ago overcome for the Great Attractor, can still an issue for subsequent studies that report much larger-scale flows.

The remaining sections of the paper are: 2. Mapping the Great Attractor; 3. Fourstar Imaging of GA Galaxies for \SBF\ Distances; 4. The Program: \SBF Distances to GA Galaxies; 5. Data Reduction; 6. How Self Calibration Makes \SBF\ Unique as a Distance Measure; 7. Results: \SBF\ Distances to 66 Great Attractor Galaxies; 8. Why \emph{Not} the Great Attractor?; 9. Mapping Cosmic Flows Near and Beyond the GA; 10. Conclusions and Future Prospects; Appendix A:
Exclusions of Centaurus Cluster Galaxies; Appendix B: Photometric Calibration

\section{Mapping the Great Attractor}
\label{sec:GA-Mapping}

Without consensus in the GA data with other studies, and with a thin sampling of galaxies in the GA supercluster by the `all-sky' 7S, and without the prospect of more accurate distance estimators, \citet{1991ApJS...75..241D} launched a survey of a steradian-sized area centered on the GA. The {Supergalactic Plane Survey} (SPS) added $\sim$1000 galaxy redshifts to the $\sim$400 in the literature.\footnote{Observations were taken over a span of 17 nights using the Las Campanas \emph{Dupont} telescope and the \emph{Modular Spectrograph} and later with the \emph{B\&C Spectrograph} plus \emph{2D-Frutti} photon-counting detector.}

\begin{figure*}
\centering

\includegraphics[scale=0.9]{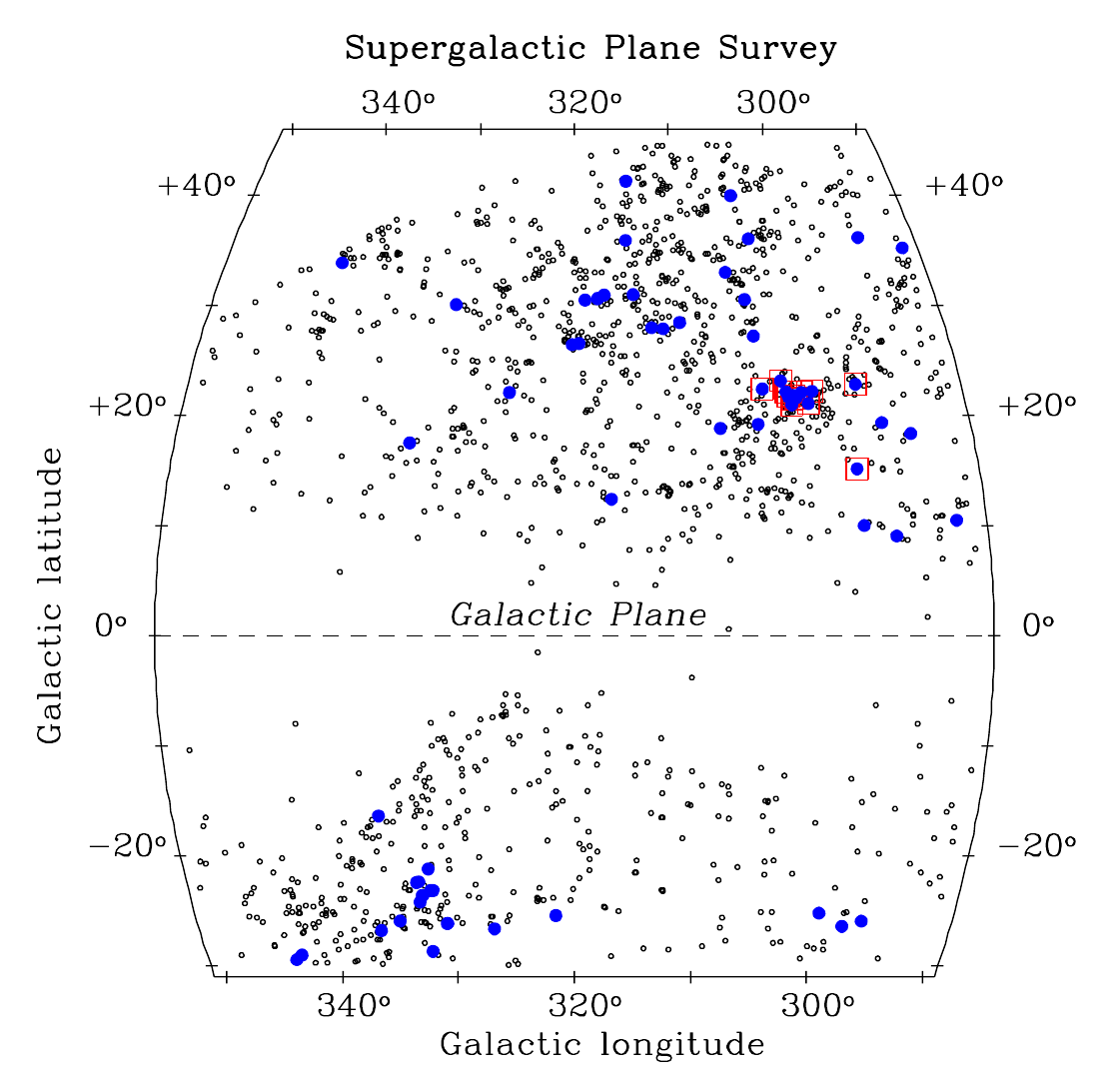}

\caption{The ``Great Attractor" region of the sky in Galactic coordinates, a region of $\sim$1 steradian (after subtracting the part obscured by Galactic dust), covering the galaxy overdensity seen in Figure 1. Small dots are 1403 galaxies, $\sim$1000 from the Supergalactic Plane Redshift Survey and $\sim$400 from the ESO catalogs \citep{1988ApJ...327..544D} that make a fairly complete sample in this volume.  The 66 larger blue dots are the new \SBF\ observations, in the velocity range
2500\kms $ < $ $V_{CMB}$ $ < $~5500\kms, including 13 galaxies in the Cen30-Cen45 groups --- the 20 sq deg clump at \emph{l} 
= 302\deg, \emph{b} = +22\deg. }

\label{fig:GA-Sky}
\end{figure*}

Figure~\ref{fig:GA-Sky} shows a map of the final sample of 1403 galaxies.  The upper left area is uncataloged, and Galactic extinction obscures much of the area of the Galactic Plane, at latitudes -10\deg$<$ b $<$ +10\deg. The ``covered'' region of the SPS survey is somewhat larger than one steradian. The larger blue dots show the 66 galaxies that were selected and observed for this \SBF\ study, discussed below.

\begin{figure*}
\centering

\includegraphics[scale=0.7]
 {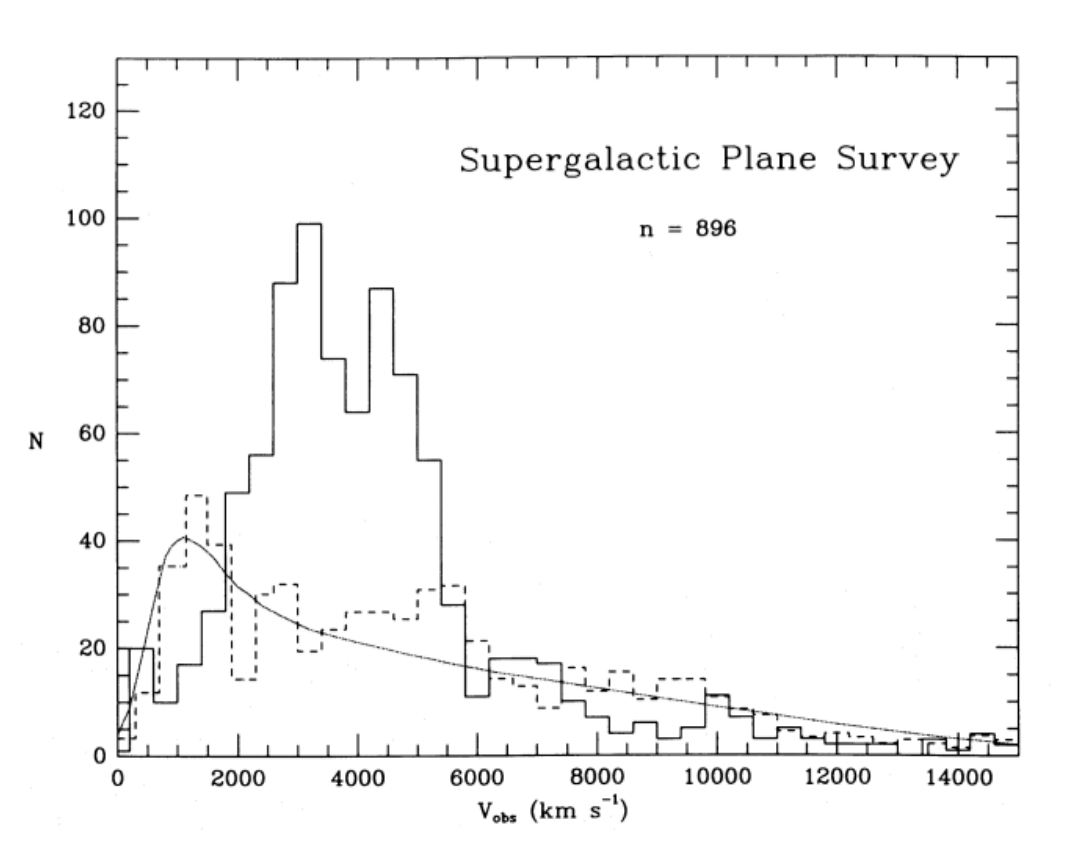}
 
\caption{A histogram galaxy redshifts from the SPS (solid line) that shows a enormous double peak at  $V\sim4000\kms$ for the `GA region' (see also middle panel of Fig~\ref{fig:Erdogu_shells}) compared
to the smoothly declining distribution (dashed line, also shown smoothed) for the rest of the southern-hemisphere sky at the same depth. As described in the text, after correction for peculiar velocity and luminosity-function selection affects, the peak at $V\approx4500\kms$ would dominate: a volume-limited, unbiased sample would be symmetric around it, in stark contrast to the dashed-line and smoothed curve for the rest of the southern sky.}

\label{fig:SPS_vel_hist}
\end{figure*}

Fig~\ref{fig:SPS_vel_hist} shows a histogram of 1403 galaxy redshifts from the SPS survey (solid line) with an enormous double peak at  $V_{obs}$~$\sim$~4000\kms\ for the `GA region' (see also middle panel of Fig~\ref{fig:Erdogu_shells}).  The SPS histogram is to be compared to the smoothly declining distribution (dashed line, also shown smoothed) for the rest of the southern-hemisphere sky to the same depth, from the \emph{SSRS} Survey \citep{1988ApJ...327..544D}.  The double peak of the GA region can be understood as a combination of galaxies at $V_{obs} \sim$2500-4500\kms\ with peculiar velocities of $\sim$1000\kms\ that are infalling into the V~$\sim$~4500\kms\ (at rest) center of the Great Attractor. Adjusting for this and the falling sensitivity of the sample between $V_{obs}$ $\sim$ 2500 and $V_{obs}$ $\sim$ 4500 \kms\ (peculiar motion removed), a plot of number density vs \emph{distance} should be dominated by the ${V_{CMB}}$ $\sim$ 4500\kms\ peak. Further,  a strongly dropping sensitivity for observing galaxies $V_{obs}$ $>$ 6000\kms\ suggests that an unbiased sample would in fact be symmetrical around $V_{obs}$ $\sim$ 4500\kms. 

\begin{figure*}
\centering

\includegraphics[scale=0.47]
 {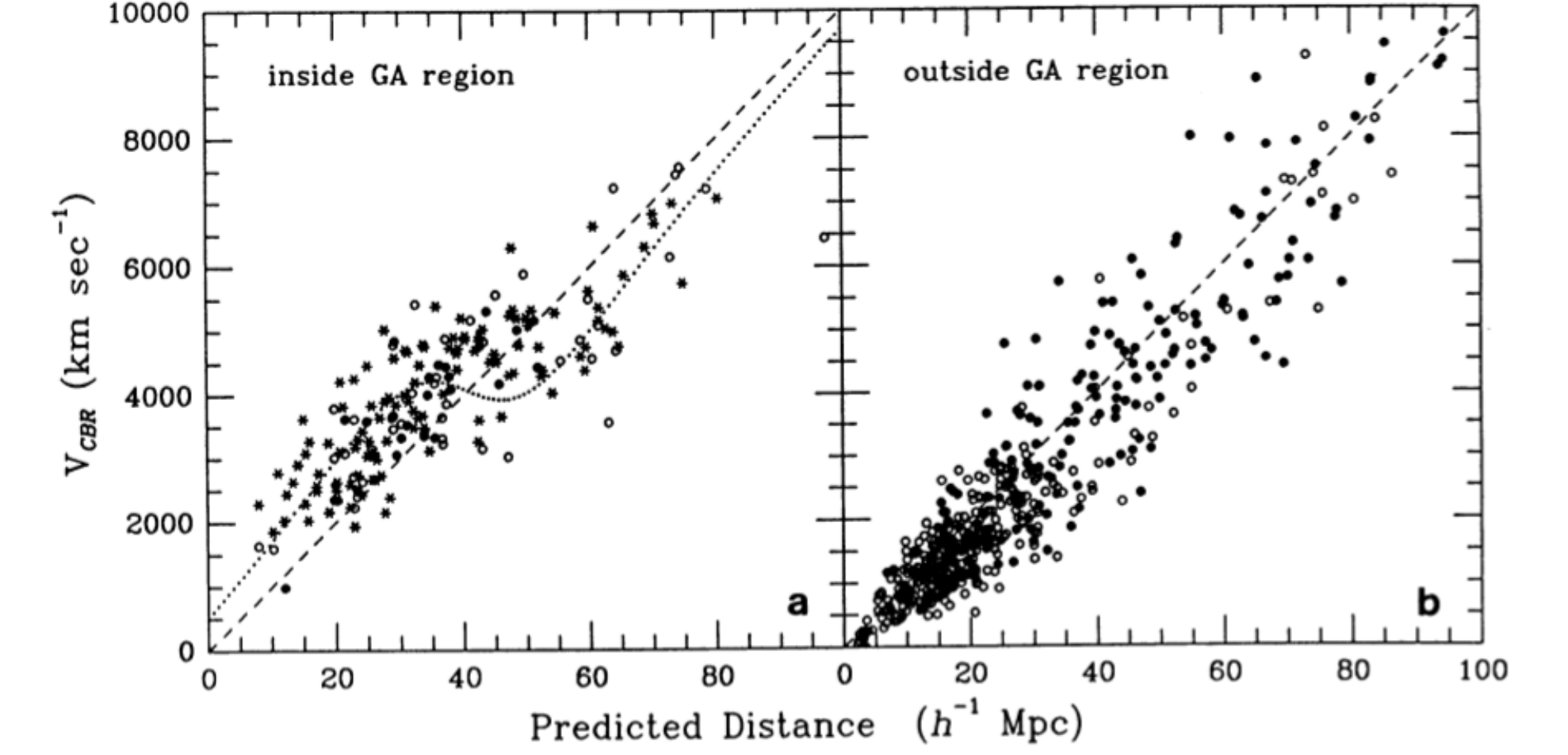}

\caption{Hubble diagrams \citep{1990ApJ...354L..45D} for 273 E, S0, and spiral galaxies in and out of the Great Attractor. (left) FP and Tully-Fisher distances vs $V_{CMB}$ for the GA region. The study confirmed the 7S result of large positive peculiar motions in the GA direction --- not seen in the rest of the sky (right), with its comparatively unperturbed, though noisy, Hubble flow. The plots also show that distance estimators with errors of $\sim$25\% are unreliable for D~$>$~50~Mpc.}
\label{fig:In_out_GA}
\end{figure*}

{Armed with this deeper survey of the GA region, \citet{1990ApJ...354...13D} obtained new velocity dispersions for 136 elliptical and S0 galaxies (see also \citet{1991ApJ...368...54D} to predict distances using the FP method, and rotation curves for  distances to 117 spiral galaxies using the Tully-Fisher method  \citep{1990ApJ...354L..45D}. In both samples and methods the basic feature of the 7S work emerged: high peculiar velocities of $\sim$1000\kms\ for galaxies in the GA area of the SPS.  Fig~\ref{fig:In_out_GA}\footnote{Figure 2 in \citep{1990ApJ...354L..45D}.} shows the Hubble diagram for these samples combined to 253 galaxies, an order-of-magnitude larger than the 7S sample in the GA region. The outflow for D $\ls$ 40 Mpc is clearly defined, but beyond, the `noise' from $\sim$25\% distance estimators starts to dominate. The GA region outflow is highlighted by comparing to the right panel that plots galaxies \emph{outside} the GA --- a unperturbed Hubble flow.  For both regions there is a tight relation for D $\ls$ 40 Mpc and increasingly large scatter beyond. Comparing ``inside GA" to ``outside GA" confirms beyond doubt the 7S result of large peculiar velocities in the GA, the general direction of the CMB dipole.}

{However, finding \emph{convergence} of the flow within the GA at 40 $ < $ D $ <$ 80 Mpc is critical to identifying the GA as the primary \emph{source} of CMB dipole signal. Failure to find convergence to near-zero peculiar velocity would implicate a further, richer supercluster or superclusters as the source. Unfortunately, measuring accurate peculiar velocities beyond $\sim$50 \Mpc\ was not possible with Tully-Fisher or FP, because distance errors of 1000--2000 \kms\ --- both statistical and systematic --- would dominate any signal.}

{The conclusion of the Dressler-Faber work was that the GA and other relatively nearby superclusters could indeed be responsible for the CMB Dipole anisotropy, in direction and magnitude, but confirming with certainty that the GA is the primary actor would require distance measurements of higher accuracy. This paper is the product of that conclusion.}

\section{\emph{FourStar} Imaging of GA Galaxies for \emph{SBF} Distances}
\label{sec:Observations}

Our project was made possible by the 2010 commissioning of a wide-field near-IR camera for the at Las Campanas Observatory, an instrument capable of producing the deep, high-resolution, and extremely uniform exposures that are essential for \SBF\ measurements.

The \emph{FourStar} is a large-field, 1.0-2.5 \micron\ (JHKs) camera on the Magellan-Baade 6.5m telescope at Las Campanas Observatory in Chile \citep{2008SPIE.7014E..2VP, 2007AAS...211.1112M}.  A collaboration of Carnegie Science Observatories (Pasadena) and the Instrument Development Group at Johns Hopkins University (Baltimore), \FourStar\ has been operational since 2010. 

The instrument uses four Teledyne HAWAII-2RG arrays to deliver a 10.9' x 10.9' field of view at a scale of 0.16\arcsec/$pix^{-1}$, producing images down to 0.30\arcsec\ \emph{FWHM} resolution. The outstanding seeing at LCO, coupled with \emph{FourStar’s} high sensitivity and large field of view, enables unprecedented ground-based near-IR surveys and very deep, targeted science programs.

\emph{FourStar} observations for our program of H-band Surface Brightness Flucuations took advantage of \emph{FourStar's} wide field, where a typical galaxy in our sample filled half of the field of one of its four detectors.  In addition, we were able to take advantage of the multi-detector format to monitor variation in H-band sky-brightness and to record and maintain image quality over an observation.  For our program, \emph{FourStar} observations were executed using a custom dither pattern which stepped the target horizontally (along RA)) at 9 locations (DEC) separated by 4\arcsec\ ($\sim$25 pixels).  At each location a single image was taken. After the 9 steps the telescope was offset to the alternate diagonal chip (1 or 3) and the 9 steps were repeated again.  The target was cycled between chips 1 and 3 ten times (five on each chip), each with a slight offset so that no two images landed on the same pixels, for a total of 90 images (9 steps x 2 chips x 5 cycles). The typical exposure time was 14.56 or 11.64 seconds (depending on the sky background level), resulting in 21.8 or 17.5 minute long sequences (36.5 and 32 minutes with readout and dithering overhead), respectively. Before each macro sequence a Shack-Hartmann test-iteration was performed to focus and tune the shape of the primary mirror, after which the Shack-Hartman was turned off. Because the exposures are relatively short, and the dither-macro does not stay in one position long enough to acquire a star, it was much more efficient to forgo guiding the telescope or making Shack-Hartmann tests during a sequence.\footnote{The Magellan Telescope as normally run in a mode of continuous, or at least frequent monitoring to correct primary mirror shape and telescope alignment, but for our program this would generate excessive overhead.  Fortunately the system is sufficiently stable to support intervals of $\sim20-30$ minutes, even obtaining seeing of better than 0.5\arcsec --- essential for this program. Completed images $\sim$0.40\arcsec\ FWHM accounted for $\sim$10\% of the data.}

The regular horizontal dither pattern was chosen to more easily remove systematics in the \SBF\ analysis. It is easier to identify and filter-out the known 1-d correlation when performing a 2-d correlation.  Alternating between chips 1 and 3 created intermediate sky background frames for sky subtraction on each chip.  Chip 1 was chosen for having the fewest bad pixels. Chip 3 (with more bad pixels) was chosen to provide the largest offset from chip 1 for sky subtraction purposes. 

In photometric conditions, these sequences included at least one short exposure sequence for photometric calibration.  A standard dice-5 macro-dither-sequence was executed with an exposure time of 4.56 seconds to avoid saturation of 
\2MASS sources. These provided the bootstrap calibration to the final coadded images. 

The raw data was corrected for linearization and dark subtraction, and divided by a flat-field.'  A first pass sky frame was created for each chip using images for chips 1 and 3 when they were not imaging the target galaxy.  Once the first pass sky frames were created, sources were identified and masked to create a second-pass sky frame with the first-pass sources masked. The individual images were processed using SCAMP \citep{2006ASPC..351..112B} to calculate their relative position offsets, image distortion, and relative photometric offsets.  The final sky subtracted frames were resampled using the outputs from SCAMP using SWARP \citep{2002ASPC..281..228B} to 0.160 arcsec per pixel (the native plate scale) and co-added (\emph{IRAF} `imcombine') to create a final stack for each macro sequence.  We chose a nearest-neighbor resampling algorithm to minimize interpolation effects.   

For each galaxy, each of the individual images were compared to the photometric sequence to determine a relative zeropoint offset.  Images meeting the following requirements were included in making the final image: seeing better than 0.75”; sky background below 35,000 ADU; zeropoint offset within 0.2 mag from standard.  Once the individual images were flagged as acceptable/unacceptable, new stacks were created for each sequence.  Each macro-sequence stack was then scaled by exposure time to have the same effective flux zeropoint, before co-adding them to make the final image stack.  As a final step, the final image stack was compared to the photometric sequence to determine the photometric zeropoint on the \2MASS system, then converted to AB magnitude using 1.364 Vega-to-AB offset.

\section{The Program: \SBF\ Distances to GA Galaxies }
\label{sec:SBF_Distances}

Planning this project began in 2015 through a series of conversations with Joseph Jensen (Utah Valley University) about the suitability of using the surface-brightness-fluctuations (\emph{SBF}) method pioneered by \citet{1988AJ.....96..807T}\footnote{Development history --- see \citet{2012Ap&SS.341..179B} and \cite{2012PASA...29..489F}.} to make distance measurements of unprecedented accuracy ($\sim$5\%) using \emph{Las Campanas Observatories'} \emph{Magellan-Baade Telescope} with its wide-field, near-infrared camera \emph{FourStar}. The specific goal was investigating the claim that an enormous supercluster of galaxies --- dubbed the Great Attractor because of its extent and richness --- is largely responsible for the striking CMB anisotropy amounting to $\sim$600\kms\ of peculiar motion of the Local Group of galaxies.

{\cite{2015ApJ...808...91J} and \cite{2025hst..prop18027J} describe an ambitious program of using the \emph{WFC/IR} camera of \emph{Hubble Space Telescope} to extend \SBF\ measurements to cosmological distances, taking full advantage of their  higher spatial resolution and near-IR sensitivity.  These lead to much higher amplitude fluctuations that will eventually supercede measurements of cosmological parameters relying on variable stars and supernovae.  The `standard candle' anchoring \SBF\ measurements is the TRGB star (tip of the red-giant branch), understood via stellar evolution to be near constant in absolute luminosity for all old ($\tau > 1$ Gyr) stellar populations, that is, without significant deviations due to metal abundance and age.}

Consultation with Jensen regarded using the FourStar IR camera (described in Section~\ref{sec:Observations}) to make ground-based \SBF\ measurements from images that would take advantage of Magellan-Baade median-seeing of 0.60\arcsec\ (H-band).  Early test images were evaluated by Jenson and Dressler to test the feasibility of the project and educate AD in the use of the powerful, complex software package. The data reduction enables the processing of direct-images into power-spectra of spatial fluctuations that arise from statistical variance in a galaxy when the seeing disk encloses tens to hundreds of (mainly) TRGB stars.

Based on galaxy brightness, seeing, fluctuation amplitude, and sky-brightness, H-band was judged to be the best of the J-H-Ks bands available with FourStar.\footnote{\emph{I}-band exposures with \emph{IMACS} were intended to supply the critical galaxy color and possibly a second \SBF\ signal, but the latter proved uncompetitive, and a superior alternative for color was \2MASS photometry with \emph{J} and \emph{H} bands through a 14 arcsec aperture, cataloged in \emph{NED} --- https://ned.ipac.caltech.edu.} Because TRGB stars are cool, the amplitude of surface-brightness-fluctuations in the near-infrared is considerably higher than in visible light. On the other hand, the sky at H-band is both brighter than in the visible and variable in brightness on time scales of minutes to hours.  This means it is insufficient to just point at a position where the target galaxy fills a large fraction of the field. Instead, we used a ``chopping" scheme where the camera is measuring the sky brightness in an adjacent area and where the entire field of the camera can be calibrated. 

As described in Section~\ref{sec:Observations}, a set of programs was developed to make such observations, with a range of integration and total exposure times.  The observations required for this project required hours of integration, but exposures would saturate the detectors in timescales of tens-of-seconds, so it was necessary to develop a scheme that read out the camera frequently and chopped on-and-off the field of the target galaxy.  Our typical observation was to take 14-sec exposures 9 times (dithering) and then move the target field to the diagonal detector for the ``chop." A full exposure had 10 of these ``chops," yielding 1260 s on target and integrating. Because of these steps, observations were not very efficient, 59\%, with 41\% of the time spent reading the detectors or moving the telescope. (With the also-common 11-sec exposures in brighter sky, the efficiency was 54\%.)

52 nights of (bright run) observing was allocated in Carnegie's proposal review process in 7 `first semester' runs from 2015 to 2021. 12 of those nights were unavailable when Las Campanas was closed because of the Covid-19 epidemic, leaving a total of 40 nights. Of these, only 35\% had more than 80\% transparency and seeing better than 0.70 arcsec, the minimum requirements for H-band \SBF\ measurements for this project. This amounted to 35\% of the allocated time, compared to $\sim$60\% expected for Magellan operations, a disappointing, unusually low yield for Las Campanas. This prolonged the program and resulted in a loss of about 20\% of the targets that would have been observed.

Table 1 records the integrated observing time, H-band seeing, and a derived S/N\footnote{S/N =  (18485/Vcmb)*SQRT(exp(hr)/FWHM(arcsec)} for each of the 66 galaxies.  A minimum S/N $\sim5$ was chose for the final sample, but the typical value is $\sim$7-8.

\section{Data reduction}
\label{sec:Data_redux}

\cite{2015ApJ...808...91J} provide a comprehensive description of the basic steps in \SBF\ data reduction, in particular for \emph{HST} data that require a more complex process than required by our program. Here we provide only a schematic description of the programs that were provided.  The basic processing of \emph{FourStar} frames discussed in Section~\ref{sec:Observations} are a head-start on the reduction Jensen \etal\ describes: the ``nodding and shuffling" technique produces a uniform exposure, sky-subtracted image that is typically several times as large as the galaxy image, and the photometric calibration --- essential for an \SBF\ reduction --- is continuous over the data gathering because each exposure contains 10-20 stars of known flux, that were calibrated when each new set of observations began.

\begin{figure*}
\centering

\includegraphics[scale = 0.585]
 {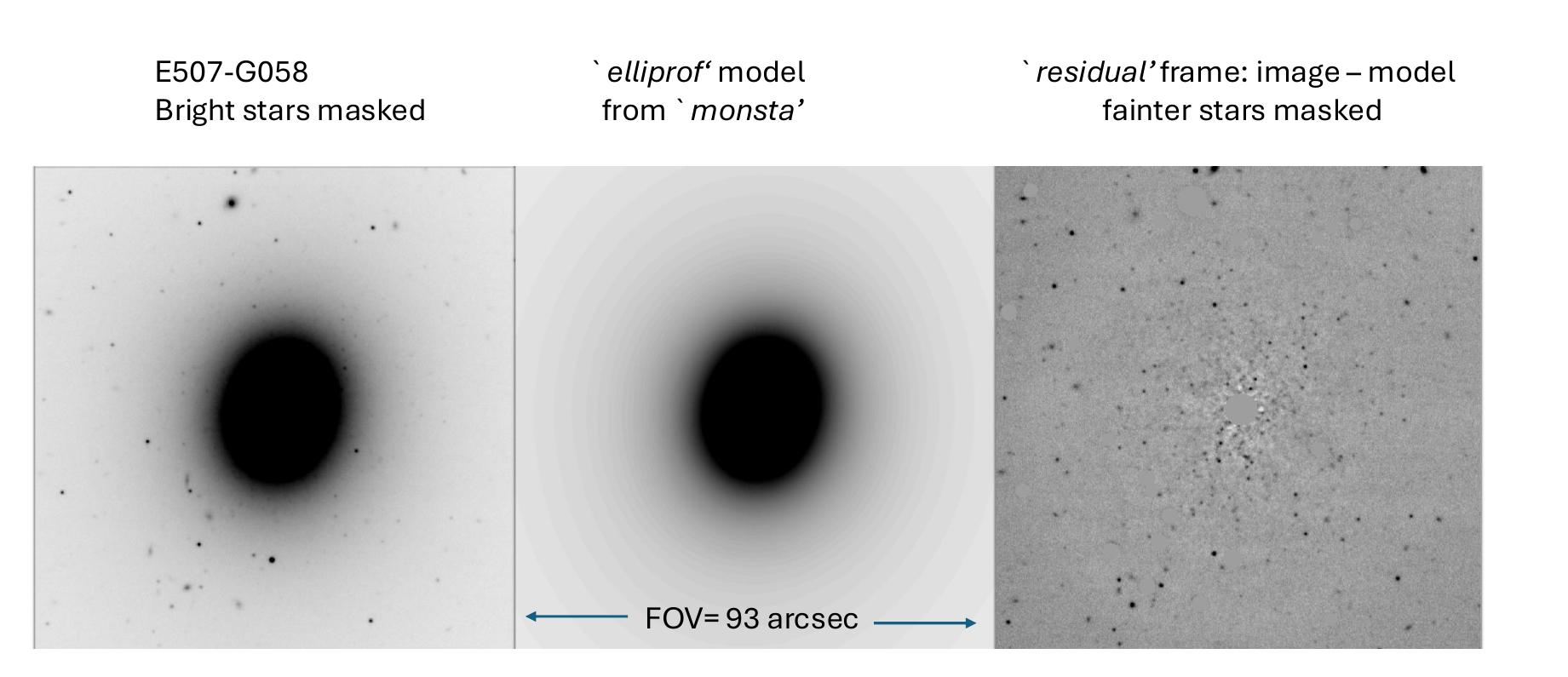}

\caption{First steps in \SBF\ processing. (left) The observed, processed H-band image of E507-G058, with $\sim$10 bright stars masked. (center)  The `elliprof' model from the \emph{monsta} set of programs has fit the galaxy to a set of elliptical isophotes. (right) Subtraction of the middle image from the left image has produced the high-frequencies-only image (right) that includes distant galaxies, globular clusters, and the surface-brightness-fluctuations that are used to calculate the galaxy's distance to $\sim$5\%.}

\label{fig:E507-G058_profile-part1}
\end{figure*}

\begin{figure*}
\centering

\includegraphics[scale = 0.60]
 {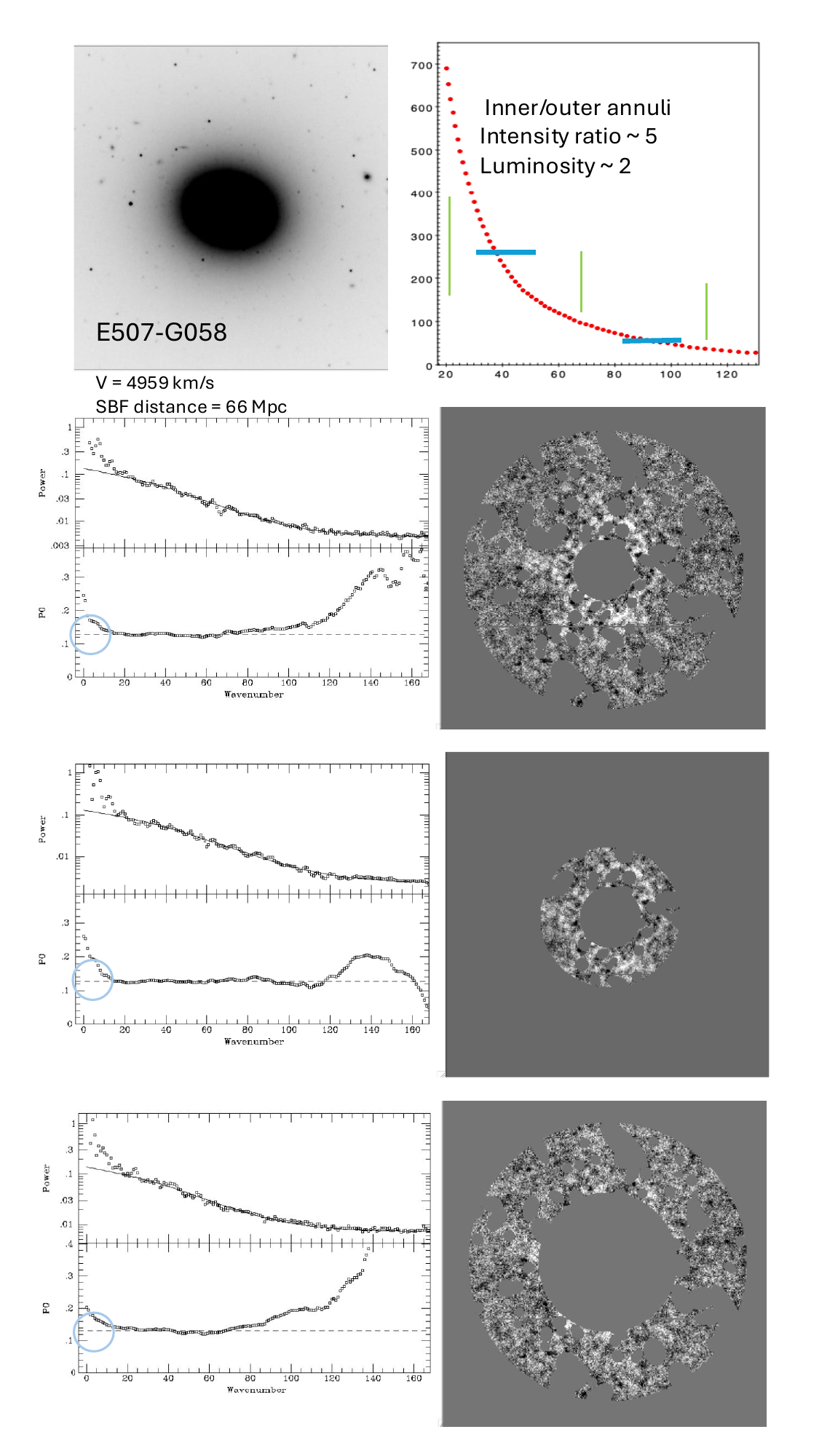}

\caption{The figure shows, for a sample galaxy E507-G058, the three (galaxy-subtracted) annuli and the brightness profile recovered by \emph{elliprof} (upper right).  As the text explains, comparison of the profiles and power-spectrum measurements for the two regions (`c0' and `c1') that comprise the whole area --- `cc', demonstrates \emph{SBF's} remarkable precision as a galaxy-distance \emph{measurements}, compared to previous `distance estimators.'}

\label{fig:E507-G058_profile-part2}
\end{figure*}

The programs used for \SBF\ data analysis were the imaging program \emph{monsta} (a combination of the data-reduction programs \emph{Mongo} and\emph{Vista}), SExtractor, and \emph{likenew6}.  This is a brief summary of the important steps.

1.  Model the galaxy: The program \emph{elliprof} fits the galaxy image with nested elliptical annuli of decreasing flux that result in a brightness profile --- the target galaxy's brightness with radius.  We used \emph{elliprof} to make a single profile, or an average of two, with different numbers of annuli, or spacing. The full range in radius is typically an inner radius of 3" and outer radius between 30" to 50". The subtraction of the model produced by \emph{elliprof} results in a ``flat" picture --- the original galaxy image with the brightness profile subtracted. Remaining small-scale variations are relative to zero. A mask is made to cover moderately bright stars.  Step 1 is shown in Figure~\ref{fig:E507-G058_profile-part1}

2. The variance of the galaxy-subtracted image is determined. This is input for SExtractor to prevent misidentification of \SBF\ fluctuations as point-sources.  Actual point sources --- stars and globular clusters --- are identified and a mask created.    

3. After SExtractor, a \emph{python} program \emph {likenew6} uses the catalog of detected sources to extrapolate the number density of \emph{very} faint sources, using input parameters --- color band, image-scale, psf, and expected distance of the target-galaxy. This helps constrain the the expected number counts with magnitude for faint galaxies (a power-law) and point sources (a gaussian) --- mostly globular clusters 

4. At least 4 clean, unsaturated field stars are chosen to define the point-spread-function and to measure the power-spectrum of a point-source. \SBF\ results using several selected \emph{psf stars} will be compared to assure uniformity.

5.  Define regions for \SBF\ measurement. For this program:
Define 4 annuli of the galaxy image --- a set of broad rings. R = 1--20 pixels (R $<$ 3.0\arcsec) is masked (not used);  full area `cc': R = 20--128 pix (3.2\arcsec--20.5\arcsec), `c0': R = 20--64 pix, `c1': R = 65--128 pix.

5. Calculate \emph{SBFs} for regions `cc', `c0, `c1'. Measure the power spectrum of the chosen \emph{psf} star. Create the \emph{FFT} and power spectrum of the data frame for each of the 3 annuli. Calculate \emph{fluc} as the ratio of the \emph{psf-power-spectrum} to the \emph{data-power-spectrum} for a chosen interval of \emph{wavenumber}, typically \emph{k} = 20 -- 50. This step is shown in Figure~\ref{fig:E507-G058_profile-part2}.

6. \emph{Python} program \emph{bestfluc} calculates \emph{mbar} for all regions chosen, then uses a calibration specific to the wavelength band to convert this to distance modulus, \emph{m-M}, and thus galaxy distance.

\section{How Self Calibration Makes \emph{SBF} unique as a distance measure.}
\label{sec:SBF_unique}

\emph{SBF} observations have a unique ability to `self-calibrate' and `confirm their accuracy' that comes from the rapidly falling brightness profile of a galaxy and how this constrains the measured \SBF\ signal.  At each radius, the fluctuation signal measures statistical (Poisson) fluctuations from the finite number of stars in a seeing disk, while the sum of the light from those same stars is read as the brightness profile.  How well these \emph{two measures} of the number of stars at each point in the profile agree is a test of \emph{SBF's} accuracy.

Figure~\ref{fig:E507-G058_profile-part2} shows the galaxy image as processed to remove the `DC-signal" --- the galaxy profile (see middle picture of Figure~\ref{fig:E507-G058_profile-part1} registering the ``number of stars", leaving the `AC-signal' --- the fluctuations from zero intensity.  The brightness profile is, however, retained, and is used to link these two measurements together.  Figure~\ref{fig:E507-G058_profile-part2} also shows the measured power spectrum (PS) for each of the three regions: the entire R = 20-128 pixel annuli, then split into R = 20-64 and R = 64-128.  In each of the three plots, the top trace is the galaxy PS and the fit of a psf star PS to it (thin line): the lower trace is the division of the two.  The top plot is for annulus `cc' --- R=20-128, the fully processed area. 

Typically --- for $k<15$ and $k>80$ --- the psf star's PS is not a good fit to the galaxy PS: too much power at low frequencies (the waviness of the image frame), and too much high-frequency noise in the galaxy image. In between, the spectrum of the galaxy is well fit by the psf star, so the bottom trace is level --- that's good.  It says that, for k-numbers 20 to 60, the light of the model-subtracted galaxy image (right image of Figure~\ref{fig:E507-G058_profile-part1}) is entirely composed of point sources: these are \textbf{surface brightness fluctuations,} statistical fluctuations of the finite number of stars in a seeing disk. The lower the value of the bottom trace (the y--intercept), the lower the fluctuation amplitude, \emph{the greater the distance to the galaxy.}

The two bottom plots below are for the annuli `c0' and `c1' that together make up `cc' (the plot top). The bottom trace for `c0' is almost as ``clean" as for `cc', while the trace for `c1' is a little less flat than `cc' and a bit bumpier. This reflects the smaller \emph{signal} in `c0' and one even smaller in `c1'. The important point, though, is all three share the same y-intercept --- 0.123, 0.122, and 0.126 --- which translates to Mbar = 30.23 +/- 0.09 and, with calibration, to (m-M) = 34.09, or D = 65.83 (+2.67/-2.79) pc. In other words, two distinct regions (annuli) of the galaxy, with a brightness ratio of $\sim$5 and a total flux ratio of $\sim$2, measure the galaxy distance to about 4\% --- superb accuracy for the distance to a galaxy independent of redshift.  This is not only for the two regions (the average), but at every point along the brightness profile: the number of stars in a seeing disk both defines both the brightness (the total number of stars) and the \SBF\ (the Poisson statistics for that finite number).  This could only be true if `standard candle' TRGB stars are responsible for both: the degree to which these two fully independent measurements (`c0' and `c1') agree is the accuracy of the distance measurement --- full stop.  This further confirms the absence of stellar population gradients, like age or metallicity, perturbing this \SBF\ measurement.

\begin{figure}
\centering

\includegraphics[scale = 0.50]
 {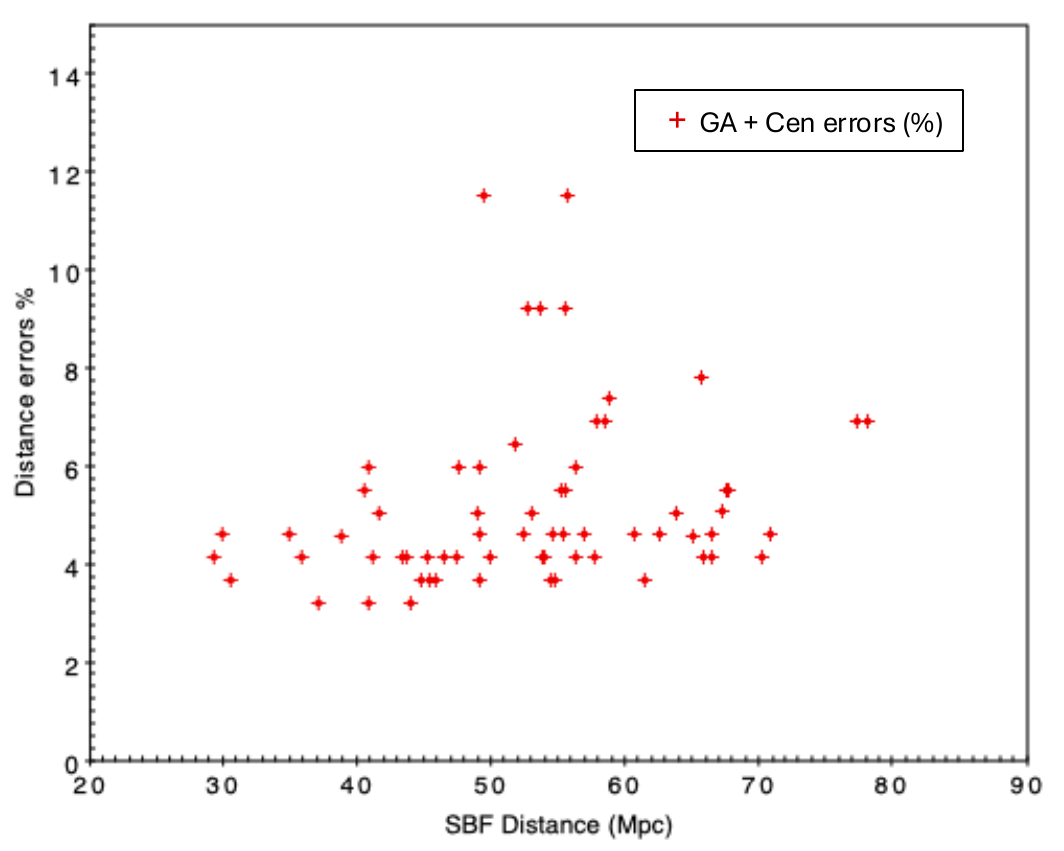}

\caption{Errors in the \emph{SBF}-Distance measurement for sample galaxies. A comparison of the distance for each of the regions `c0' and `c1' records the ability  of \SBF\ to compute distances over a wide range of surface brightness, only possible because both the galaxy profile and the Poisson fluctuations for \SBF\ are linked by the number of stars in each seeing-resolved patch.}

\label{fig:Distance_Errors}
\end{figure}

We show in Figure~\ref{fig:Distance_Errors} the distance error from the comparison of `c0' and `c1' for each galaxy in our sample.  80\% of the galaxies have distance errors between 3\% and 6\%. This `feature' make the distances derived from surface-brightness-fluctuations the most accurate of any technique relying on a galaxy's stars.

\section{Results: \SBF\ Distances to 66 Great Attractor Galaxies}
\label{sec:Results}

\begin{figure*}
\centering

\includegraphics[scale = 0.60]
 {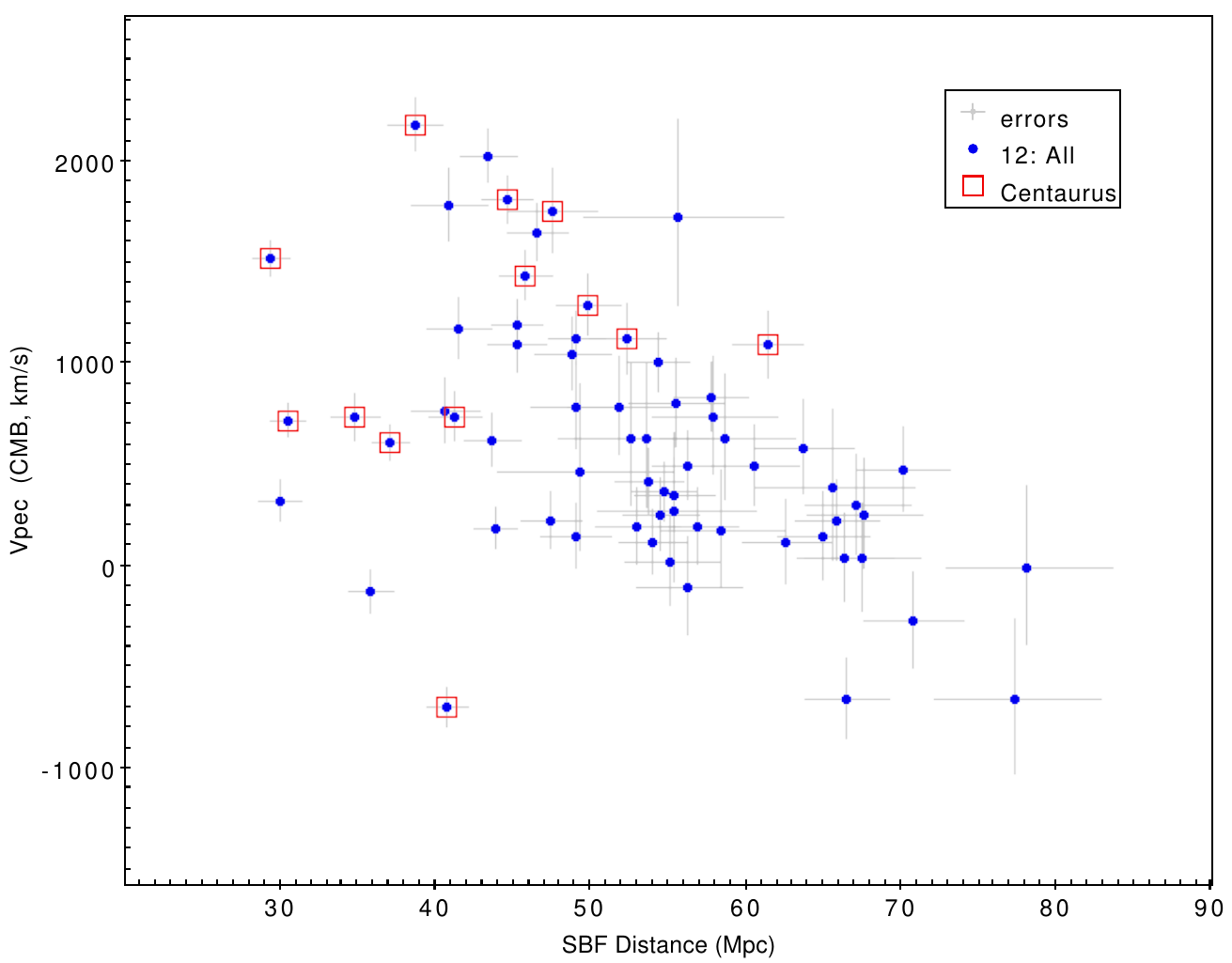}

\caption{The plot of \emph{SBF}-Distance versus $V_{pec}$ for the full 66 galaxy sample.  Error bars are `real' --- derived from the data themselves (see Section~\ref{sec:SBF_unique}).  The red boxes are Centaurus galaxies, 11/13 within a 20-square-deg area on the sky.  The `noise' in $V_{pec}$ they display is evidence for a deep gravitation potential --- one or more clusters, or a filament mainly along the line-of-sight.  The trend of falling peculiar velocities, $40<D<60$ Mpc seen in the 7S data and in \citet{1990ApJ...354...18B} is confirmed beyond doubt.}

\label{fig:GA+Cen_all}
\end{figure*}

Results from the data obtained and processed in this study are shown in Figure~\ref{fig:GA+Cen_all}, a plot of the distance to each of the 66 galaxies (blue dots) in the GA region (Figure~\ref{fig:GA-Sky}) versus their peculiar velocities (in the CMB reference frame) for H$_0$ = 72 km sec$^{-1} Mpc^{-1}$.  The light gray error bars of $\sim$5\% were calculated, galaxy-by-galaxy, as described in the previous section. Also in Figure~\ref{fig:GA+Cen_all} are blue dots within a red square box: these are the 13 Centaurus galaxies.\footnote{E322-G008, E268-g004, E322-G056, E322-G066, E322-G075, E322-G091, E322-G101, E323-G003, E323-G011, E323-g016, E323-G023, E323-G036, E323-G063)} 11 of the 13 are confined to a region of only $\sim$20 sq deg in the Centaurus constellation, compared to an area of$\sim$4000 sq deg for the GA area --- less than 1\%.

The remaining 53 galaxies cover the entire region of Figure~\ref{fig:GA-Sky}. Those with distance 40-50 Mpc have peculiar velocities ranging from $500-1500 \kms$ at their \SBF\ distances, compared to the Local Group's peculiar motion (CMB) of $\sim$600\kms.  This shows clear acceleration (infall) towards the Great Attractor center. Beyond a distance of $\sim$50 Mpc peculiar velocities clearly decline, at $\sim$70 Mpc reaching a point where galaxies are essentially \emph{at rest} with respect to the CMB. This is the same distance `center' seen in Figure~\ref{fig:SPS_vel_hist} --- the plot of the number of SPS (GA) galaxies with distance --- when corrected for incompleteness. Focusing on the 4 most distant of the 53, there is even evidence for negative peculiar velocities. This \emph{might be} GA \emph{backfall}, something that has remained elusive --- beyond the reach of the `Fundamental Plane' and `Tully-Fisher' methods that have $\sim$20-25\% accuracy (see Section~\ref{sec:GA-Mapping}. 

{These four `most-distant' \SBF\ distances and peculiar velocities average to $D_{SBF}$ = 74.08 $+/-$ 2.11 \Mpc, $V_{pec}$ = -122 $+/-$ 238 \kms\ (error in the standard deviation) --- plotted as a green cross in Figure 12 (see Appendix A).  Therefore, this `zero-crossing' is significant evidence that the Great Attractor is in fact the primary source of the CMB dipole: a larger-scale bulk flow could not produce this result! Evidence for such further contributions, like the Shapley Supercluster (aka \emph{Giant Attractor}) \citet{1989Natur.338..562S}) should show up in studies that extend measurements of $V_{pec}$ to $\sim$100 Mpc --- dozens of \SBF\ distance measurements --- within reach of \SBF\ imaging with \emph{JWST} --- would be decisive.  However, we consider the search for ``backside infall" to be somewhat of a \emph{red herring}, since it is the zero-crossing that secures the GA's priority as a CMB-dipole generator.}

\begin{figure*}
\centering

\includegraphics[scale = 0.60]
 {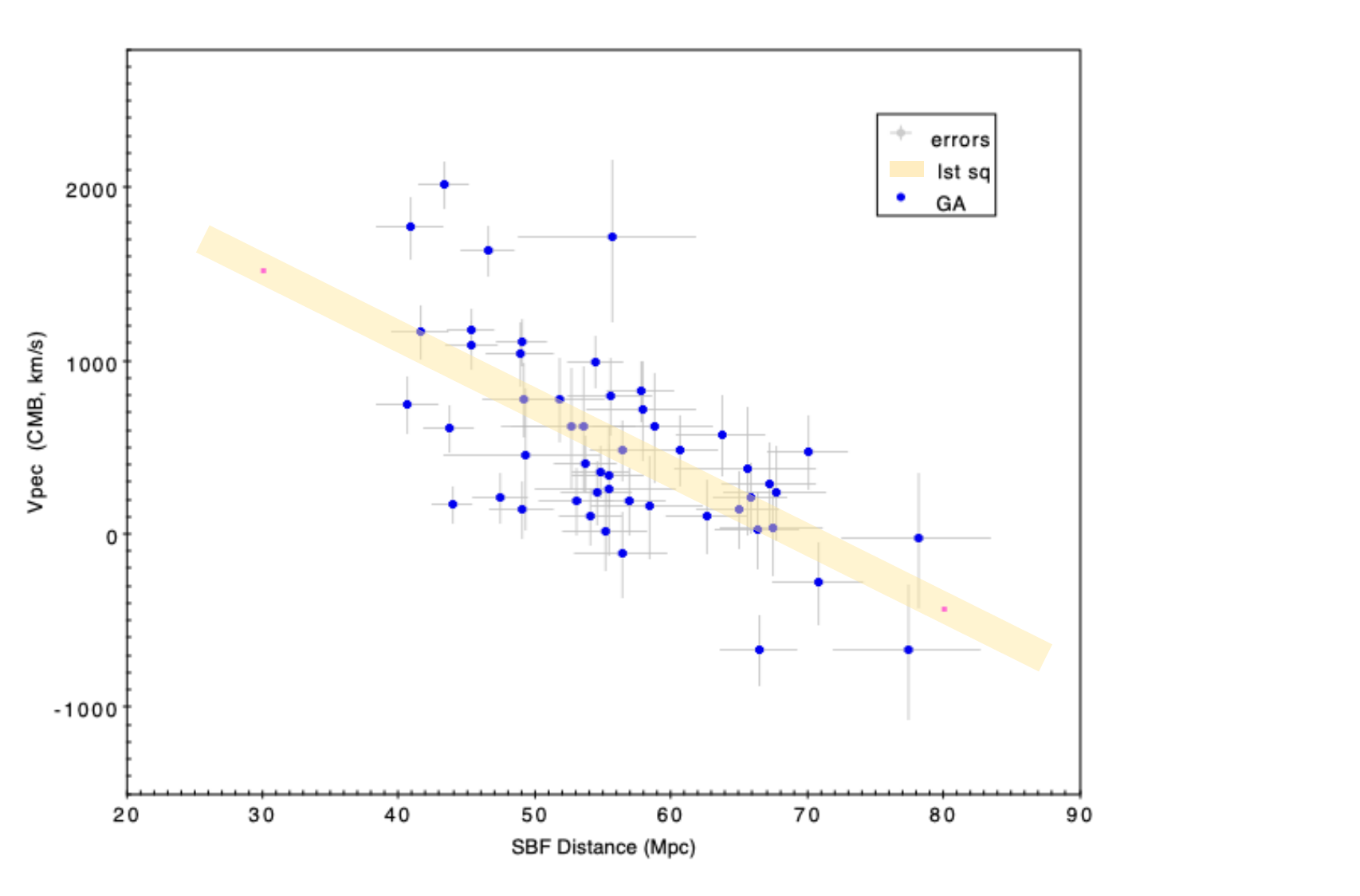}
 
\caption{The main result of this paper: The \emph{SBF}-Distance vs $V_{pec}$ for the 53 GA galaxies (Centaurus excluded).  The new data confirm \citet{1990ApJ...354...13D} in finding a peculiar velocities of $\sim$1000 \kms\ --- with a factor of $\sim$4 lower noise, clearly declining beyond R$\sim$45~Mpc and reaching $\sim$zero at R$\sim$70~Mpc.}

\label{GA-sample_D-vs-Vpec+LSQ}

\end{figure*}

The 13 peculiar velocities labeled `Centaurus' in Figure~\ref{fig:GA-Sky} are crowded into a very small area.  For decades these galaxies have been identified as part of a rich cluster of galaxies, or perhaps two clusters falling together, that account for the wide range of peculiar motions.  But the accurate \SBF\ distances, ranging from 30-60 Mpc, suggest instead a rich \emph{filament} undergoing tidal disruption by the GA, of a region that had perhaps been \emph{destined} to be a single cluster.

{Because our aim was to characterize peculiar motions over a steradian-sized area of the GA, including these Centaurus galaxies would be counterproductive: clearly their dynamics are not of a ``large-scale flow."  Appendix A shows how the best fit of the motion of these 13 alone is diverging rather than converging. Nor could a single point represent the average flow at the Centaurus barycenter: the very large dispersion confirms that dynamics in Centaurus track its own dark matter. Finally, even if the Centaurus galaxies were \emph{included}, the combined flow shows only a small difference in slope, and \emph{no change the decline in peculiar motion toward zero} --- our main result.}  

It is now more clear that Centaurus is only a minor player in the Great Attractor story, a confusion that has unfortunately misled several studies that did not grasp what and \emph{where} the GA is, and therefore failed to cover its full extent, to sad effect --- \eg \citet{2000ApJ...530..625T, 2003A&A...410..445M}. 

Figure~\ref{GA-sample_D-vs-Vpec+LSQ} plots \emph{SBF}-Distance \emph{vs.} peculiar velocities for the remaining 53 galaxies, showing a much more orderly distribution with Centaurus excluded: a strong decline in peculiar velocity with increasing distance, from $V_{pec}\sim$~1000--2000~\kms\ to approximately zero.  A least squares fit of $V_{pec} = -39.0 \times D_{SBF}$  + 2700 \kms\ confirms this steep slope with high certainty. $\chi^2$ is somewhat greater than 1, but well within what should be expected when the gravitationally perturbed motions \emph{within} the GA are considered.  The vast region in sky this GA covers must include galaxy-density fluctuations that themselves generate peculiar motions of hundreds of \kms. 

\begin{figure*}

\hspace{1.0in}  
\includegraphics[scale = 0.60]
 {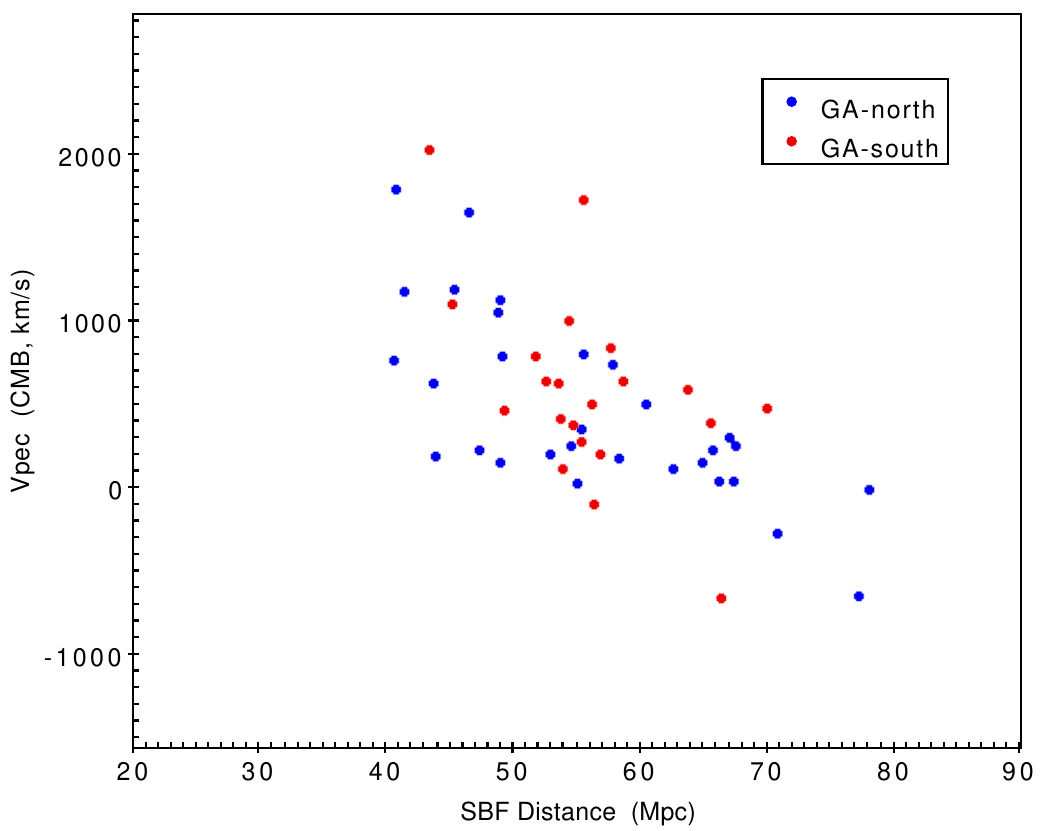}
    
\caption{Comparison of GA sample, for north and south Galactic latitude. (Centaurus and E184-G012,G014 omitted).  The main result of this paper -- the declining $V_{pec}$ with increasing distance -- is seen for galaxies both north and south of the Galactic plane. The GA is more than a sterdian-sized-swell in galaxy-density, a SUPERcluster.}

\label{fig:GA-north-south}

\end{figure*}

Figure~\ref{fig:GA-north-south} is a final demonstration that the GA is a coherent structure that extends over at least a steradian of the sky. The sample of Figure~\ref{GA-sample_D-vs-Vpec+LSQ} is split into the galaxies north and south of the Galactic plane.  It is clear that the same gradient in $V_{pec}$ is seen in both samples, the north sample centered at Galactic latitude $b\sim$~+30\deg\ and the south sample centered at $b\sim$~-25\deg\ (see Figure~\ref{fig:GA-Sky}).  The GA supercluster is a coherent, enormous structure in the local universe.

{Our goal here has been to show convincingly that the center of the GA at a distance of D~$\approx$~70 \Mpc\ that is moving slowly, if at all, with respect to the CMB. With this alone, we confirm that a overdense region of \emph{infalling galaxies} extends all the way to the Local Group.  {\bf This in turn means that the CMB dipole must be largely generated by the Great Attractor, the largest such `structure' within R~$\sim$~70 \Mpc\ of the Local Group, as Figures~\ref{fig:Lahav-sky} and Figure~\ref{fig:Erdogu_shells} have long suggested.}  Larger structures beyond might be moving all the galaxies \emph{within 70 Mpc of the GA center} to some extent, perhaps even to a contribution of a few hundred \kms, but as yet there is little evidence of this. What we have found and shown here runs counter to the idea that more-distant overdensities ({\eg Giant Attractor}) pulls the GA as a whole, and that they account for most of the CMB dipole signal.  If that were so, peculiar velocities in the region we mapped should be at least constant, and more likely rising rather than falling as we find.  A `bulk flow' of $\sim$1000\kms\ over an area of order 200 \Mpc\ diameter would be expected if even more distant, huge `attractors' were mainly responsible. As was thought to be the case for the GA itself in the 1980's, such a result would likely violate what is arguably the best understood feature of our universe: the CMB power spectrum.}

\section{Why \emph{NOT} the Great Attractor?}
\label{sec:NOT_GA}

{As early as the first 7S papers about large-scale streaming of galaxies in the local universe (\citet{Dressler1987a}, \citet{1988ApJ...326...19L}, \cite{1990ApJ...354...18B}, there was considerable skepticism within the cosmology community regarding the discovery of large amplitude peculiar (non-Hubble) flow generated by a huge ``local" ({\i.e.,}~within 100 Mpc or the Local Group) overdensity that had been largely unrecognized as a coherent structure, one substantially obscured by the dusty disk of our own Galaxy.  As discussed earlier, part of this was the preference for a ``critical density" universe that allowed only flows of lower amplitude and/or smaller extent, for example, the well-accepted flow of the Local Group and its neighbors toward the Virgo Supercluster --- comfortably within $\Omega_{\emph{m}}= 1 $  constraints.}

{Concerns were raised that this so-called \emph{great attractor} might be due to errors in measuring peculiar motions. (1) Could there be systematic changes of zero-points in distance estimators in different environments, or errors in Galactic absorption? (2) Were the results for this \emph{great attractor} model consistent with galaxy motions within and between other galaxy concentrations? (3) Was this \emph{great attractor} the same for different kinds galaxies, \eg spirals and ellipticals?  \cite{1990ApJ...354...18B} investigated these issues and dismissed them: not only were detections of large peculiar motions robust, but they had also pointed to a major, previously unrecognized, mass concentration.}

{Another major objection was the magnitude and size of coherent motions in the Great Attractor and its claimed contradiction to the favored cosmological model. As previously mentioned, this problem faded away as the cosmological model changed to an $\Omega_{\emph{m}} \sim 0.20 $  universe.}

{In \S{VII} of Burstein \etal, 7S proponents puzzled that the lack of enthusiasm for the Great Attractor extended as far as abandoning the orthodoxy that the Cosmic Background Radiation, CBR, is a standard-of-rest for the local universe.  After all, adding a carefully directed arbitrary velocity vector of 300-400\kms\ to the CMB dipole, would slow the large-scale GA flow, bringing it to agreement with the favored cosmology. But, Burstein et al. noted that this would include non-zero amplitudes of high-order moments (\eg quadrupole) of the CBR (not seen), and/or very special initial conditions or special location for the Local Group (unpopular) in the early universe. Moreover, why then would measured gravity-induced motions match --- even to an in order-of-magnitude --- the velocity inferred from the CMB dipole anisotropy, and why should its direction closely match the Great Attractor flow and point at the largest, nearest, and most significant overdensity in the nearby universe --- achieved by chance only 7\% of the time.  Finally, other early evidence, including the early detection of a possible decline in the Great Attractor flow beyond a distance of 40 \Mpc\ from the Local Group, along with other distant samples, confirmed the idea that a single large perturber was mainly responsible.}   

{Yes.~These ``coincidences" and ``oddities" could avoided by abandoning the CBR as the standard or rest, but this would ``generate far more questions than answers," they observed.}   

{In following years, the SPS survey and larger sampling of the flow toward the Great Attractor firmed up all this evidence, and the opportunity to simply dislike or disbelieve dried up.  In its place, astronomer skeptics looked far beyond for further ``Attractors" as the source of the CMB dipole, and physics skeptics invented dozens of explanations for why the CMB dipole was in fact not ``peculiar motion generated.''}

{In this paper we add a critical piece of evidence: reaching the CMB-dipole-velocity $\approx$ zero for the GA center.  Many more, and better observations can and should be made, but it is our belief that they can only improve our knowledge and understanding --- and not substantially revise it.}

\section{Mapping Cosmic Flows Near and Beyond the GA}

\label{sec:Mapping_Cosmic_Flows}

\subsection{Extensive Mapping of Peculiar velocities and Large-scale Mass concentrations in the 1990s}

{In their pioneering study of 96 Sc I spiral galaxies around the sky, {\citet{1976AJ.....81..719R} suggested a substantial departure from pure Hubble --- a `peculiar velocity' of $\sim$450 \kms\ of the Local Group.  That result rested on assuming the spiral galaxies were ``standard candles,' but with better tools --- distances to spirals via the Tully-Fisher relationship and to ellipticals via the Fundamental Plane --- such data became critical for studies of cosmology \citep{2022-book}. The demonstration by {\citet{Aaronson1982} of infall into the Virgo Supercluster via TF, and the 7S team's full-sky mapping of FP distances to elliptical and S0 galaxies, opened the door to many studies in the 1990's. Guided by a remarkable review by \cite{1995PhR...261..271S} of the formidable power of redshift and peculiar velocity surveys for cosmological theory, these programs aimed to learn more about large-scale galaxy flows out to the GA scale of $\sim$80 \Mpc\ ($\sim$6000 \kms), and eventually to more ambitions programs out to 150 Mpc and beyond. The thorough, lucid discussion in \S{7.1} of Strauss and Willick provides additional perspective and detail for the 7S work and similar-depth `flow studies' in other regions of the sky at R~$\lesssim$~80 \kpc, a sample of which we now describe.}

{Around the time of \citet{Dressler1987a} and \citet{1988ApJ...326...19L}, other programs were measuring peculiar motions in different parts of the sky.  An interesting early example was the \citet{1982Natur.297..191H} study that measured distances for only 84 spiral galaxies.  It led them conclude that the Local Group's motion approximately matched the CMB Dipole direction and amplitude, that it must arise `locally' rather than from some distant gravitational pull, and that $\Omega_{matter}$ was between 0.15 and 0.50, not the value of 1.0 that was prevalent in the world of physics at that time. Even now, these conclusions are not fully accepted, but they may all be correct.} 

{Many groups focused on the infall to the Virgo Supercluster, in this way comparing different distant-estimator techniques and their sensitivities to galaxy samples and to various methods of data reduction and analysis \citep{1980AJ.....85..801S,1981ApJ...246..680T, 1981ApJ...248..395D, 1985ApJ...294...81T, 1984ApJ...281..512D}. Values for Virgo infall ranged from 200 -- 450 \kms\ --- considerably less, and in a markedly different direction, from the CMB dipole velocity of the Local Group. Several studies  noted that maps of Local Group infall to the Virgo Supercluster implied a larger contribution from the Hydra-Centaurus Supercluster  \citet{1984ApJ...280..470S, 1986ApJ...307...91L} to match the velocity implied by the CMB dipole.} 

{
Fortuitously, the all-sky 7S study included this region.  Its original aim was not to measure peculiar motions, but its ability to do so made it possible to expand this important work of understanding the source of the CMB diplole.}

{SPS studies by Dressler and Faber in the early 1990's that followed 7S were the first to map a particular region --- the GA, with hope of progress through a dense sampling in the one streaming flow --- possibly converging --- in the approximate direction of the CMB dipole. As we reviewed in Section~\ref{sec:GA-Mapping}, TF and FP distances for 253 GA targets in a new denser sampling of E, S0, and spiral galaxies within $D \lappr$50 \Mpc, strongly confirmed the 7S result of a large-scale flow of peculiar velocities of $\sim$1000~\kms.  However, it was also clear that --- for galaxies with predicted distances D$\gappr$50 \Mpc\ --- the limitations of distance estimators with $\sim$20-25\% accuracy (here, FP and TF) prevented any reliable measurement of `convergence' --- the crucial `zero-crossover' --- let alone `backside infall.'} 

{
In short order a study of the peculiar motions of 1355 spiral galaxies, 657 of them in the the GA area, by \citet{1992ApJ...389L...5M} provided a check and challenge for \citet{1990ApJ...354L..45D}.  Their study confirmed the outflow velocity of $\sim$1000 \kms\ for galaxies D$\lappr50$ Mpc that defined the Great Attractor flow, but took issue with the claim that backslide infall had been detected.  Though a previous paper \citep{1990ApJ...354...13D} had been cautious about the reality of the ``S-wave" shape, owing to concerns of larger random errors and the likelihood of systematic ones (\eg Malmquist bias), \citet{1990ApJ...354L..45D} went too far, as  Mathewson et al. said.  In this paper we have demonstrated a zero-velocity `crossover' \emph{is} seen at the much greater distance of D $\approx$ 70 \Mpc, owing the much greater precision of \SBF\ distances. However, we note that the peculiar velocities in Figure 1a of Mathewson \etal\ look basically the same as in Dressler and Faber's Figure 2a: the error of that paper was due to misinterpretation, rather than bad data.}

{
More critically, the \citet{1994ApJ...434L..39M} study of 2400 southern spirals measured flow toward the GA and to an orthogonal control sample that reached to 11,000 \kms, concluding ``The flow cannot be modeled by a Great Attractor at 4300 \kms\ or the Centaurus clusters at 3500~\kms\ " and ``...the density maps derived from the redshift surveys of `optical' and IRAS galaxies" bring into question ``the idea that peculiar motions are infall into dense regions." On the contrary, Mathewson \etal\ claimed they may have detected flow ``to the north of GA to their redshift limit of 8000 \kms\ to the Great Wall" (!).  These claims were based on using a $\sim$25\% distance indicator at great distances, \emph{averaging} peculiar motions to `beat down the noise.' We believe this approach is likely to fail when individual measurements have very low S/N.}\footnote{That is, when S/N $\equiv$ (expected-peculiar-velocity) / (distance in \kms) $\lappr$ 1.0, as opposed to incorrectly defining S/N $\lesssim$ (method-distance) divided by (method-error)  $>>$ 1. The former are subject to large systematics errors that cannot be overcome by very large samples.}

{
\citet{1990ApJ...351L...5W} focused his study of 320 spirals galaxies with Tully-Fisher distances and concluded that ``the principal feature of the velocity field in the Perseus-Pisces region is a coherent streaming toward the Local Group...similar in amplitude and direction to those of elliptical galaxies in the southern sky, which have large positive peculiar velocities to distances of at least 4000 \kms, suggesting a full scale of $\sim$100 \Mpc\ for the Great Attractor."  However, Willick also speculated that ``It is unlikely that coherent motion on this scale can arise entirely from the Great Attractor; rather, it may be due in part to low-amplitude density fluctuations on much larger scales."  Supporting this was a following study by \citet{1993ApJ...412L..51C} that used a much larger sample of spirals to argue that a significant fraction of Perseus-Pisces streaming of $\sim$360 \kms\ arises from galaxies within a sphere out to at least V~$\sim$ 6000 \kms\ in radius around the Local Group.}

{
Continuing in this vein, \citet{1997MNRAS.291..488H} made FP peculiar velocity measurements for 103 galaxies in 7 rich clusters in Perseus-Pisces and 249 galaxies in 9 widely-distributed clusters from the literature, finding ``that the full 16-cluster sample has a mean CMB-frame bulk motion of 420 ± 280 \kms...consistent both with no bulk motion in the CMB frame and with the $\sim$350 \kms\ bulk motion found by \citet{1993ApJ...412L..51C}."}

{
Other studies specifically measured `bulk flow' over the
full scale of the GA. \citet{1998ApJ...505L..91G} observed 1289 field galaxies with 21-cm TF distances.  Out to $\sim$2000 \kms, the measured peculiar velocities agreed with previous studies that found flows into the Virgo Supercluster, but out to $\sim$6000 \kms\ ``the LG reflex motion asymptotically approaches $V_{CMB}$, both in amplitude and apex direction."  \citet{1998AJ....116.2632G} carried out an analogous study by measuring distances to galaxies in 24 clusters with radial velocities between 1000 and 9200 \kms. When measured in a reference frame in which the Local Group is at rest, ``the clusters at cz $>$ 3000 \kms\ exhibit a dipole moment in agreement with that of the CMB,'' both in amplitude and apex direction, ``while the bulk flow of a sphere of 6000 \kms\ radius in the CMB reference frame is between 140 and 320 \kms."  

{Finally, the TF project \emph{SHELLFLOW} by \citet{2000ApJ...544..636C} assembled a sample of 276 Sb-Sc galaxies distributed over a shell of R $\approx $ 4500-7000\kms, with a special focus of minimizing systematic errors --- for example, between observing runs and telescopes and spectrographs --- that might produce spurious bulk flows. Their measurement of $V_{bulk}$ = 70$_{-70} ^{+100}$ \kms\ strongly indicated that the shell was `at rest' --- V $<$ 300 \kms\ at the 95\% confidence level, and further suggested that higher bulk flows found in some earlier studies might indeed have been the result of non-uniformities in their subsamples. Best of all, the placement of the shell indicated that \emph{convergence} of any local flow had been reached, at V $\sim$ 6000 \kms\ (R  $\sim80$ Mpc) compared to our \SBF\ result of convergence at R $\approx$ Mpc --- very good agreement from two very different methods and independent results}.

{
Based on these many measurements of low bulk velocity for \emph{entire} GA-sized spheres centered on the Local Group\footnote{dominating the strong flow in the steradian-sized 'corridor'.} it is sensible to conclude that the CMB dipole amplitude of $\sim$600 \kms\ arises `locally,' a notion that --- for many --- had prevailed for a decade. Nevertheless, more ambitious programs became the focus of the field --- complex and difficult, over larger volumes and beyond the local sphere --- to explore the possibility that it does not.}

{
\subsection{Pushing Beyond the GA using 20-25\% distance estimators} }
\label{sec:Mapping_Further}

{
Many studies in the 1990's succeeded in sampling the GA ``sphere" --- inside the GA itself and out to $\sim$8000 \kms, a sample of which we will now describe. The ``out of the GA'' sampling in particular (Fig~\ref{fig:In_out_GA}) showed that --- with only the FP or TF methods --- there would be little reason to begin new flow studies at a distance comparable to  that of the GA center.  Also, at distances D$\gappr$50 Mpc, results became suspect, as systematic errors associated with the large random errors at such distances --- for example, measurement errors, and the Malmquist effect --- would likely challenge and frustrate the endeavor.}

{
Although the Great Attractor model was more-or-less confirmed by these 1990's studies, questions remained about contributions to ``local" flows from galaxies, clusters, or superclusters beyond $\sim$100 Mpc. For this reason also, we think, there was little follow-up of what was (and still is) the most promising `generator' of the CMB dipole. Instead --- and ironically --- attention turned to more difficult tasks of extending distance measurements hundreds of megaparsecs to find more distant flows that might be significant contributors to the CMB dipole.}  

{
Today, with the possibility of \SBF\ distance measurements  for the entire local volume, there is potential to replace earlier measurements of peculiar motions with ones accurate to $\lappr$300 \kms\, with the hope of settling the matter altogether.  But, for distances well beyond the local volume, even \SBF\ would be challenged.}

{
Many of the studies reviewed in the previous section came to similar, but quantitatively different conclusions about the role the GA plays in generating the CMB Dipole, namely, that it was responsible for a large fraction, but that equal or even more of the dipole comes from more distant overdensities.  This has encouraged researchers to push further out, to hundreds of megaparsecs, to look for direct evidence that large, more massive overdensities --- for example, the Shapley concentration --- contribute significantly to the CMB dipole.}

{
Beginning in the mid-1990's, researchers studying cosmology via galaxy distributions developed new methods for making peculiar velocity maps through less direct measurements of peculiar motions. One alternative has been to average measurements of galaxy distances by rules that incorporate them into larger structures, including corrections due to selection for linewidth, inclination, Malmquist effect, etc., and throwing out discrepant points in an iterative procedure. Smoothing the map is a way of lowering the spatial resolution, but also lowering the errors in peculiar velocity through the combination of distance measurements that have mostly random errors.  Another technique involves using redshift maps or other galaxy-counting and mass-weighting schemes to find mass fluctuations (also a combination of discreet measurements) and to derive the resulting peculiar motions by determining expected accelerations.}

{
For example, an all-sky, deeply sampled survey study by \citet{2007A&A...465...71T} included with 3126 \emph{spiral galaxies} out to V$~\approx$ 8000 \kms\ and covering the full sky. Tully-Fisher distances based on radio observations measured linewidth --- the original TF technique.  This study went deeper than 7S and SPS, but it too showed substantial deterioration in distance accuracy at the same V $\sim$ 5000 \kms\ (see their Figure 10); these problems were mitigated by careful association and averaging of peculiar motions. With \2MASS near-IR photometry and radio spectra to measure line-widths, Theureau \etal\ made maps of peculiar motions around the sky that show the locations of galaxies and adjacent areas of inflow and outflows, indicating how well these structures are `detected' by gravity-induced flows. For example, Figure 12b shows a projection of the Great Attractor flow (in agreement with 7S and SPS) and, on the other side of the Local Group, the Perseus-Pisces (P-P) Supercluster (see Figure~\ref{fig:Erdogu_shells} of this study) also showed strong infall (in agreement with and the P-P mapping described earlier), all using with very different data, analysis, and method. The gravitational pull of P-P, only about 20\% further from the Local Group than the GA, is seen largely screened by a well-known void in its direction (also seen Figure 12b of Theureau \etal).  The void `screens' the effect of P-P from the Local Group, resulting in the CMB-dipole pointing in the other direction, instead of the GA and P-P in a `tug of war.' Mass concentrations are detected, cosmological parameters are extracted, and measurements of bulk flow are made.  Unfortunately, Theureau \etal\ did not attempt a large-scale modeling of the peculiar velocities, one that would predict the magnitude and the direction of contributions to the the CMB-dipole.}

{
In another example of this new approach the source of the CMB dipole is directly calculated from an analysis of the mass distribution of galaxies and dark matter out to $\sim$300 \Mpc.  \citet{2006ApJ...645.1043K} used X-ray selected clusters of galaxies used as a tracer of mass and adopted the methodology of using the dipole moment of a mass-tracer distribution to calculate the peculiar velocity that sample would induce on the Local Group. They measured the contributions to the dipole moment for this X-ray cluster sample, extending as far as 300 Mpc, but finding the highest amplitudes came at 40 and 150 Mpc --- roughly, the distance from the Local Group to the Great Attractor and further to the the Shapley Supercluster (SCC). In terms of the contribution to the Local Group’s motion reflected by the CMB dipole, this study concludes that 44\% of that signal is coming from the GA, with 30\% coming from the SCC and 26\% distributed over other distant galaxy over-densities. Although the GA plays a major part in Kocevski \& Ebeling's model of generating the CMB dipole, our direct measurement of the zero-crossing at $\sim$70 \Mpc\ is inconsistent with their finding that the majority of the CMB dipole arise from more distant overdensities.  Again, of course, their study does not actually measure peculiar velocities, but infers them from the distributions of galaxies, in this case, collected into clusters and selected by their X-ray  luminosities.}

{Because detailed study of flows even on the scale of 5000$\kms$ would require more accurate distance measurements, attention shifted --- somewhat paradoxically --- to investigating volumes many times larger, V$~\gappr$~15000\kms, to find other, more distant `attractors' --- especially those that might play a bigger role than the GA in the CMB Dipole.  The general approach is mapping galaxy motions over vast volumes to characterize ``bulk flows'' --- the magnitude and direction, in other words, its dipole --- with respect to the CMB.}

{
The range of issues and programs involved in understanding the variety, complexity, and reality of large-scale bulk flows, and their implications for cosmology, are legion.  Readers who are eager to follow the trail of this endeavor are referred to \cite{2025arXiv251005340T} --- an excellent place to pick up the scent.}

{
Alas, the results of the present paper mitigate against these ambitious and tempting excursions.  If Figure~\ref{GA-sample_D-vs-Vpec+LSQ} is not convincing that the CMB Dipole arises from local overdensities, then surely a more intense application of the \SBF\ method with \emph{JWST-NIRCam}, filling in other directions in the sky, could persuade.  For example, good \SBF\ distances towards Perseus-Pisces, to confirm the reported outflow via TF beyond the ``void," and towards the ``zero-peculiar-velocity" direction of the GA model, should be decisive, one way or the other.}

\vspace{0.2in}

\section{Conclusions and Future Prospects}
\label{sec:Conclusions and Future}

{
The Hubble diagram for the GA region that we present in this paper is the strongest evidence to-date for the theory that the CMB Dipole is largely a reflection of the flow of galaxies toward an extensive overdentiy of galaxies named the Great Attractor. The GA was first recognized in the late 1980's in an all-sky survey of elliptical and S0 galaxies for which `peculiar motions' were measured. In the intervening three decades, some work has been done to confirm that model, but more has been done to find something other that is harder to do and more difficult to explain. Unfortunately, progress along these lines has not cleared the air, rather it has muddied the waters, all the way to rejecting the very obvious conclusion that the CMB dipole is the result of gravity from `local' galaxies.}

{
We believe we have taken the penultimate step in proving that the CMB dipole anisotropy rises principally from local galaxy structures, in particular, overdensities that stretch from the Virgo Supercluster to a vast overdensity --- the `Great Attractor' whose center is at a distance of $\sim$70~Mpc and within 20\deg\ of the CMB dipole direction. Although substantial evidence was found decades ago, limitations on the accuracy of estimators of galaxy distances provided only persuasive evidence, and not everyone has been persuaded.  A newer method, \SBF, based on the ``standard candle" of TRGB stars as the major contributor to the light of early type galaxies, in particular --- elliptical and S0 galaxies --- is well-demonstrated as accurate to $\sim$5\%, verified internally by the fact that their light generates not only the decreasing luminosity profile of a galaxy, but also a ``point source fluctuation" signal that must simultaneously conform with the luminosity profile.  With a sample of 53 GA galaxies, we confirm the high-velocity flow of galaxies toward the GA center, rising from its value of $\sim$600\kms\ at the Local Group to over $\gappr1000$\kms\ halfway to the GA center and falling to approximately zero at the GA center: {\bf at rest with respect to the cosmic microwave background.}}

{
Because of long-continuing uncertainty about the role of the GA in producing the CMB Dipole anisotropy, alternative explanations and associated puzzles have emerged from studies of flows on much larger scales, as well as a large literature of physics-grounded models of the CMB dipole unrelated to large-scale structure   Therefore, it is important that the strong evidence for the GA model presented here is followed by further \SBF\ measurements throughout the remaining GA `sphere of influence.' A generous sampling should not only confirm the GA's dominant role, but also reveal a more detailed map of the many mass concentrations within.}

{
Fortunately, there is now a tool --- measuring \emph{surface bright fluctuations}, and vehicle -- the \emph{James Webb Space Telescope} --- that can finish the job. We hope and expect that this much improved method of measuring galaxy distances will resolve 35 years of conflicting observations and interpretations, maturing our understanding of
large-scale-structure and structure-growth in our local corner of the universe.}

\begin{acknowledgements}
    
Alan Dressler thanks Joe Jensen, Ed Ajhar, and John Blakeslee for helping me set up this program and providing the \SBF\ reduction programs, and for extensive tutoring on how to use them properly.  
AD is forever grateful to the staff and workforce at Las Campanas Observatory, for 50 years of remarkable support, guidance, and friendship in his years as a research scientist.  Carnegie's southern outpost to the heavens has been a lifetime experience all to itself, bringing a different perspective to the crowded, but isolated life most of us lead.  And, of course, AD acknowledges the Carnegie Institution of Washington, `nay,' of Science, `nay,' Carnegie Science, for offering a dream job --- a lifetime of exploration, searching for how it all began.

\end{acknowledgements}

\newpage
\clearpage

\begin{figure}
\centering

\includegraphics[scale = 0.90]{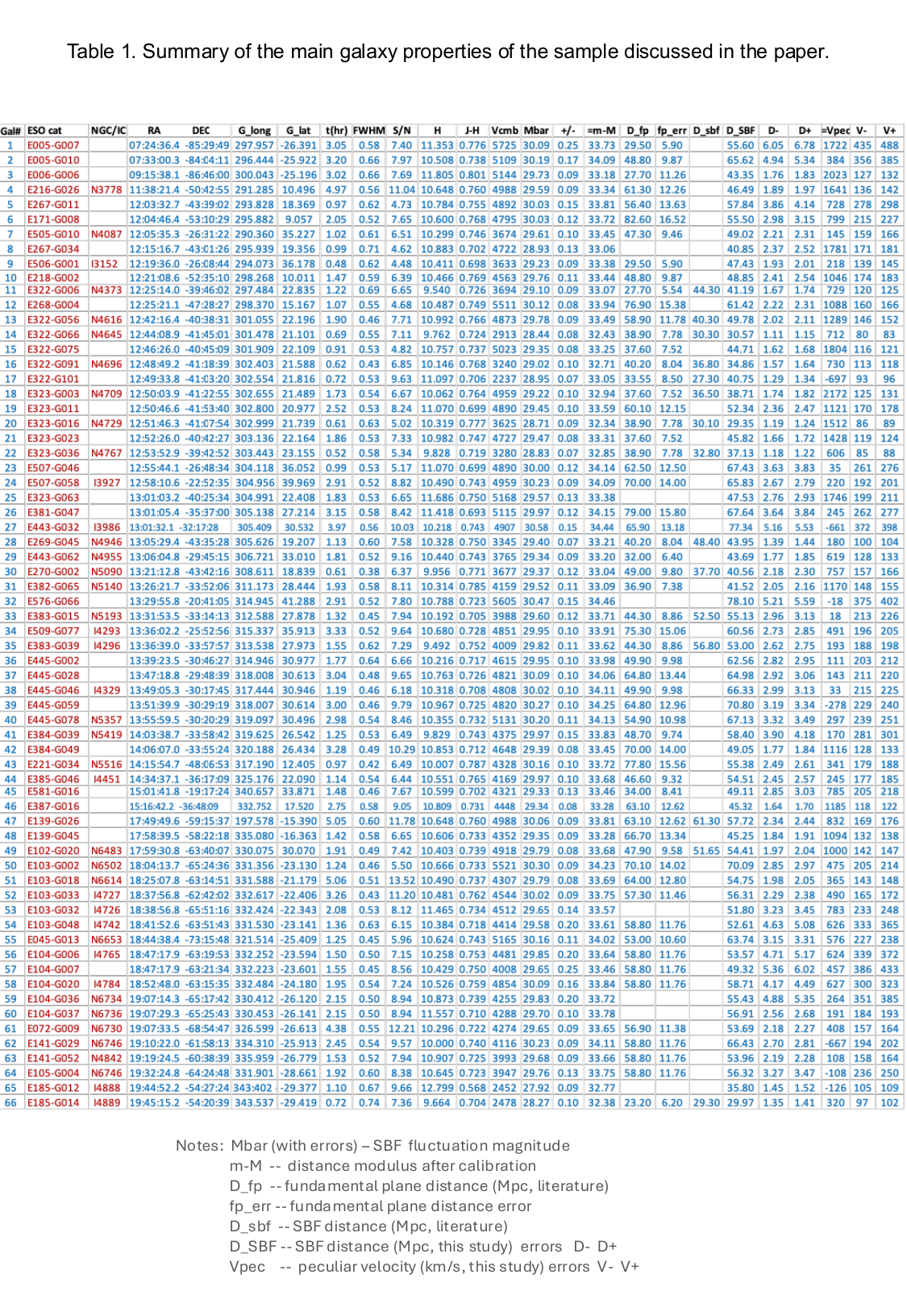}

\end{figure}

\clearpage

\appendix

\section{Exclusion of Centaurus galaxies}

\begin{figure}[h]

\centering

\includegraphics[width=1.0\linewidth]
{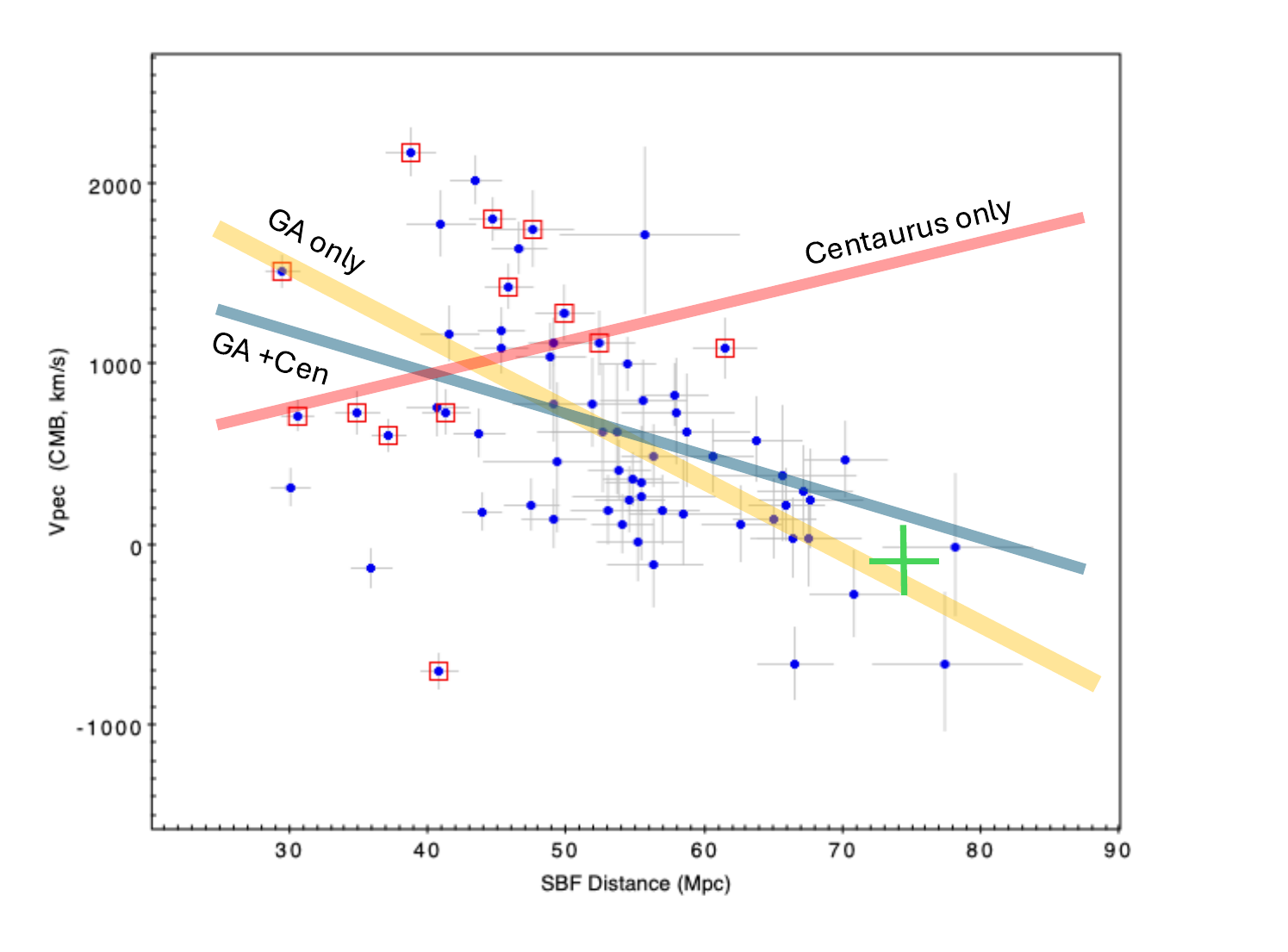}
 
\caption{Repeat of Figure 10 with Centaurus galaxies included in the analysis.}

\label{fig:Centaurus-GA_analysis.pdf}

\end{figure}

{
Figure 3 shows that a concentration of galaxies in Centaurus, traditionally described as one or two clusters, contribute $\sim$20\% to the sample studied here. However, inclusion in the large-scale flow over the entire GA region is problematical, both from the point of view of fairly sampling the steradian-area of the survey, and the problem in determining non-Hubble velocities for these galaxies.  Our project aimed at covering the entire area in Figure 3, so the sample galaxies are meant to be test particles of the large-scale velocity field.  Emphasizing this small area of the field --- less than 1\% of the area --- is counter to the goal of the program.  Of course, it would be acceptable to include Centaurus as a single peculiar velocity, but Figure 9 shows how these galaxies span more than 30 Mpc in depth and a velocity dispersion of order 1000 \kms: this is not consistent with a single or even a two-cluster structure, so no single or representative non-Hubble motion can be assigned.  This figure shows how --- with all 13 Centaurus galaxies added to the analysis --- the qualitative conclusions of this study do not change: the blue line shows the same trend as the yellow line for the 51 distributed GA-only galaxies.  But, the poorer fit to the peculiar velocities for D $>$ 60 Mpc (beyond all but one of the Centaurus galaxies) compromises the most important part of our sample: where the GA peculiar velocities are scattered around zero. Furthermore, the red line shows why this happens by fitting the just the 13 Centaurus galaxies: they contribute only a non-sensible \emph{rising} peculiar velocity for D $>$ 60 Mpc. However, we include the distances and velocities of the Centaurus galaxies  in Table 1, as they should be helpful in understanding the structure of the Centaurus overdensity, perhaps the most prominent \emph{filamentary} structure in our neighborhood.\ }

\vspace{1in}
\section{Photometric Calibration}

\begin{figure}[h]

\centering

\includegraphics[width=0.6\linewidth]
{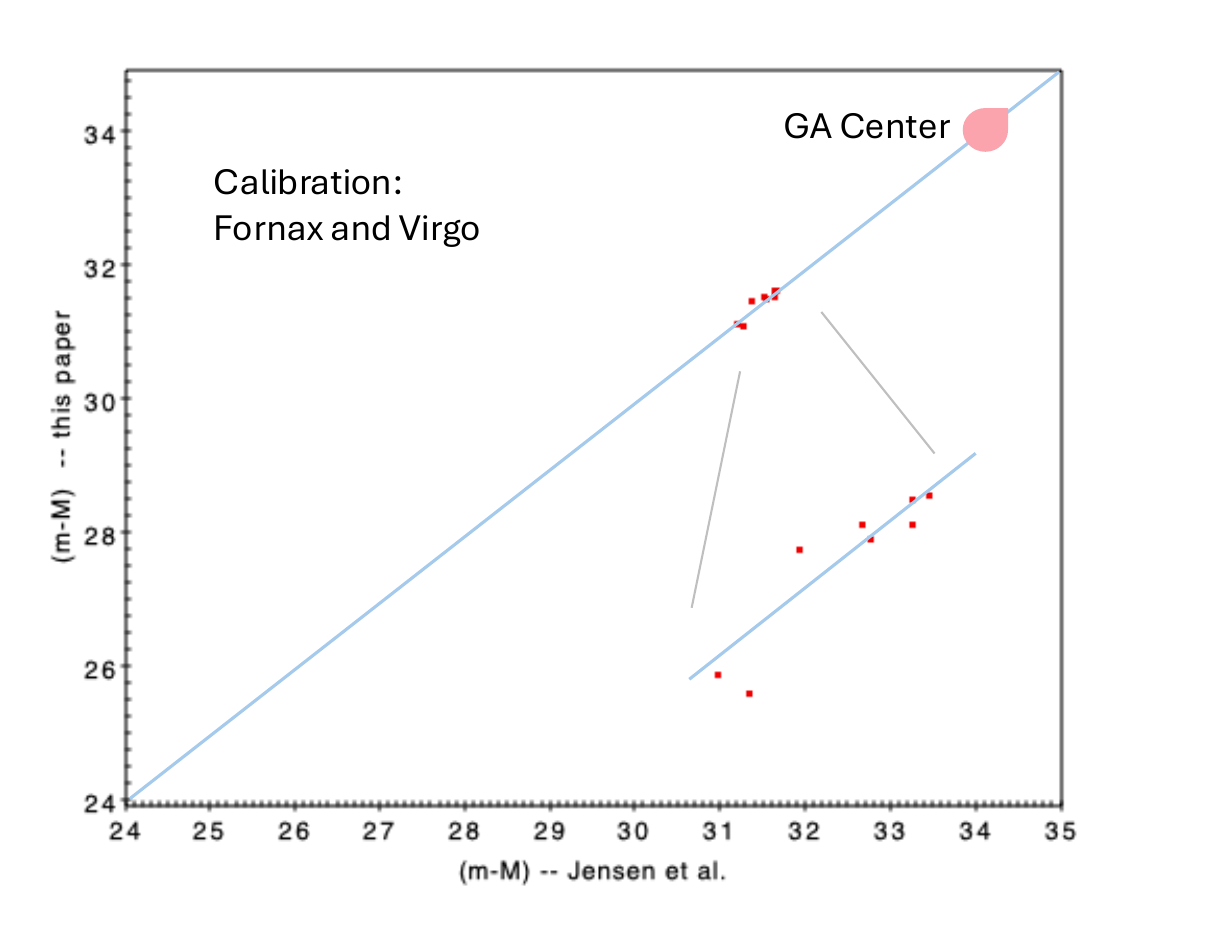}
 
 \caption{Galaxy distances from this study to 6 Fornax and 2 Virgo galaxies, compared to results of \citet{2015ApJ...808...91J}.  Photometric calibration within the sample was carried out for each observation using \emph{J}- \& \emph{H}- band photometry from the \2MASS Catalog (see Section~\ref{sec:Observations}).}

\label{fig:calibration_plot}

\end{figure}

\newpage

\begin{figure*}

\bigskip
\bigskip
\vspace{0.3in}
\centering{D. PROCESSED DATA IMAGES AND POWER SPECTRA}

\vspace{0.2in}

\includegraphics[scale = 0.185]
 {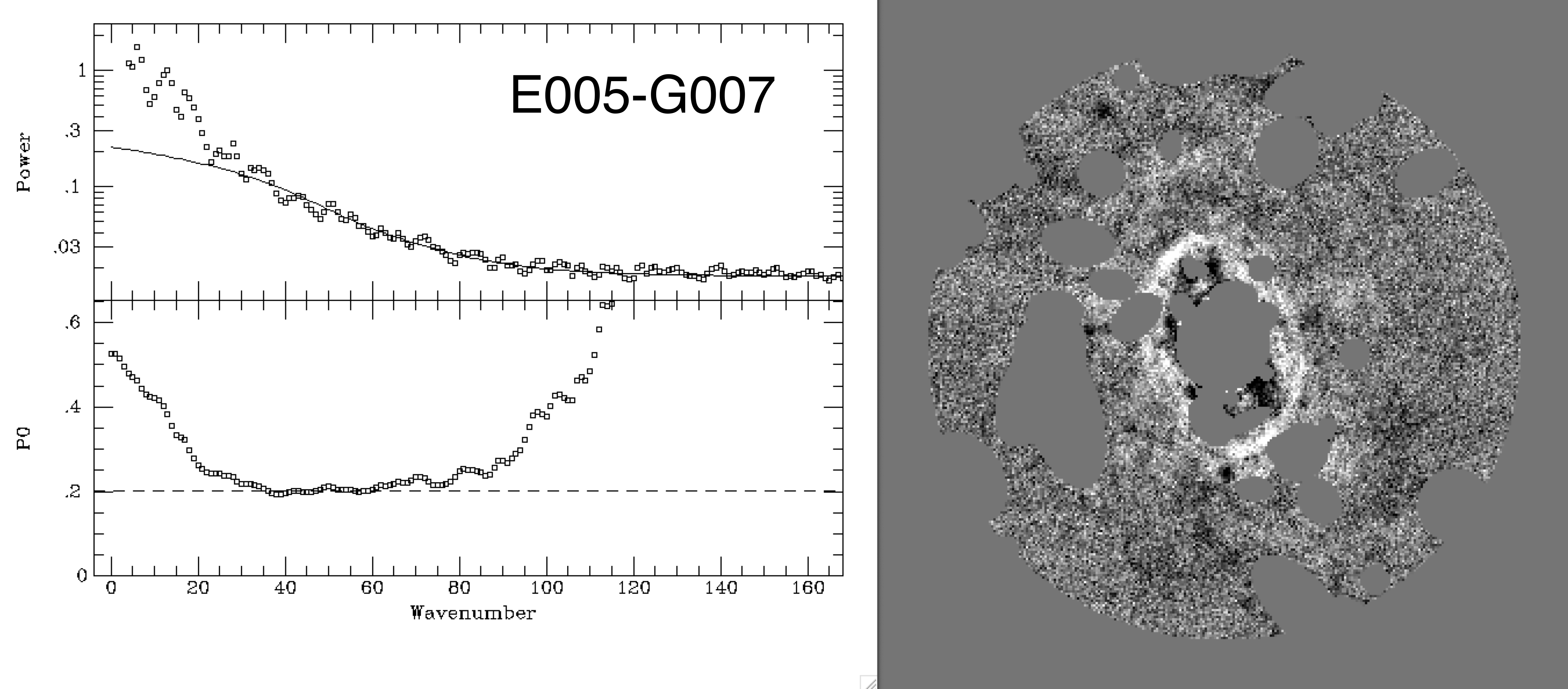}
\includegraphics[scale = 0.185]
 {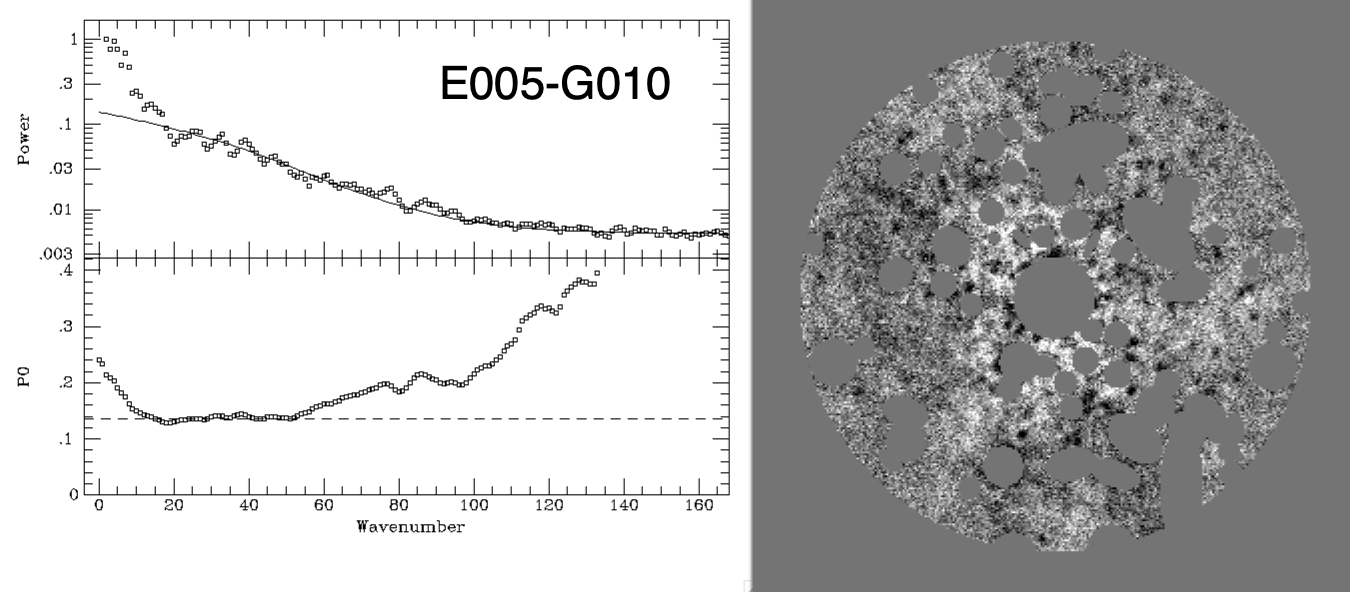}
\includegraphics[scale = 0.185]
 {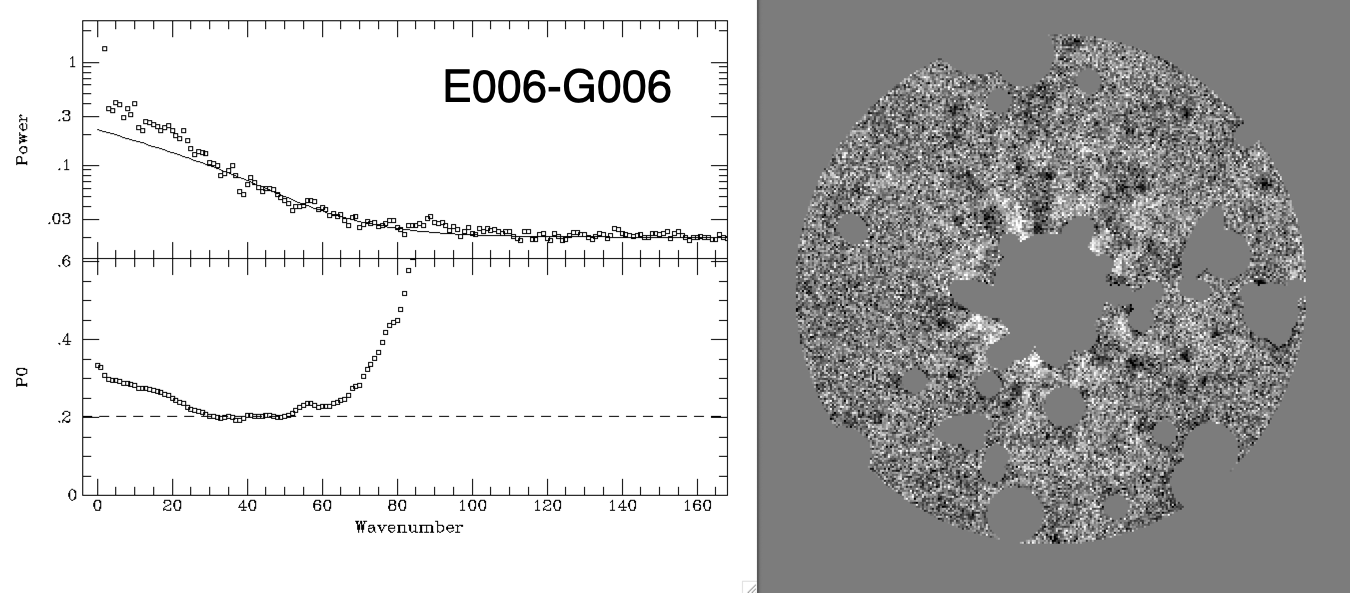}
\includegraphics[scale = 0.185]
 {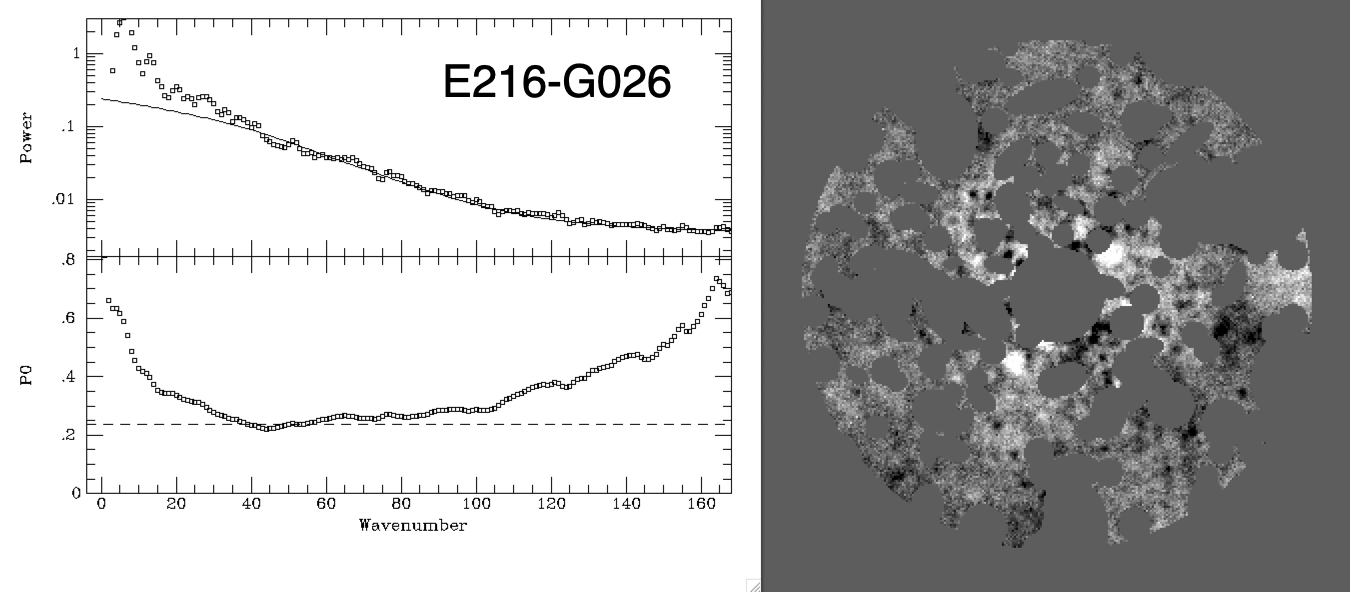}
\includegraphics[scale = 0.185]
 {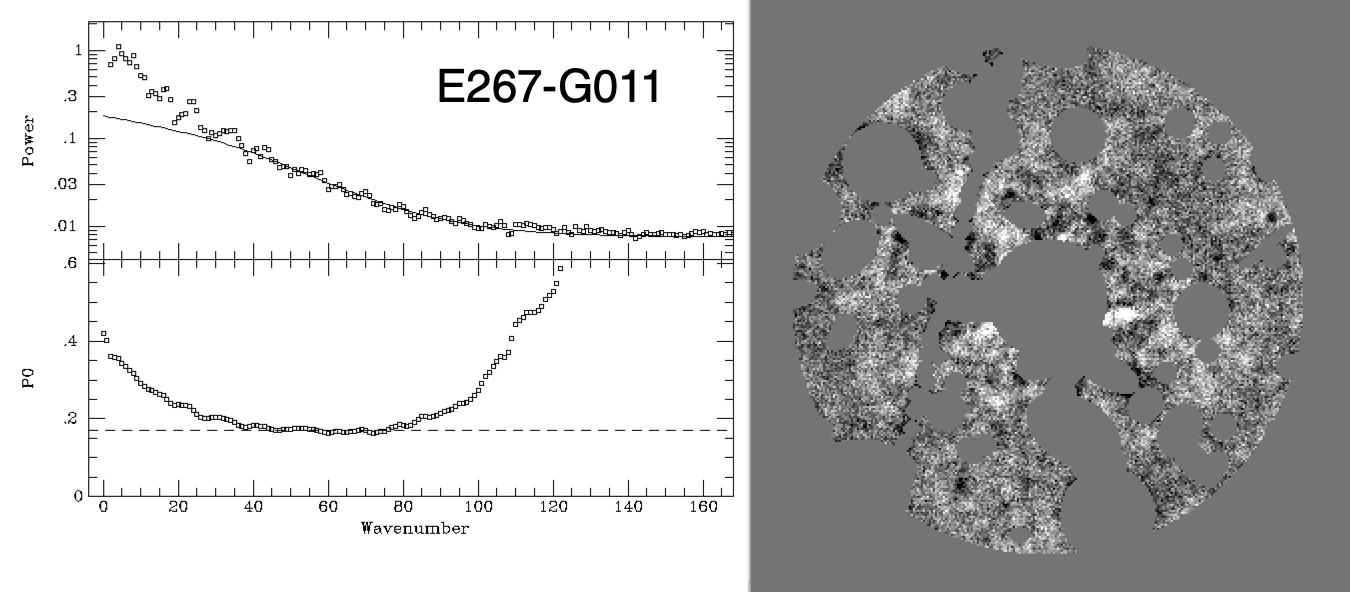}
\includegraphics[scale = 0.185]
 {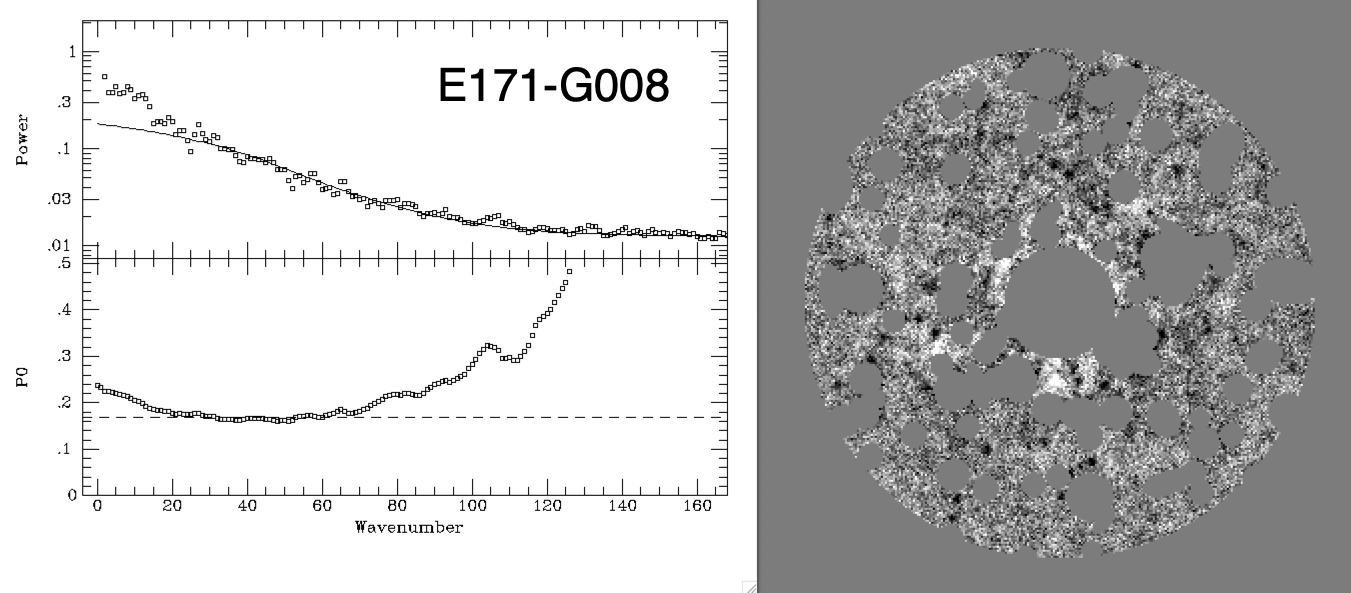}
\includegraphics[scale = 0.185]
{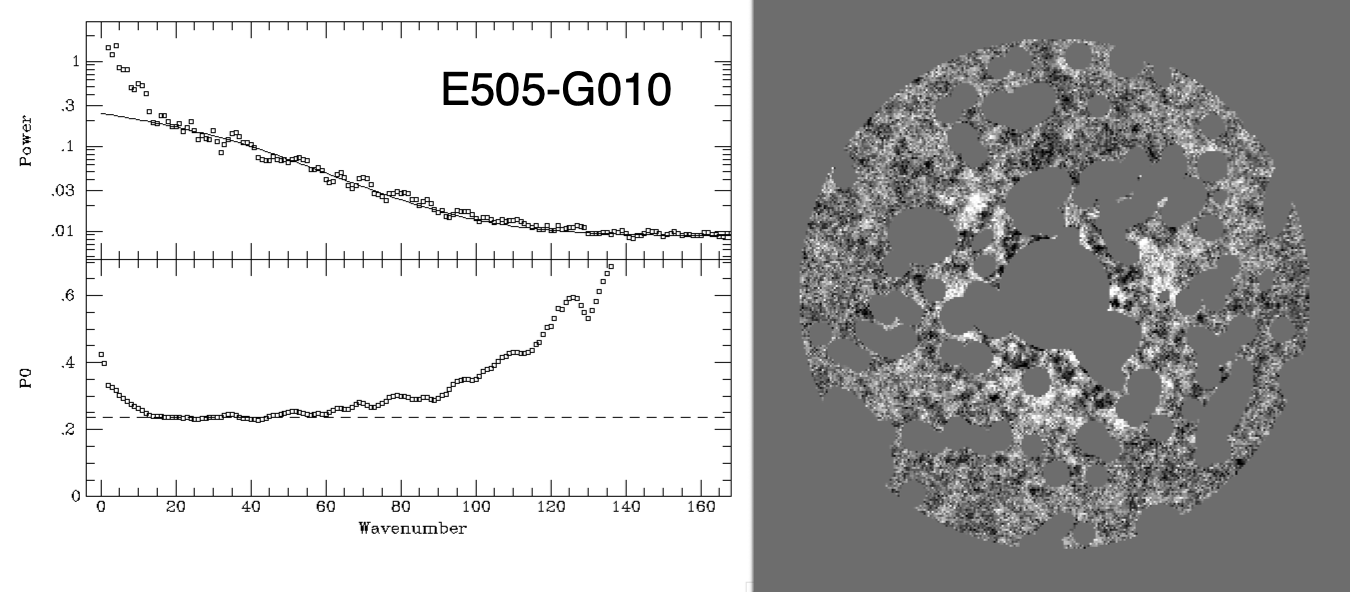}
\includegraphics[scale = 0.185]
 {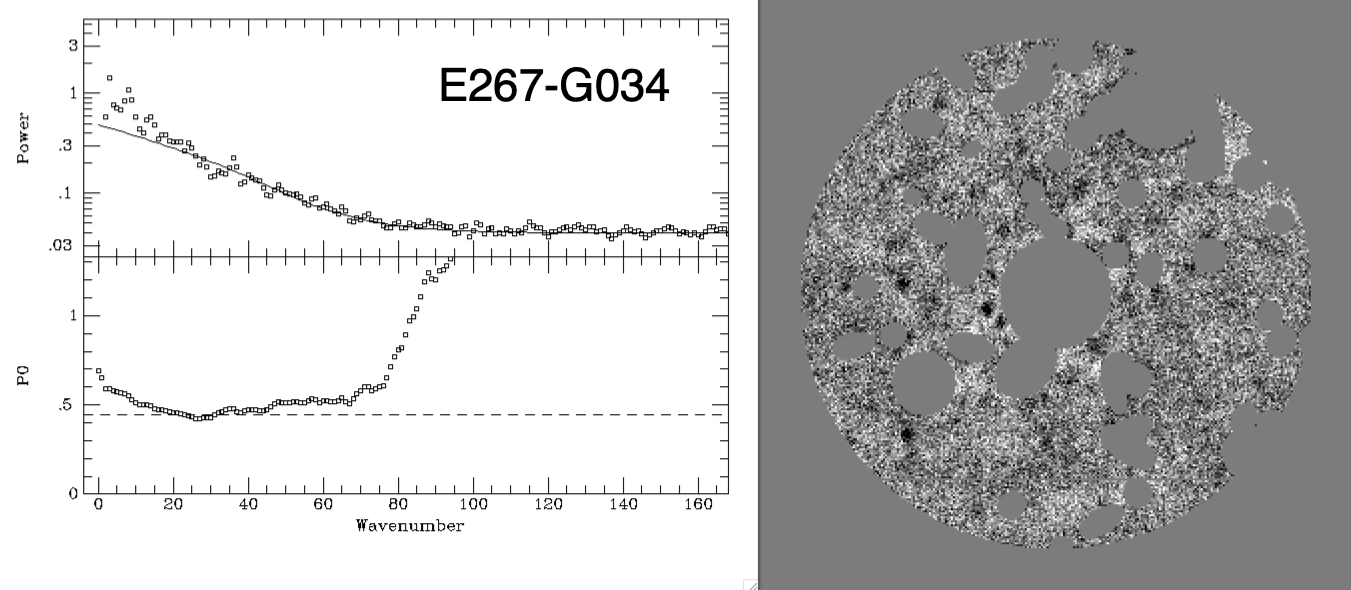}
\includegraphics[scale = 0.185]
 {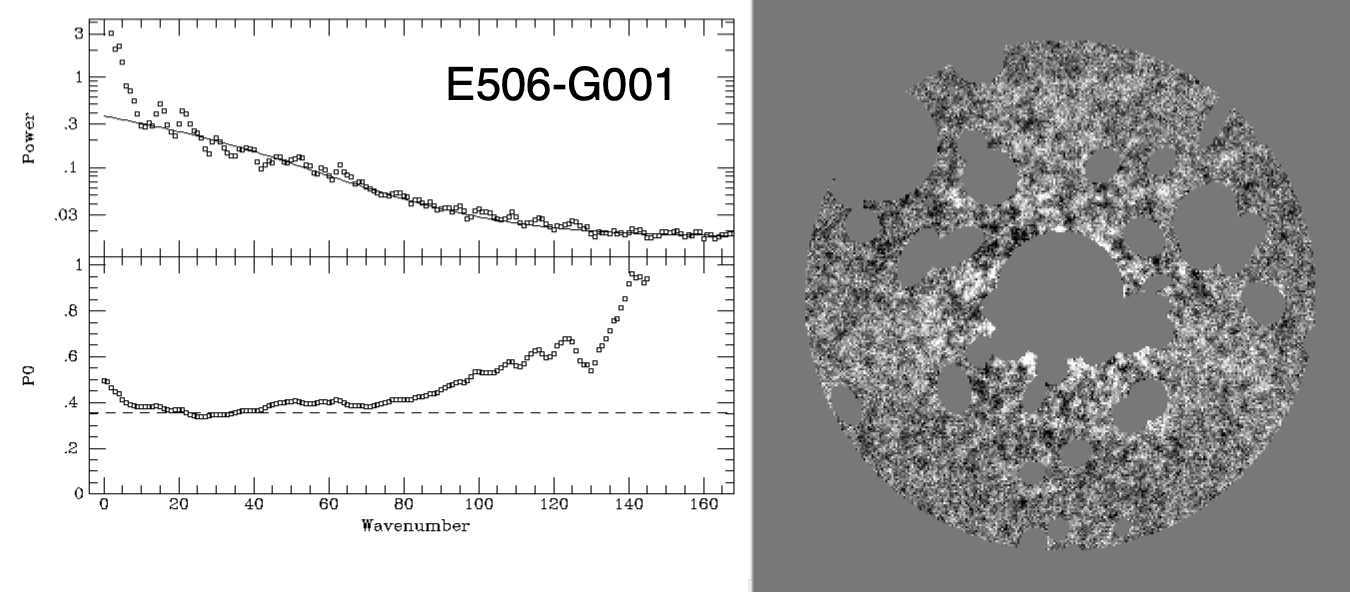}
\includegraphics[scale = 0.185]
 {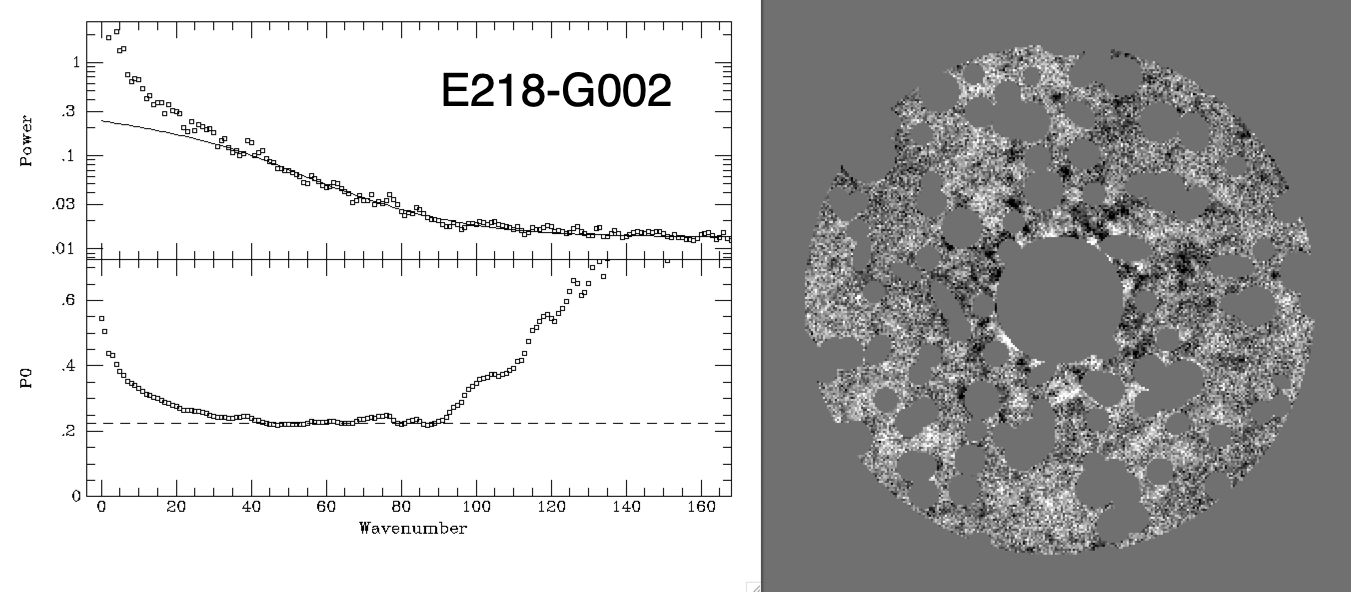}

\caption{SBF sample:1-10}

\label{fig:sample_1-10}
\end{figure*}

\begin{figure*}
\centering
\includegraphics[scale = 0.185]
 {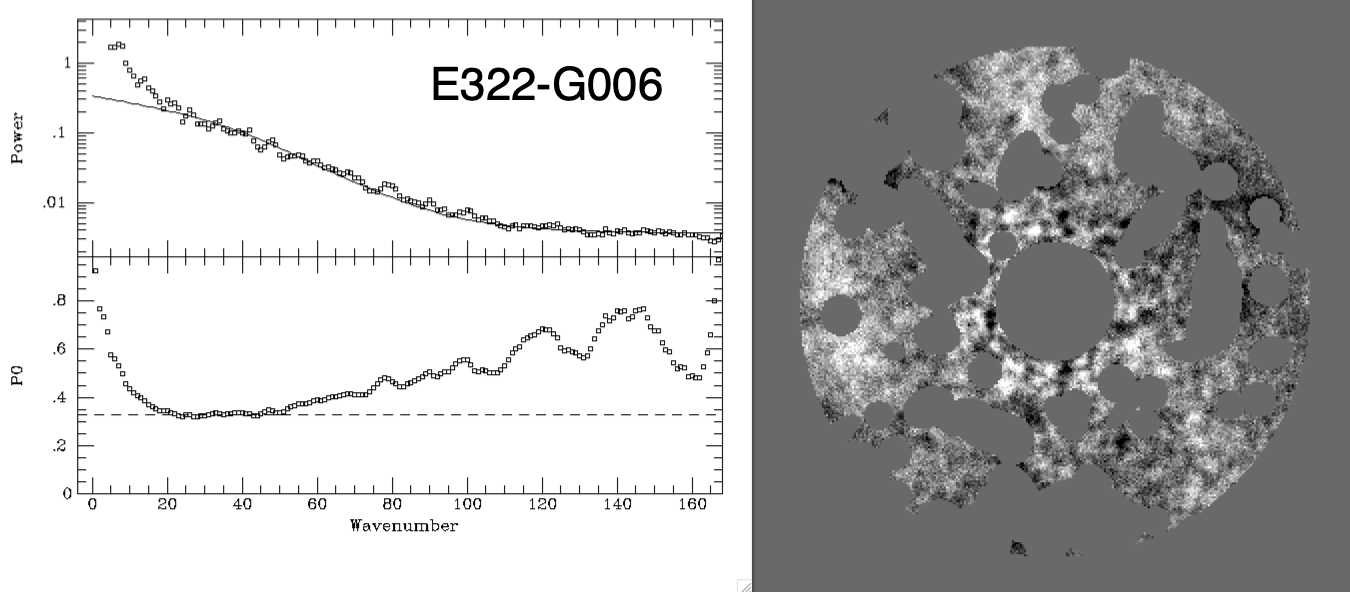}
\includegraphics[scale = 0.185]
 {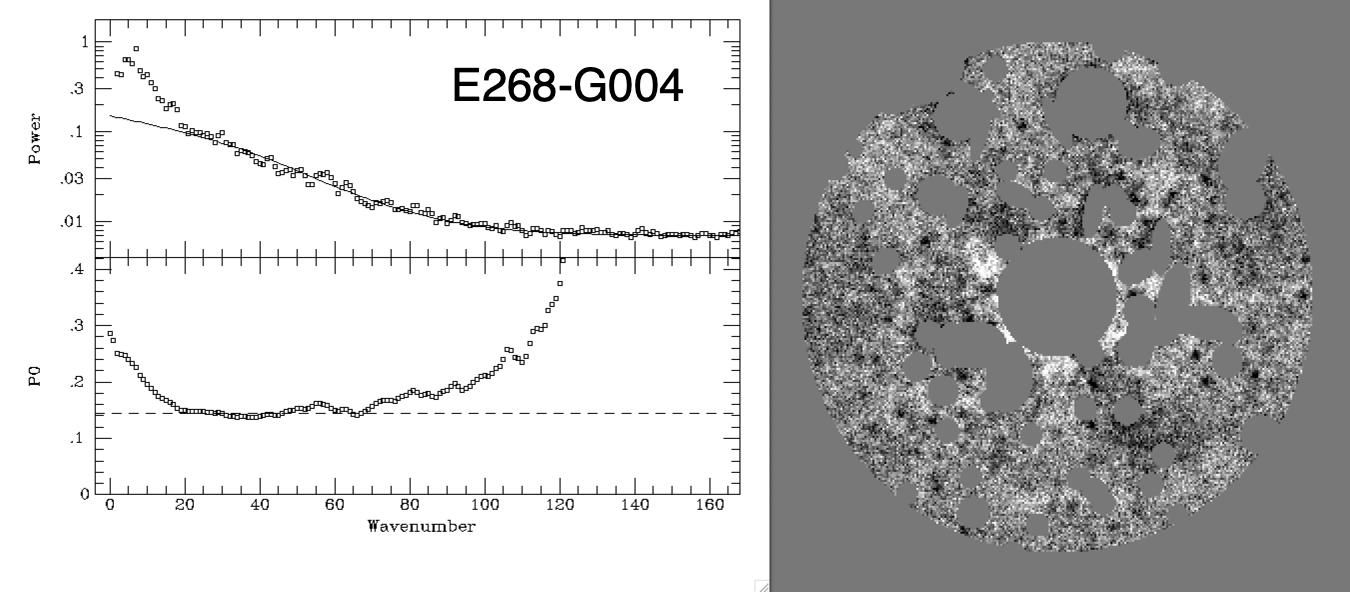}
\includegraphics[scale = 0.185]
 {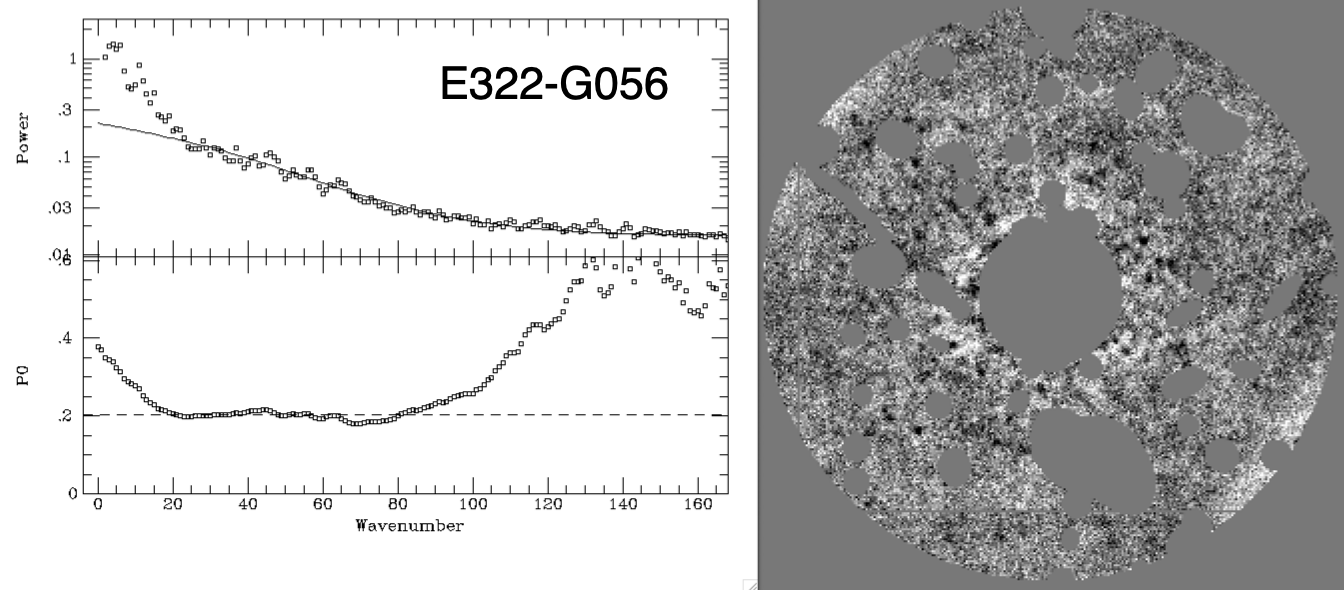}
\includegraphics[scale = 0.185]
 {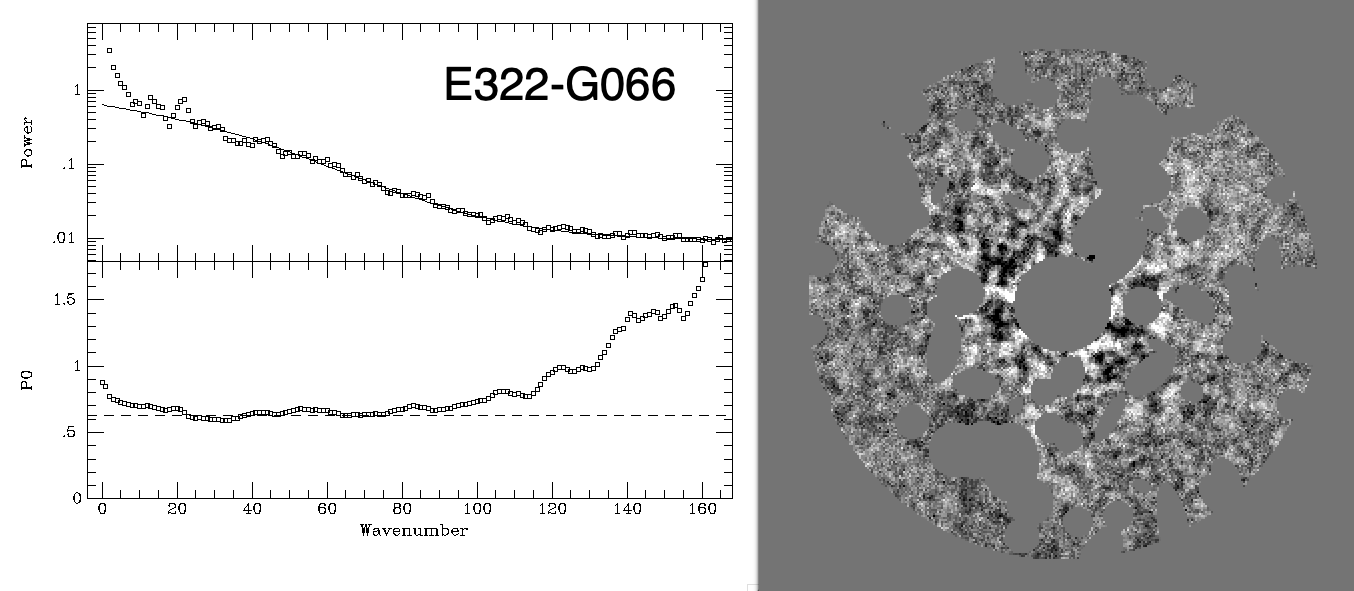}
\includegraphics[scale = 0.183]
 {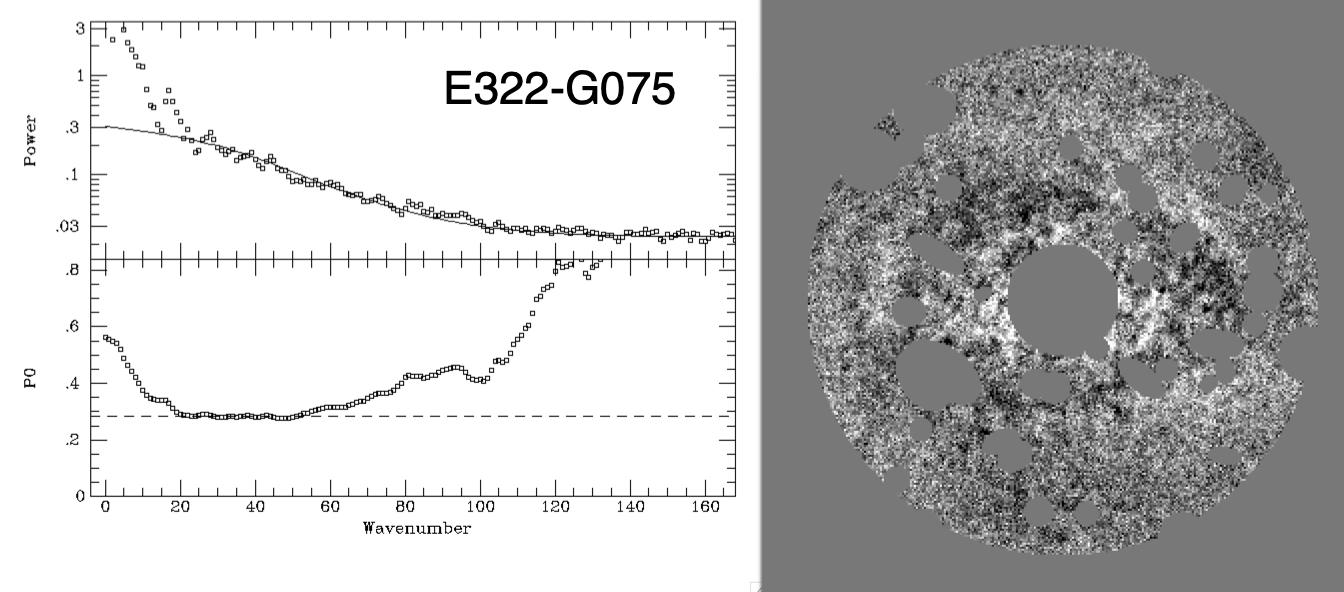}
\includegraphics[scale = 0.185]
 {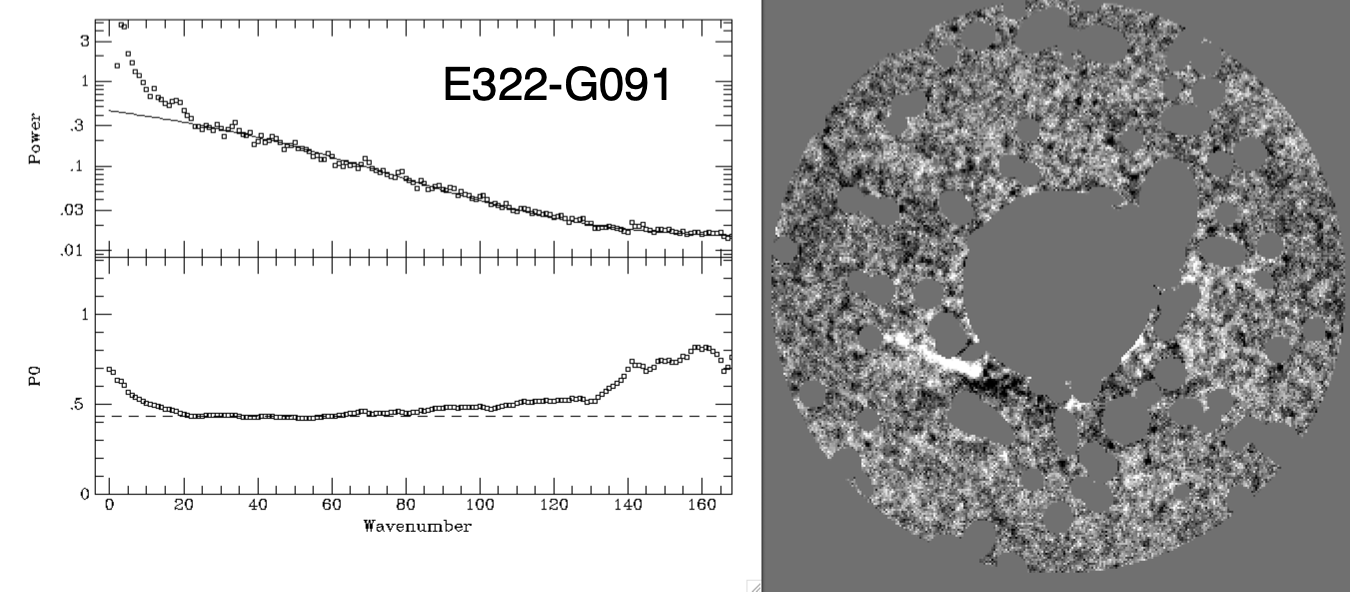}
\includegraphics[scale = 0.185]
 {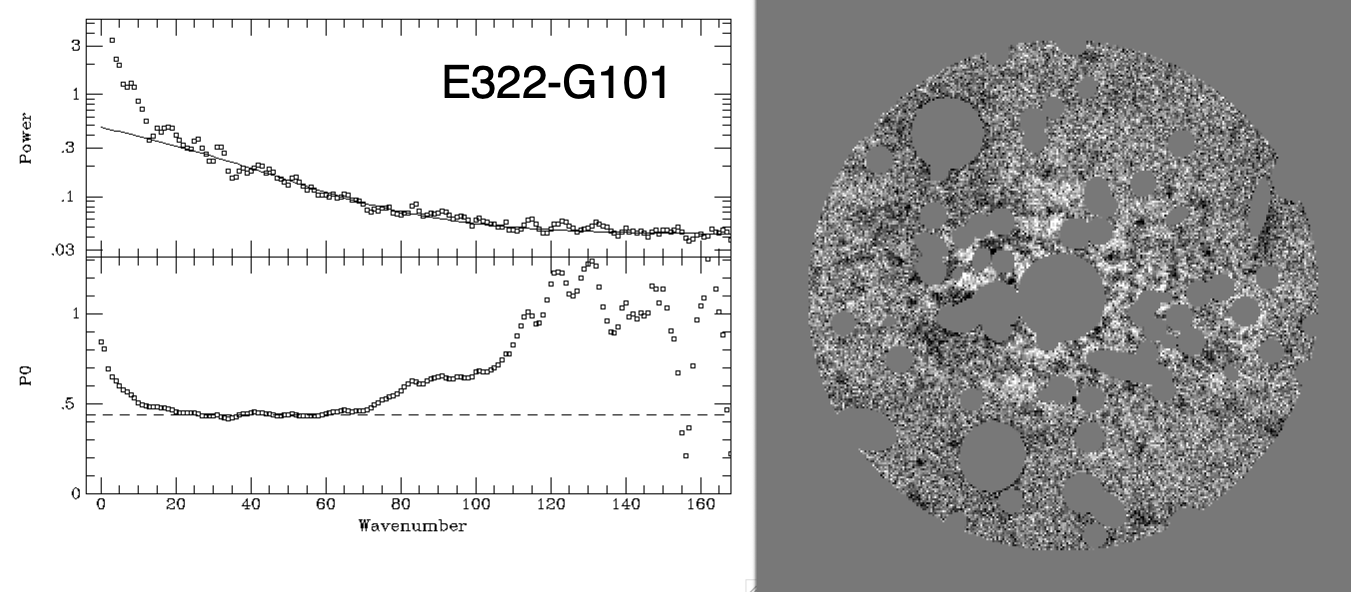}
\includegraphics[scale = 0.185]
 {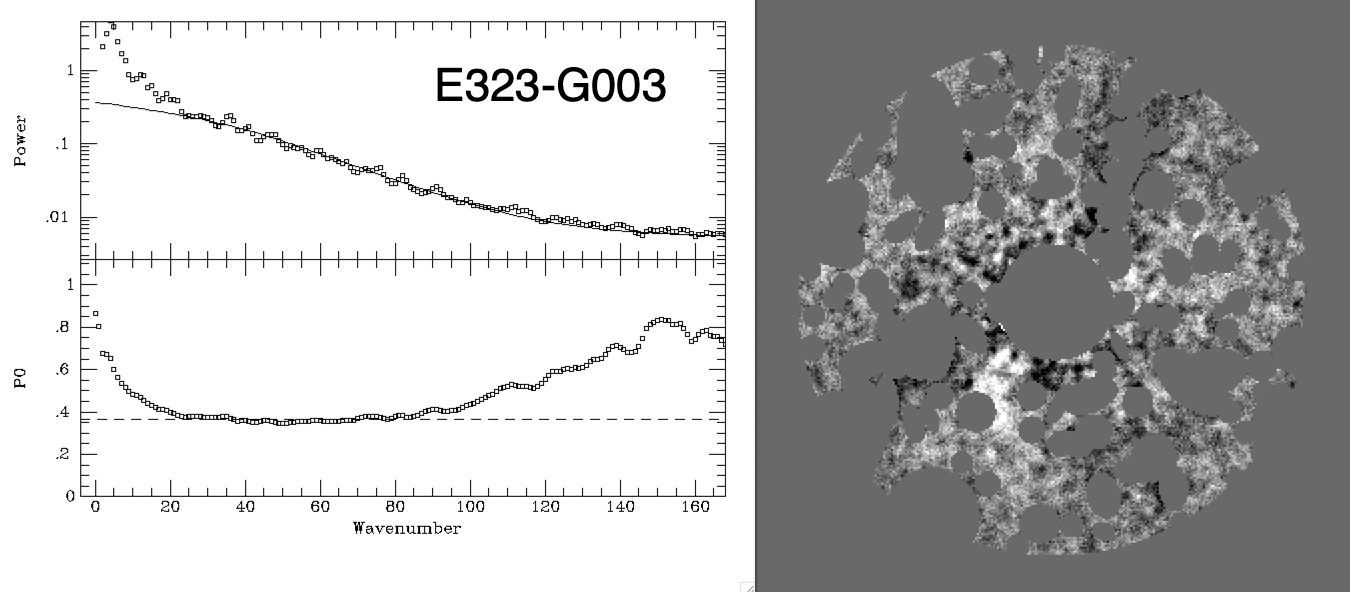}
\includegraphics[scale = 0.185]
 {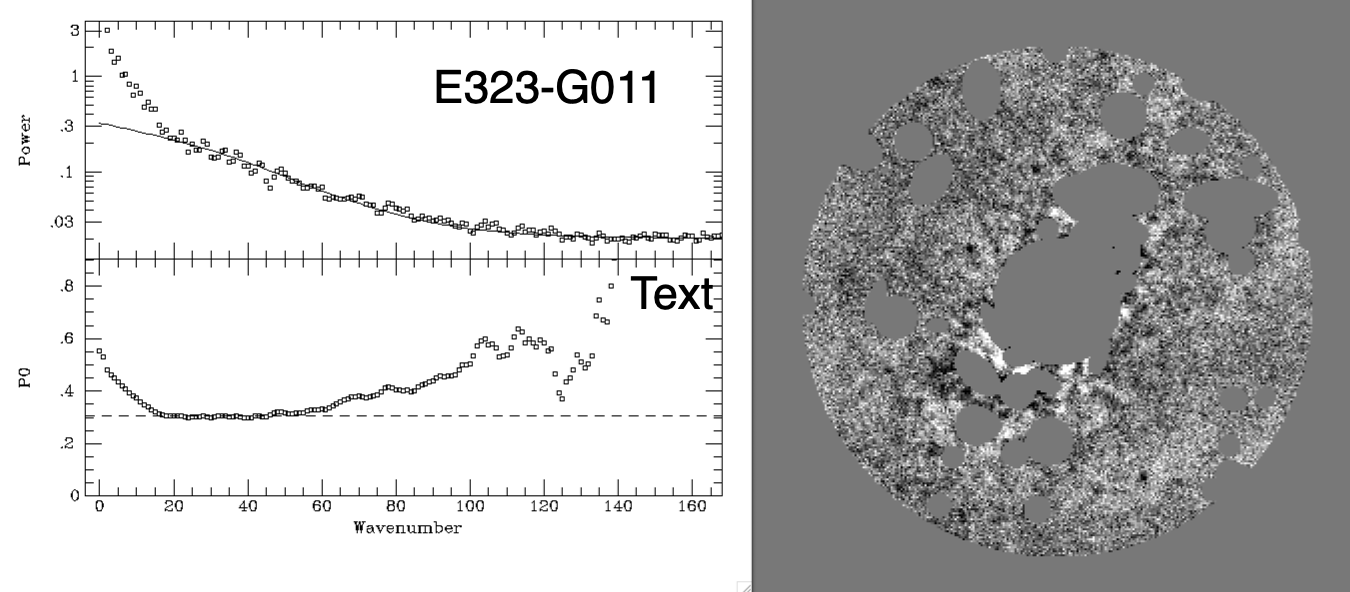}
\includegraphics[scale = 0.185]
 {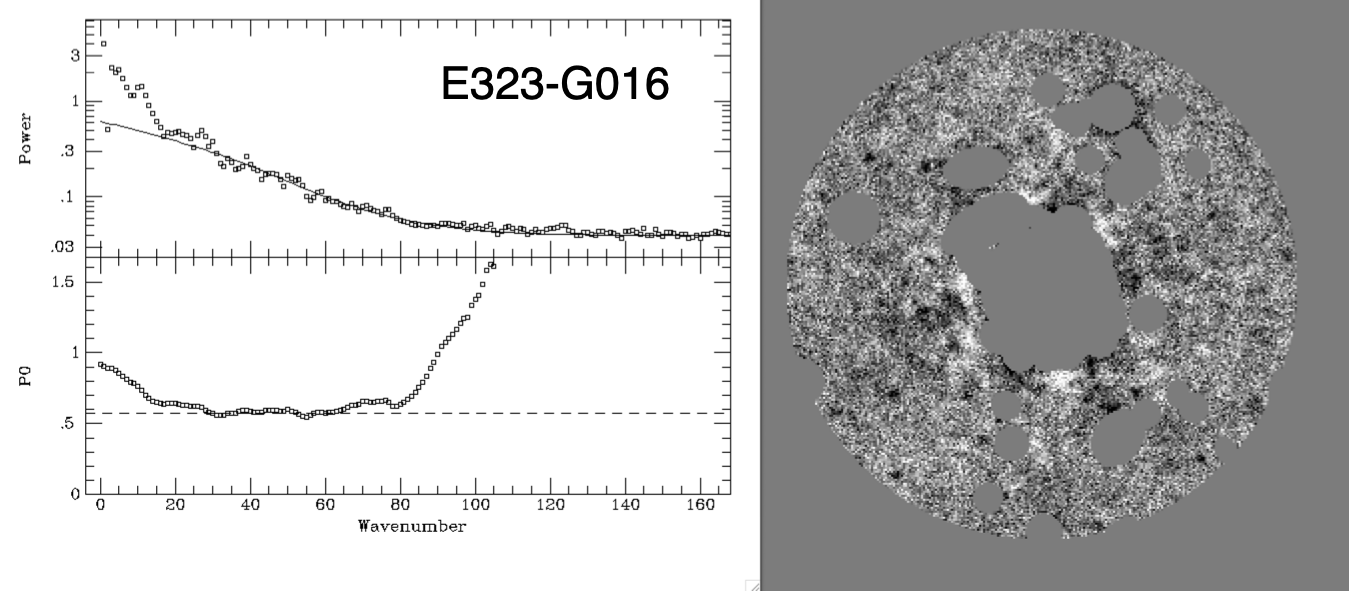}
\includegraphics[scale = 0.185]
 {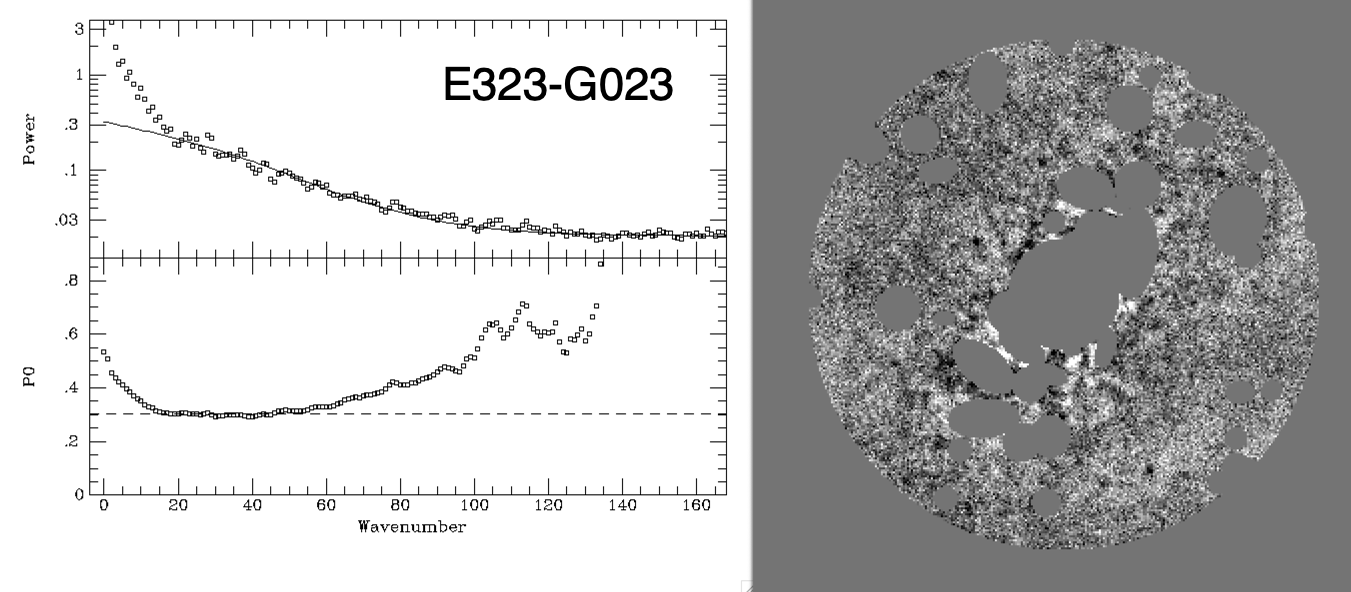}
\includegraphics[scale = 0.185]
 {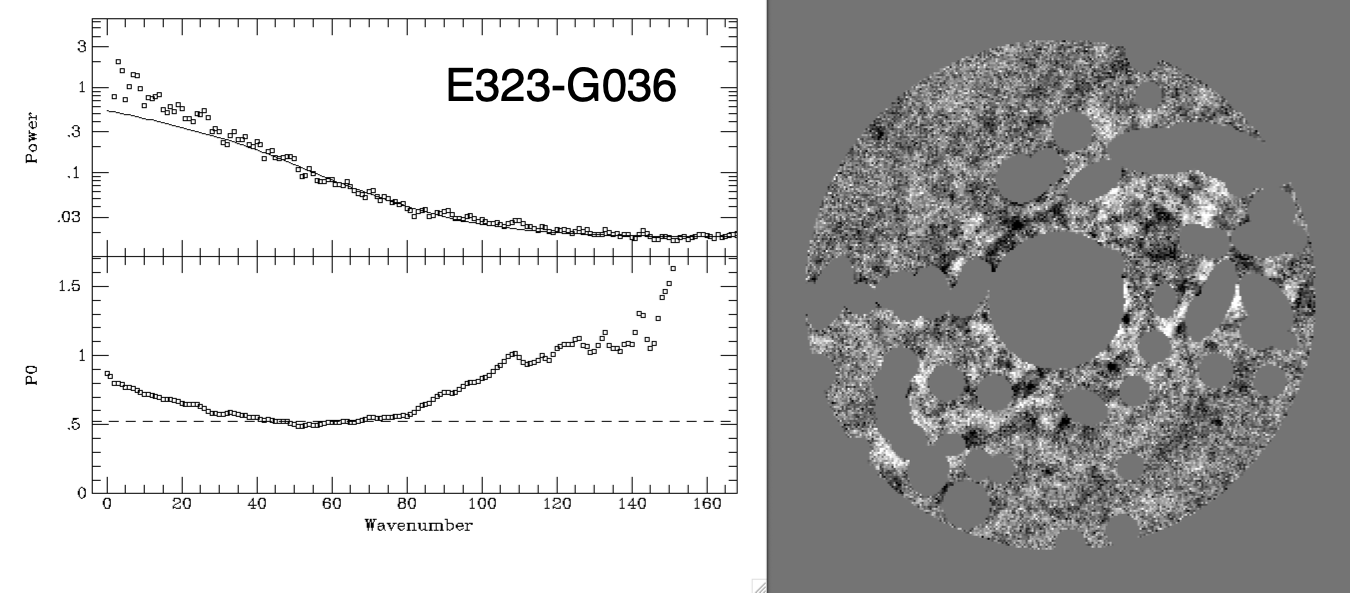}

\caption{SBF sample:11-22}

\label{fig:sample 11-22}

\vspace{-60pt} 
\end{figure*}

\begin{figure*}
\centering
\includegraphics[scale = 0.185]
 {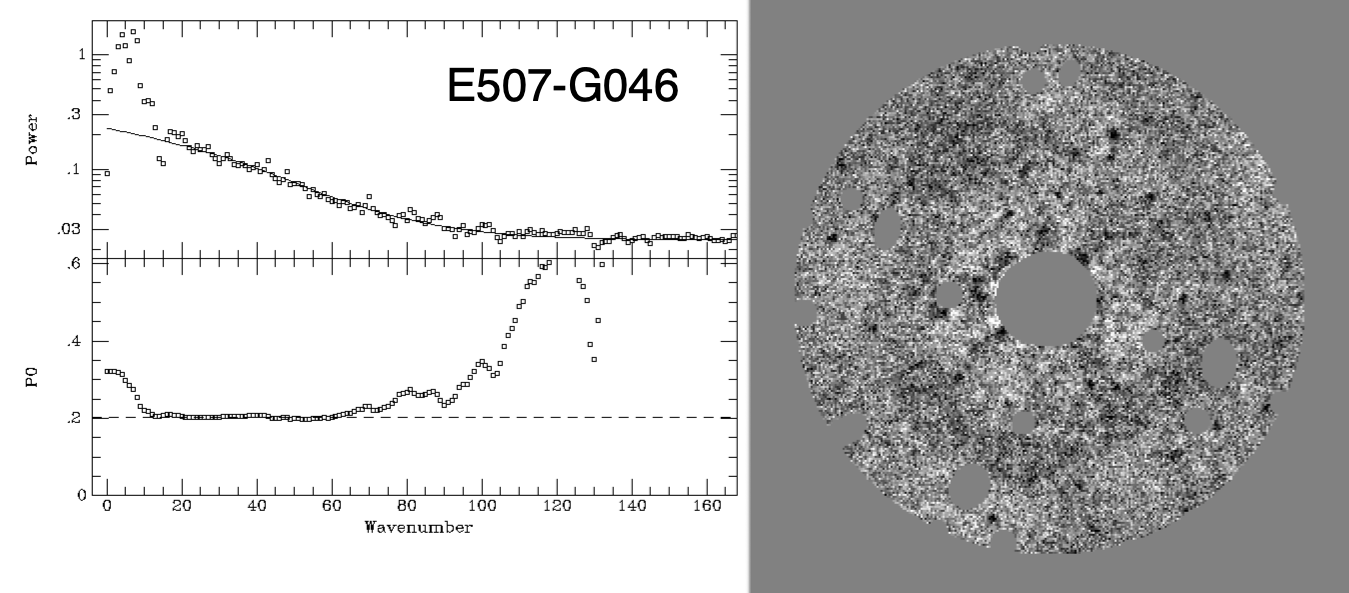}
\includegraphics[scale = 0.185]
 {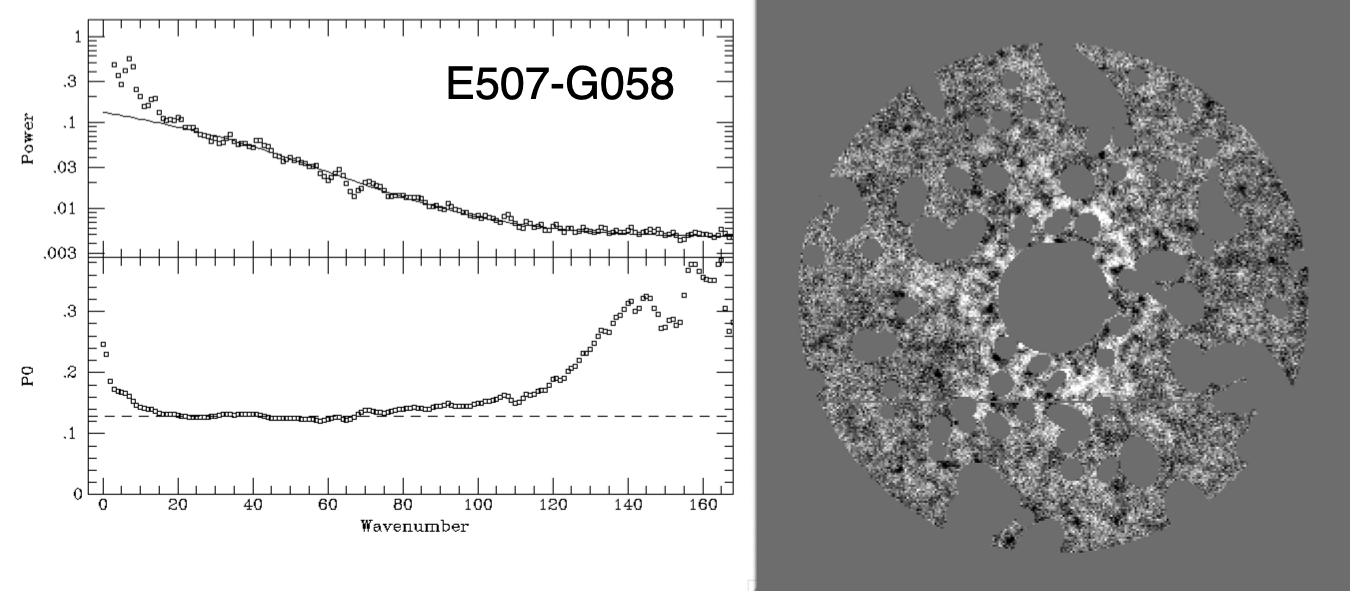}
\includegraphics[scale = 0.185]
 {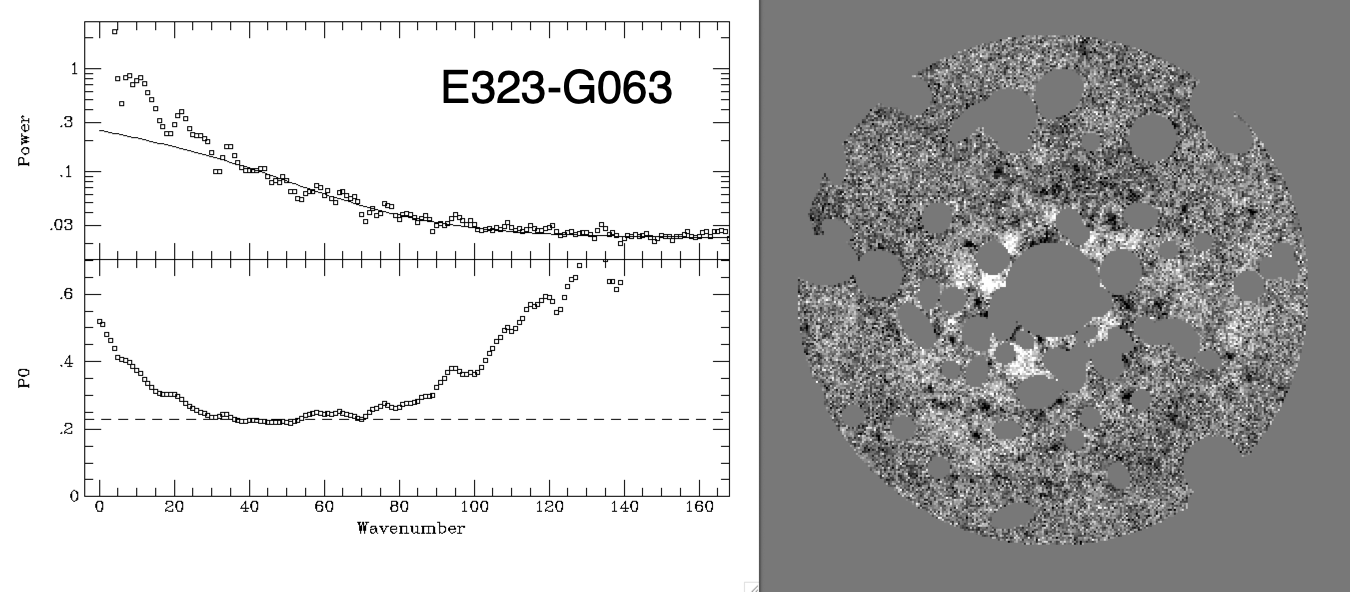}
\includegraphics[scale = 0.185]
 {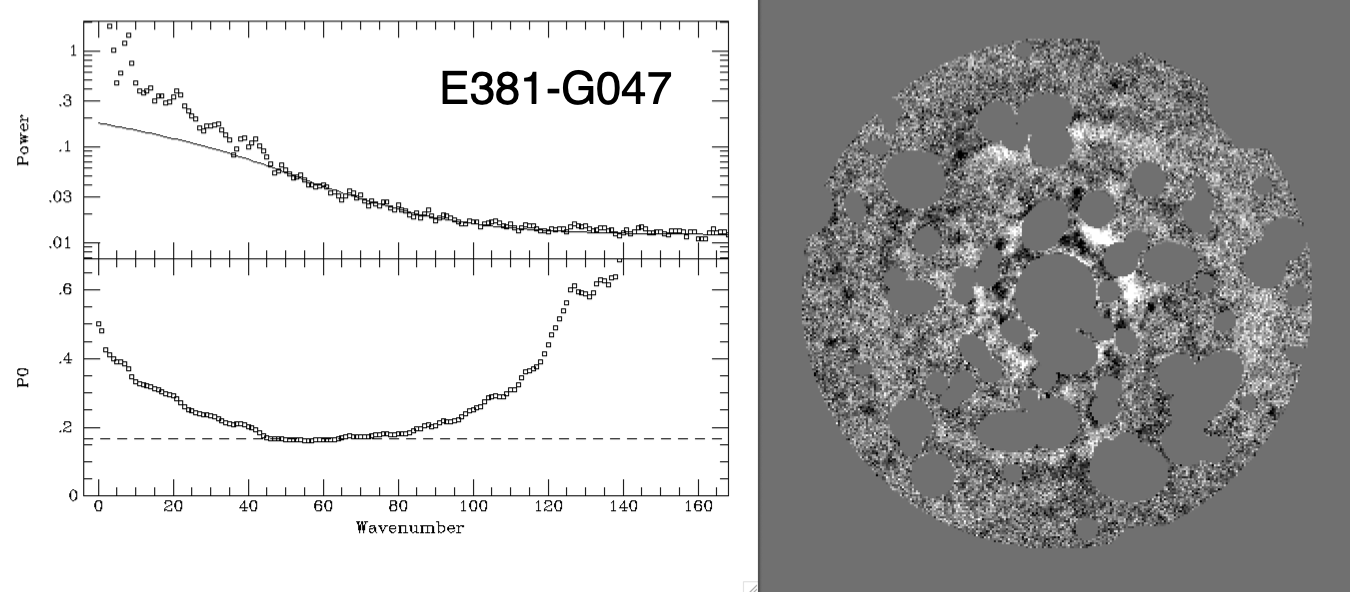}
\includegraphics[scale = 0.185]
 {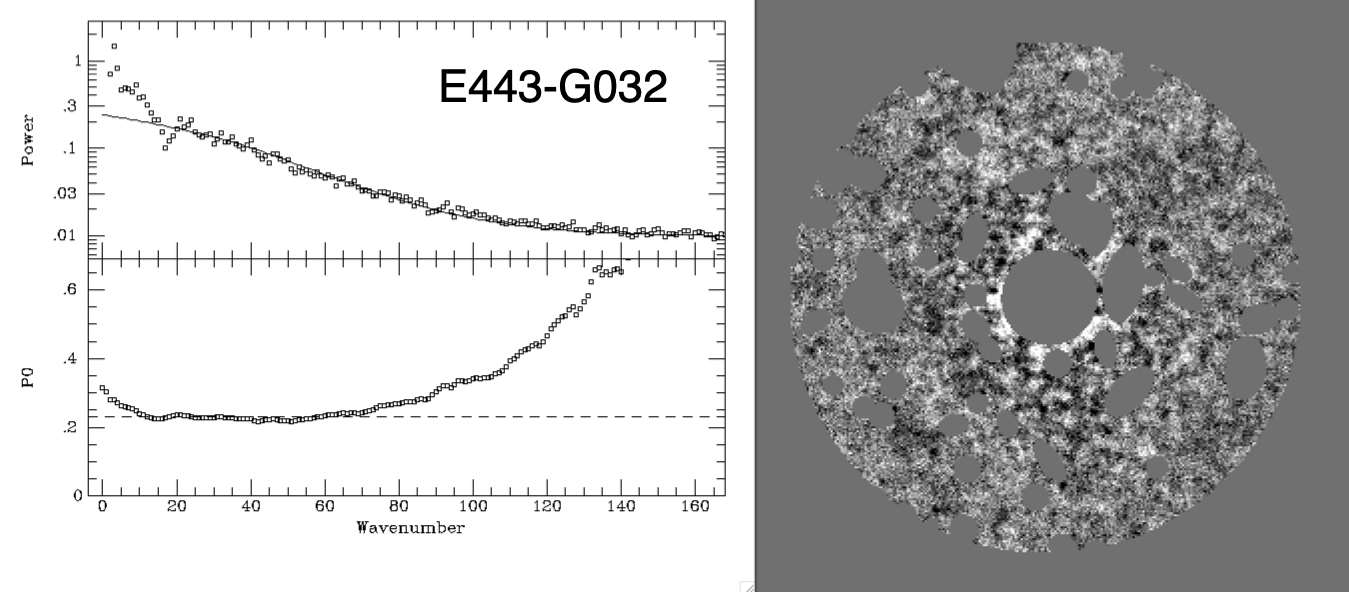}
\includegraphics[scale = 0.185]
 {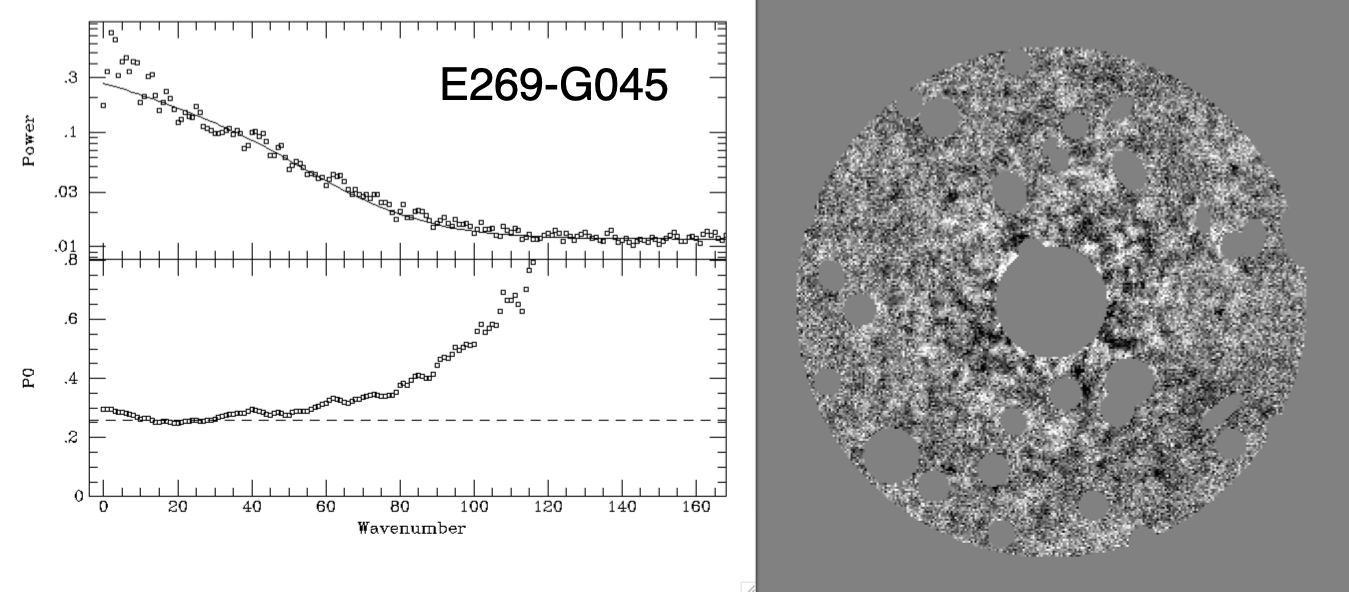}
\includegraphics[scale = 0.185]
 {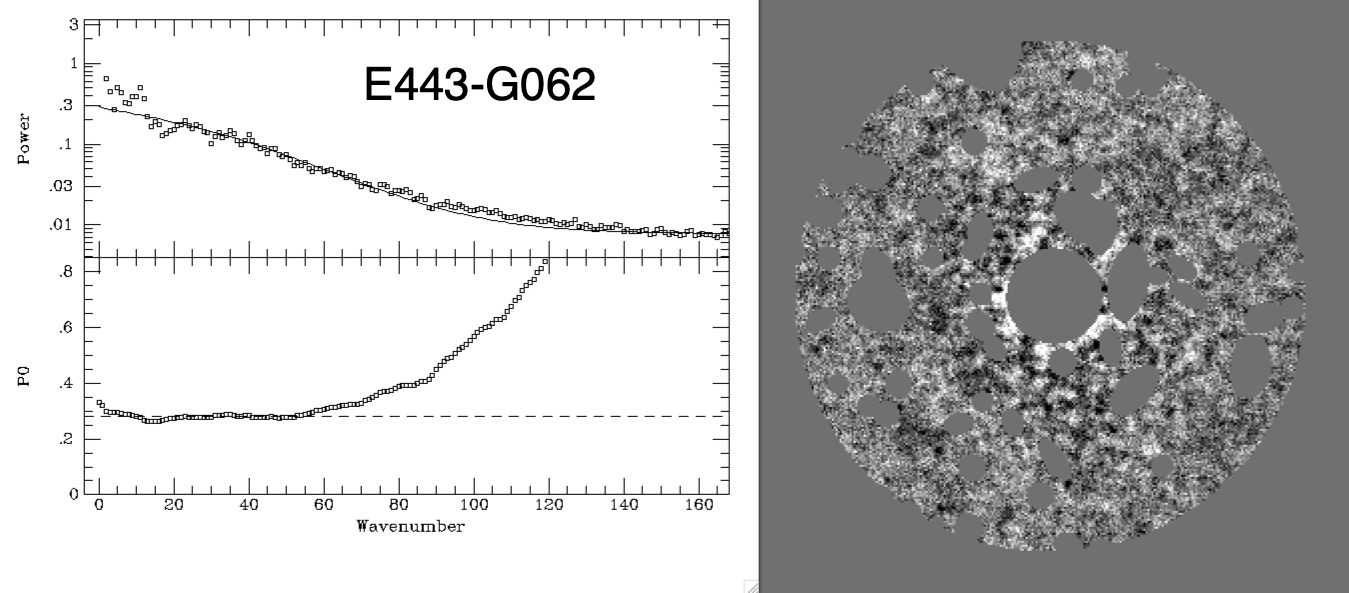}
\includegraphics[scale = 0.185]
 {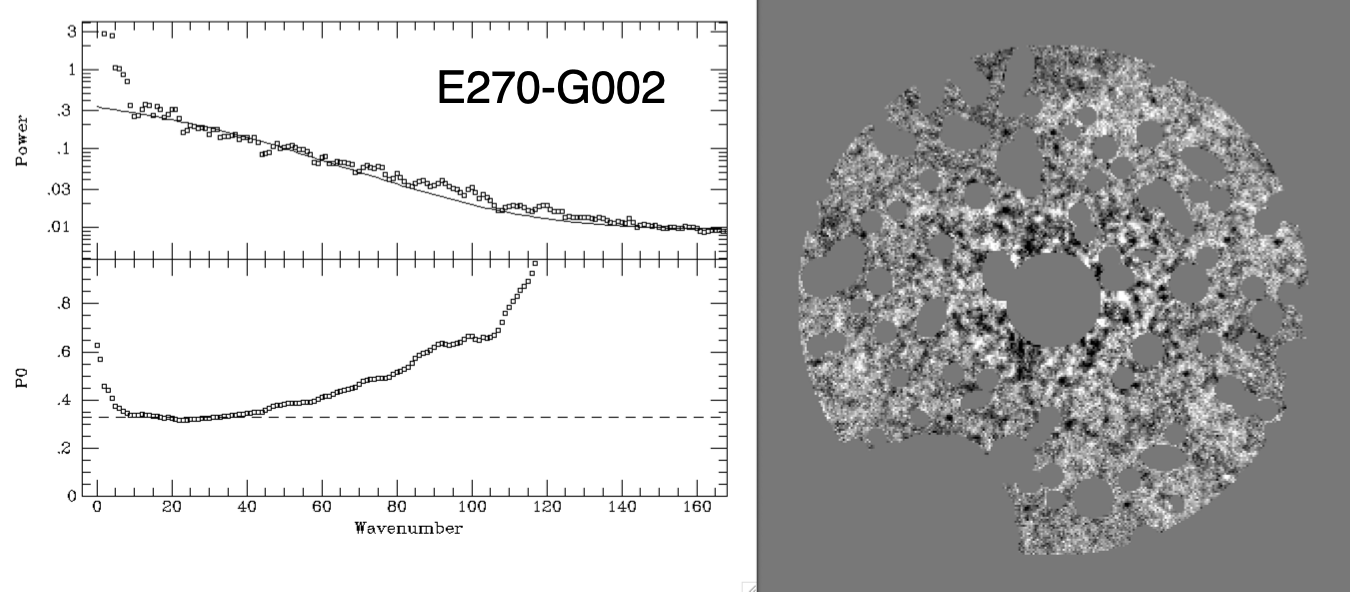}
\includegraphics[scale = 0.185]
 {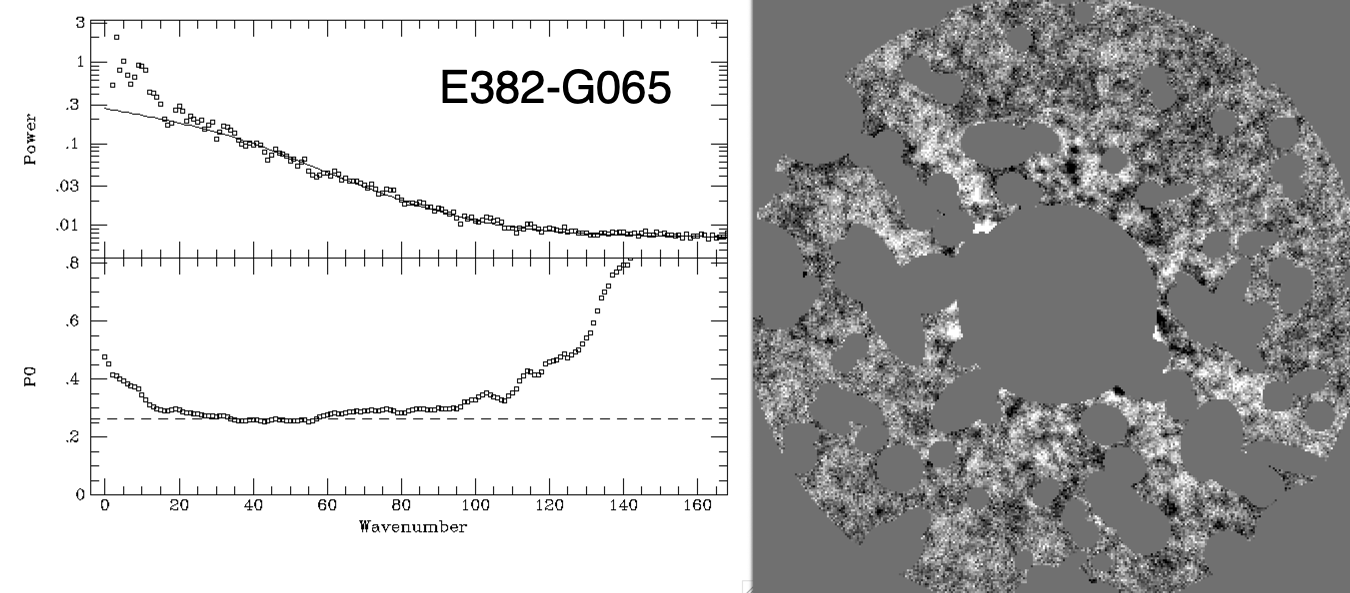}
\includegraphics[scale = 0.185]
 {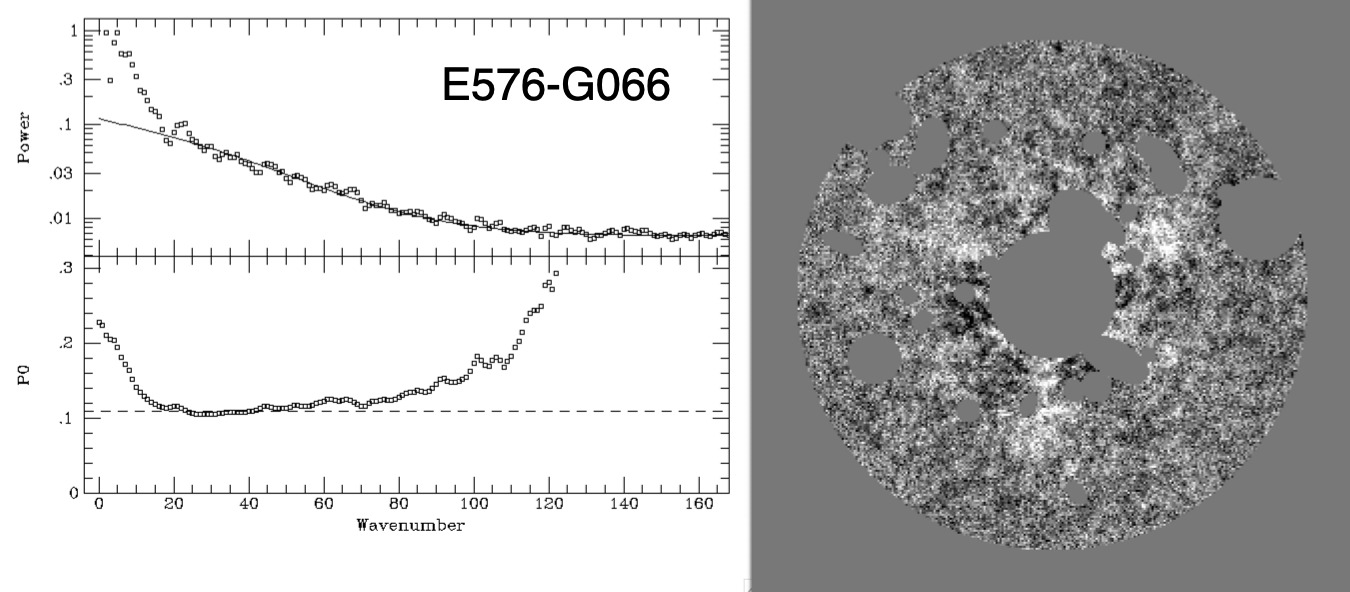}
\includegraphics[scale = 0.185]
 {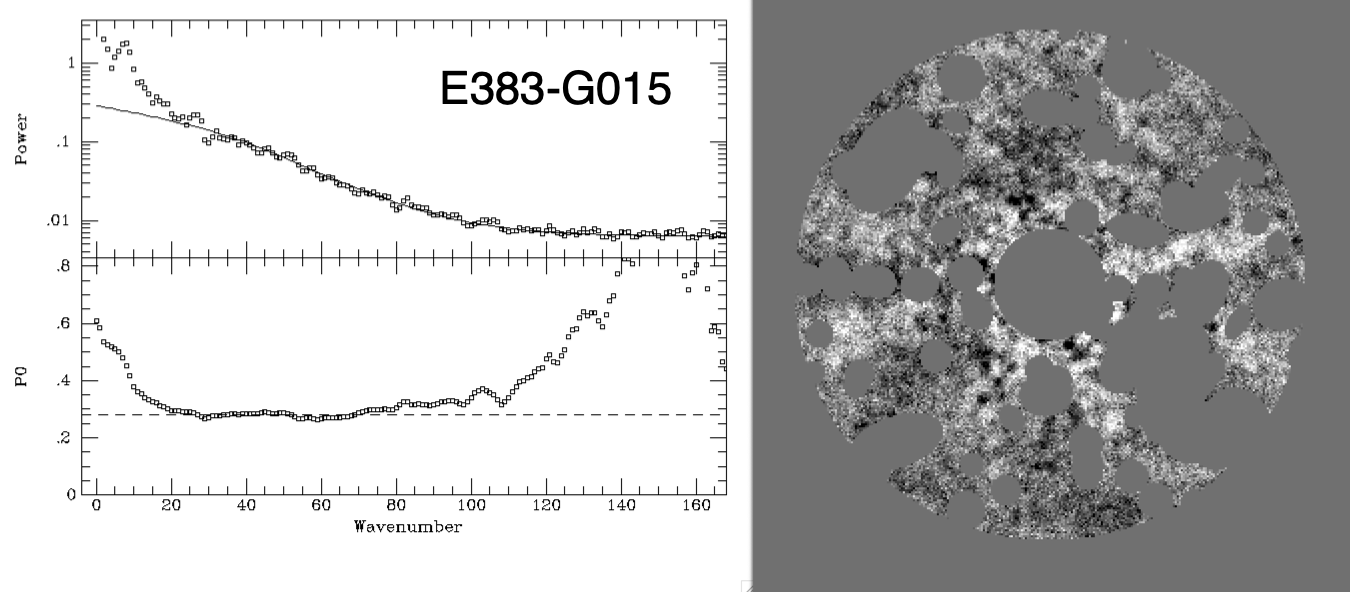}
\includegraphics[scale = 0.185]
{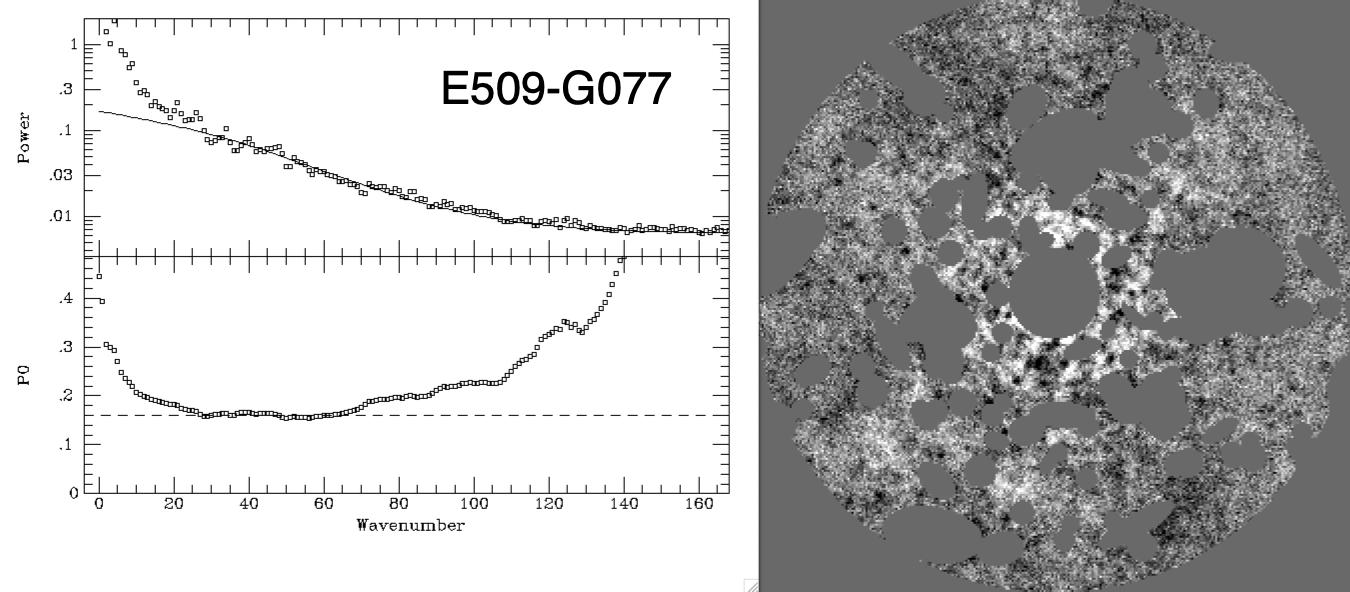}

\caption{SBF sample: 23-35}

\label{fig:sample_23-35}

\vspace{-60pt} 
\end{figure*}

\begin{figure*}
\centering

\includegraphics[scale = 0.185]
 {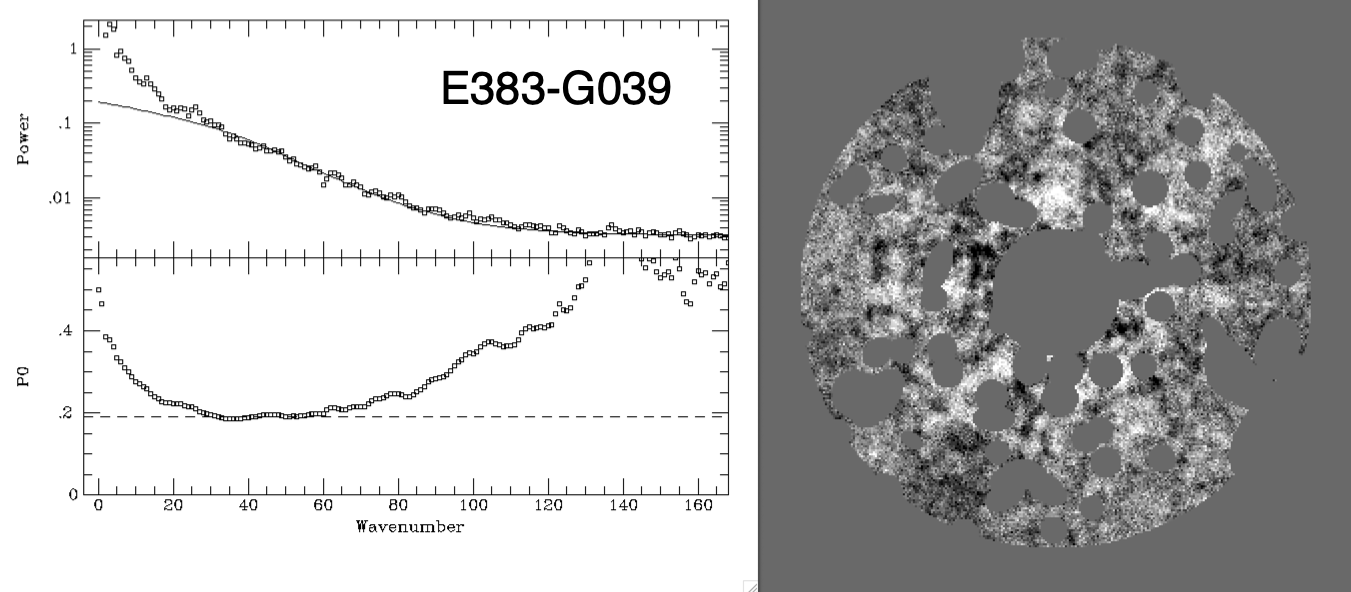}
\includegraphics[scale = 0.185]
 {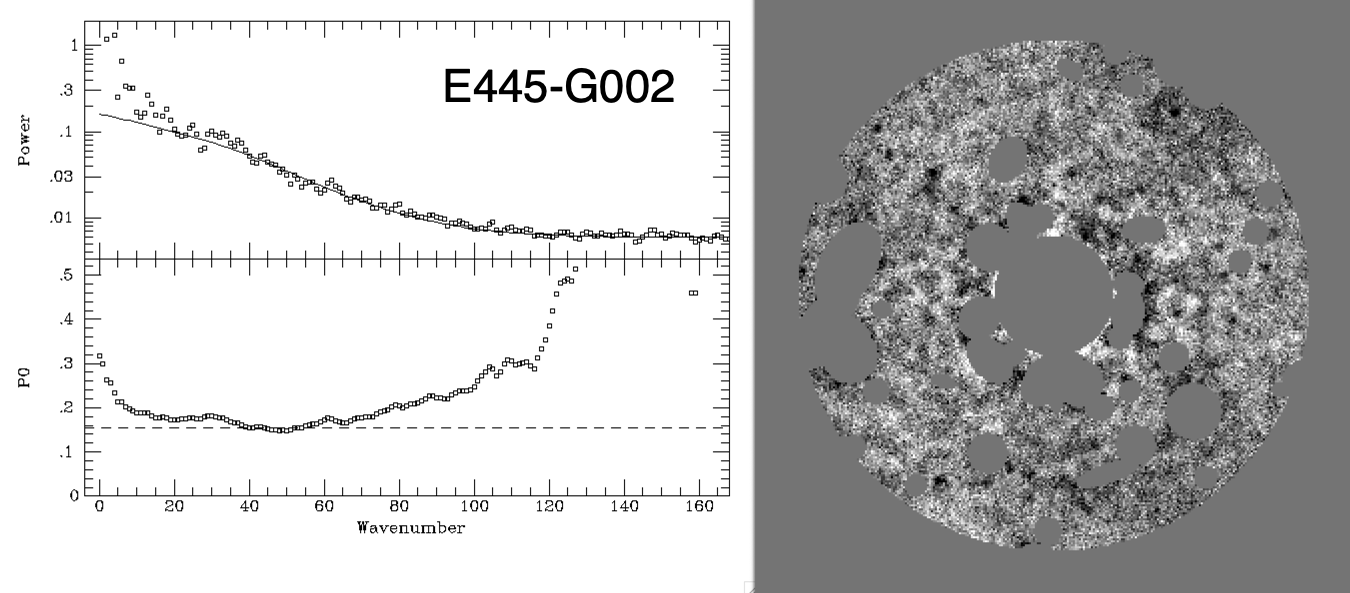}
\includegraphics[scale = 0.185]
 {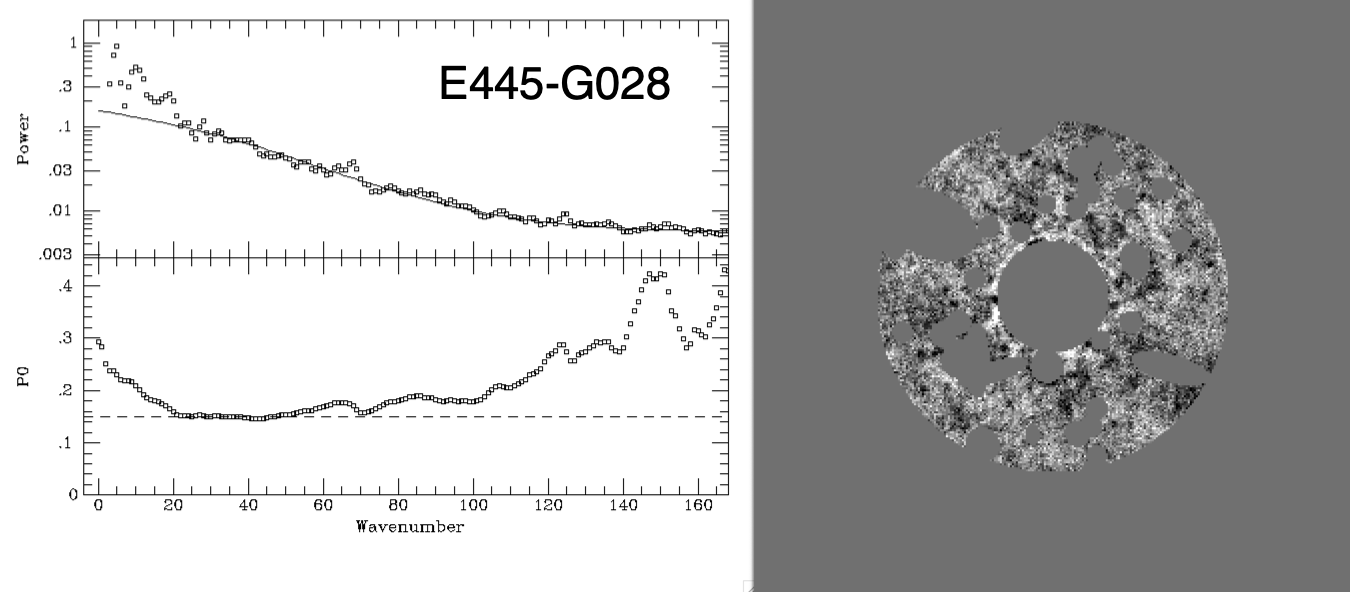}
\includegraphics[scale = 0.185]
 {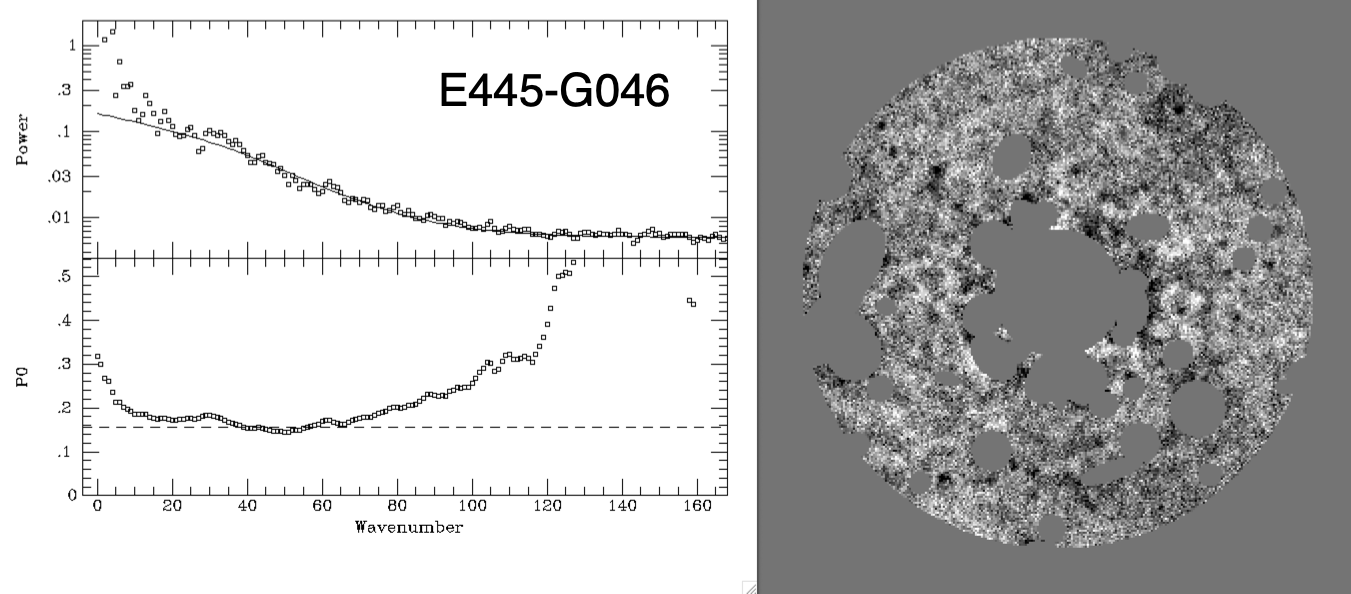}
\includegraphics[scale = 0.185]
 {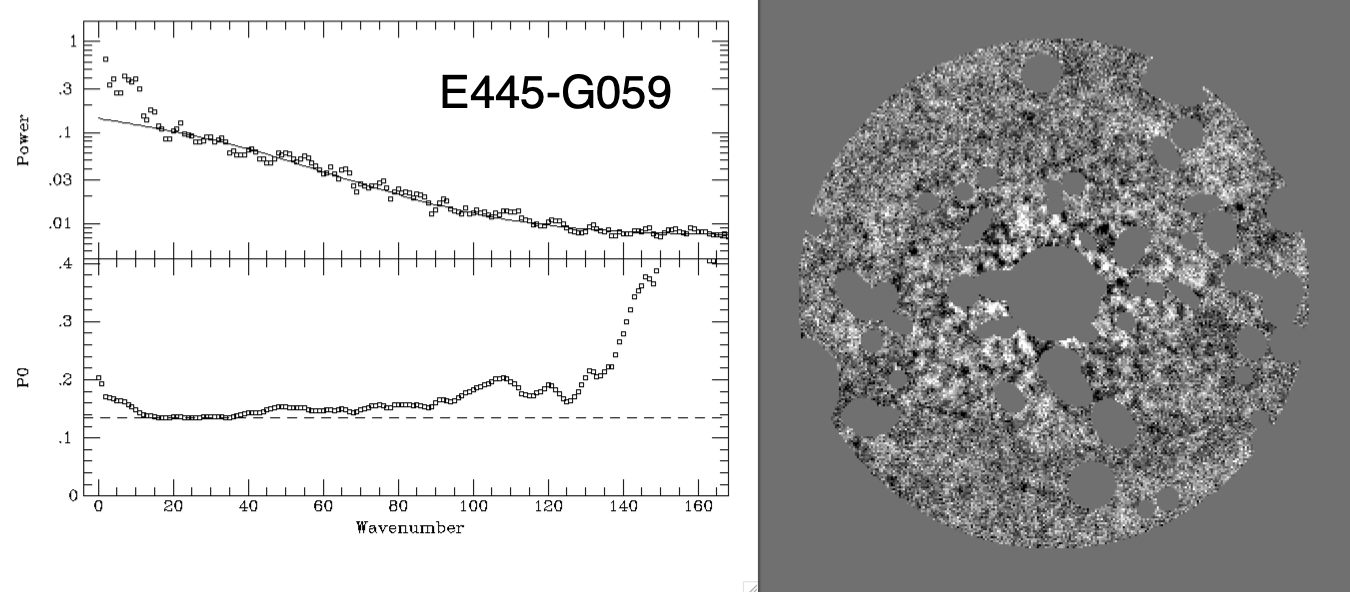}
\includegraphics[scale = 0.185]
 {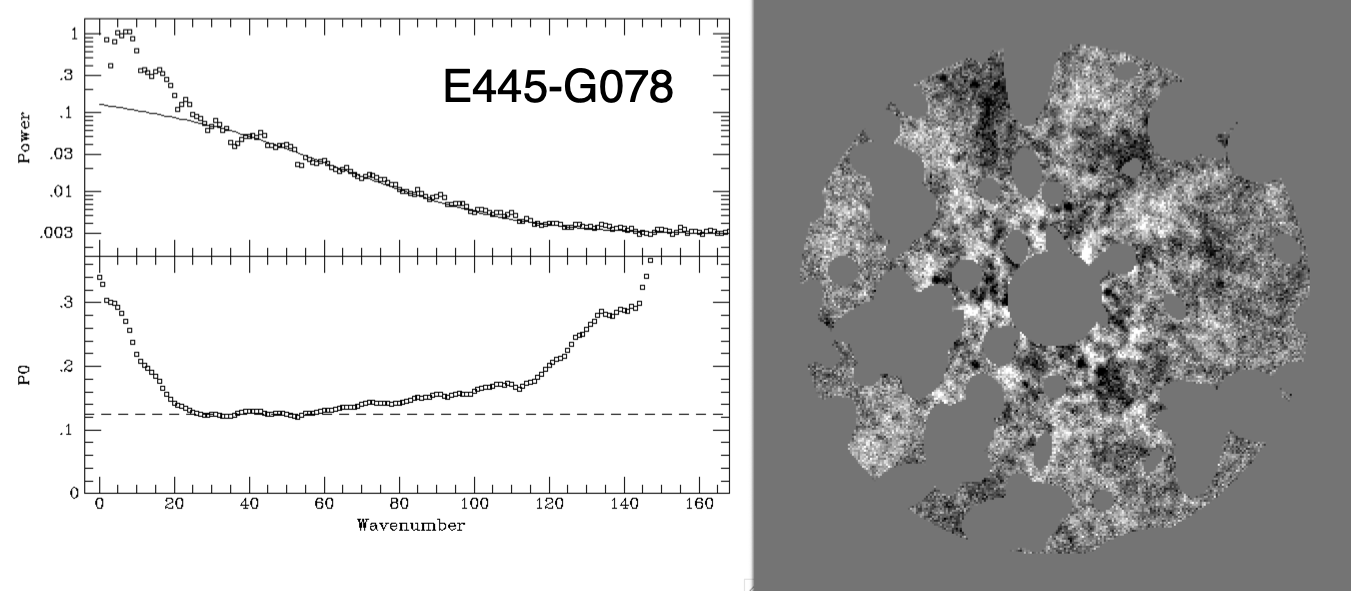}
\includegraphics[scale = 0.185]
 {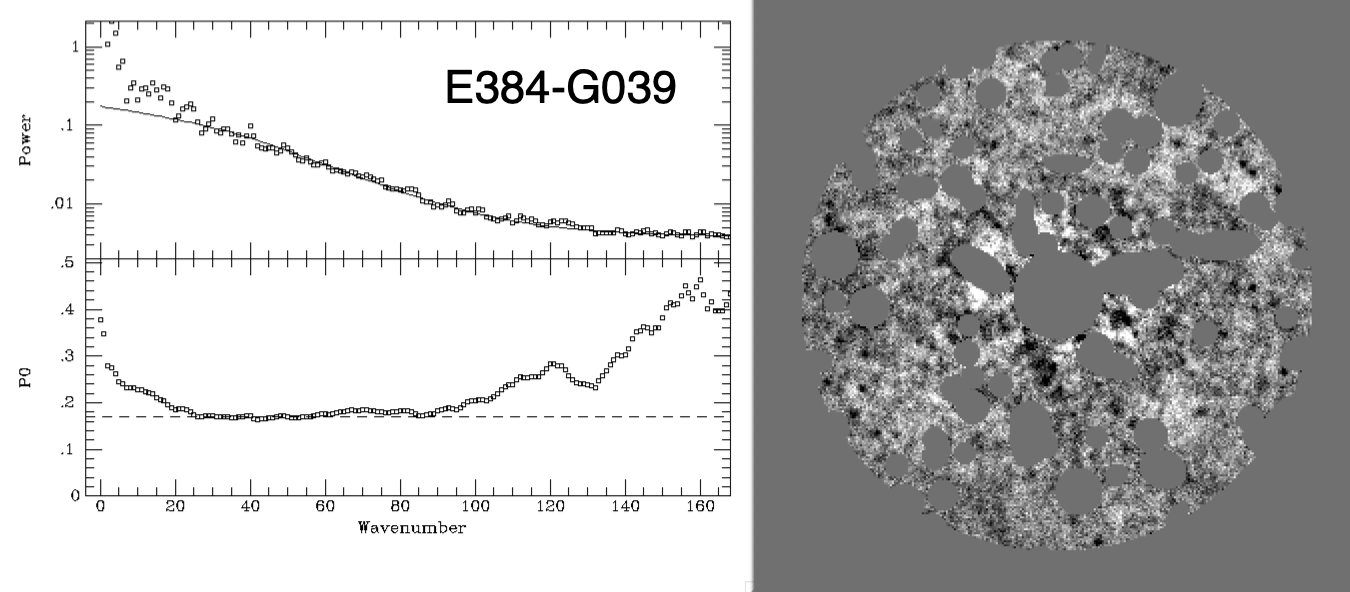}
\includegraphics[scale = 0.185]
 {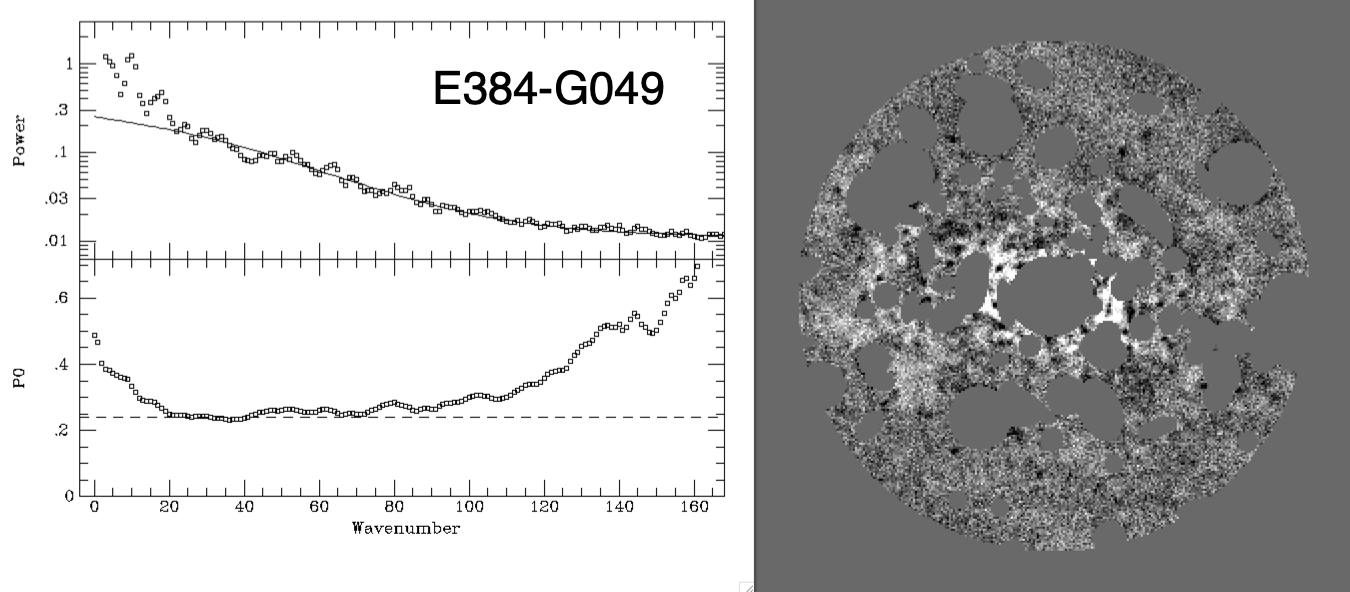}
\includegraphics[scale = 0.185]    
 {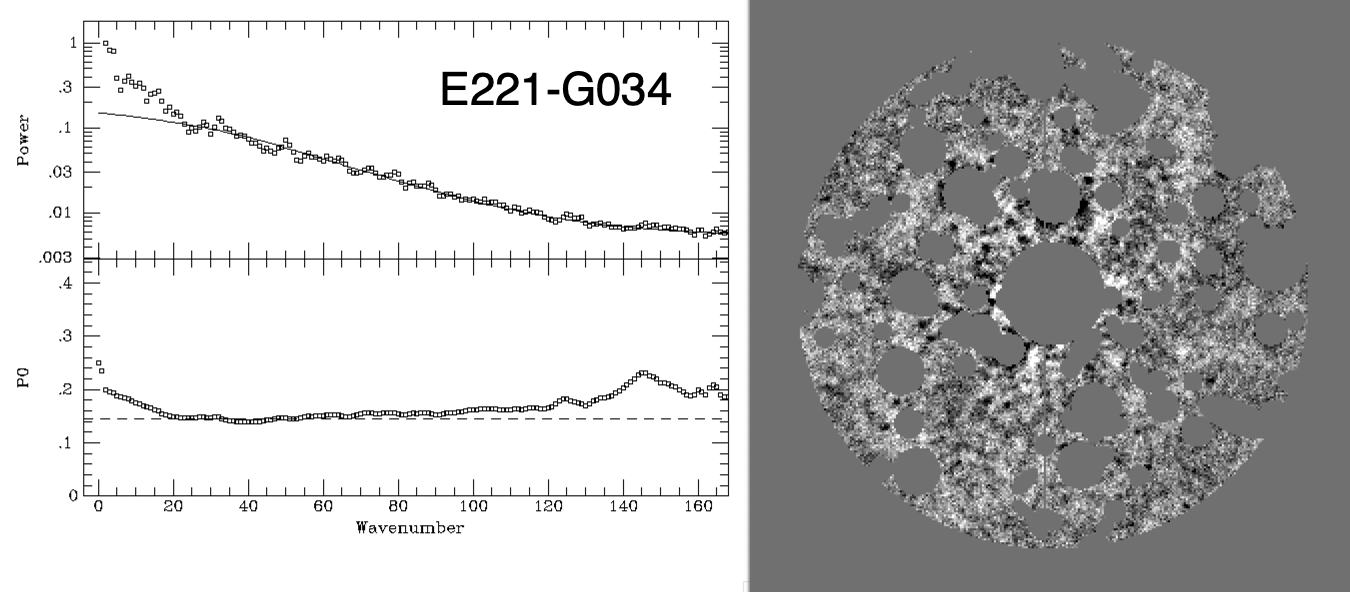}
\includegraphics[scale = 0.185] 
 {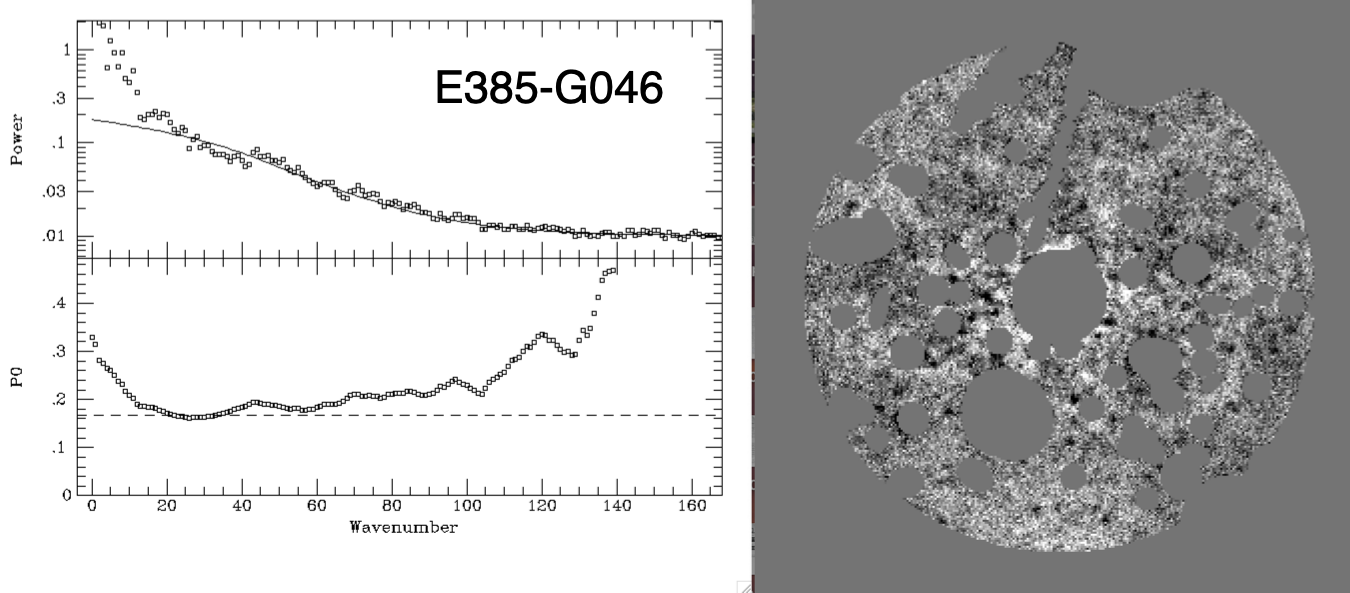}
\includegraphics[scale = 0.185] 
 {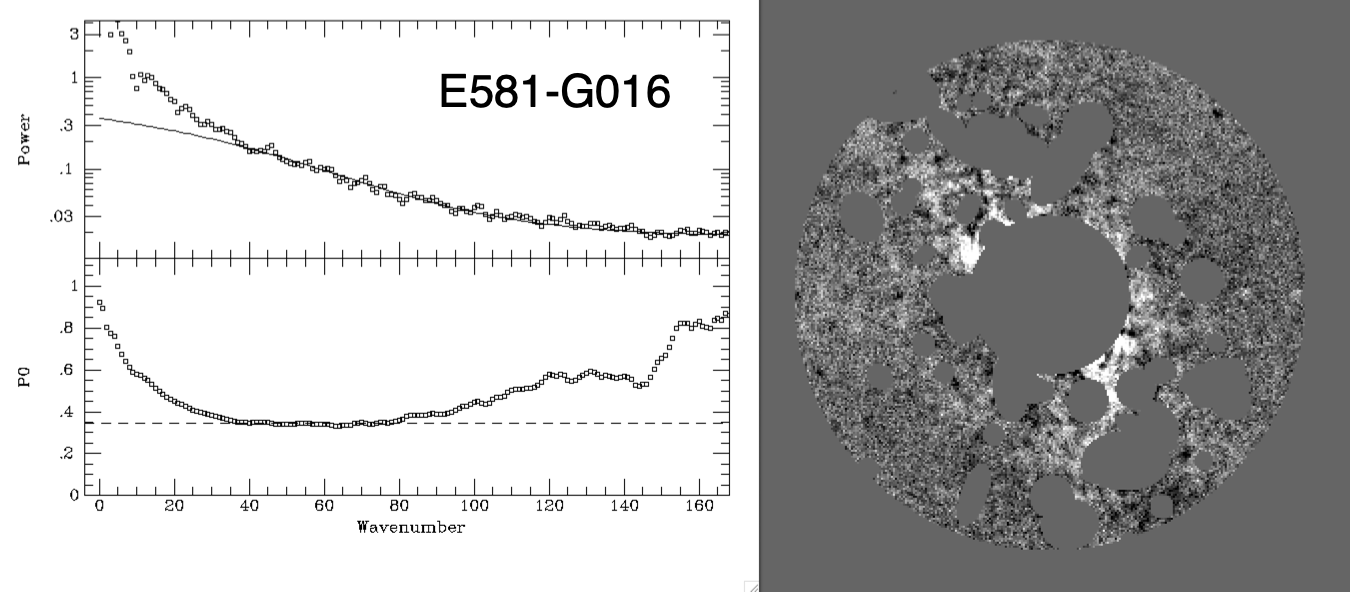}
\includegraphics[scale = 0.185]
 {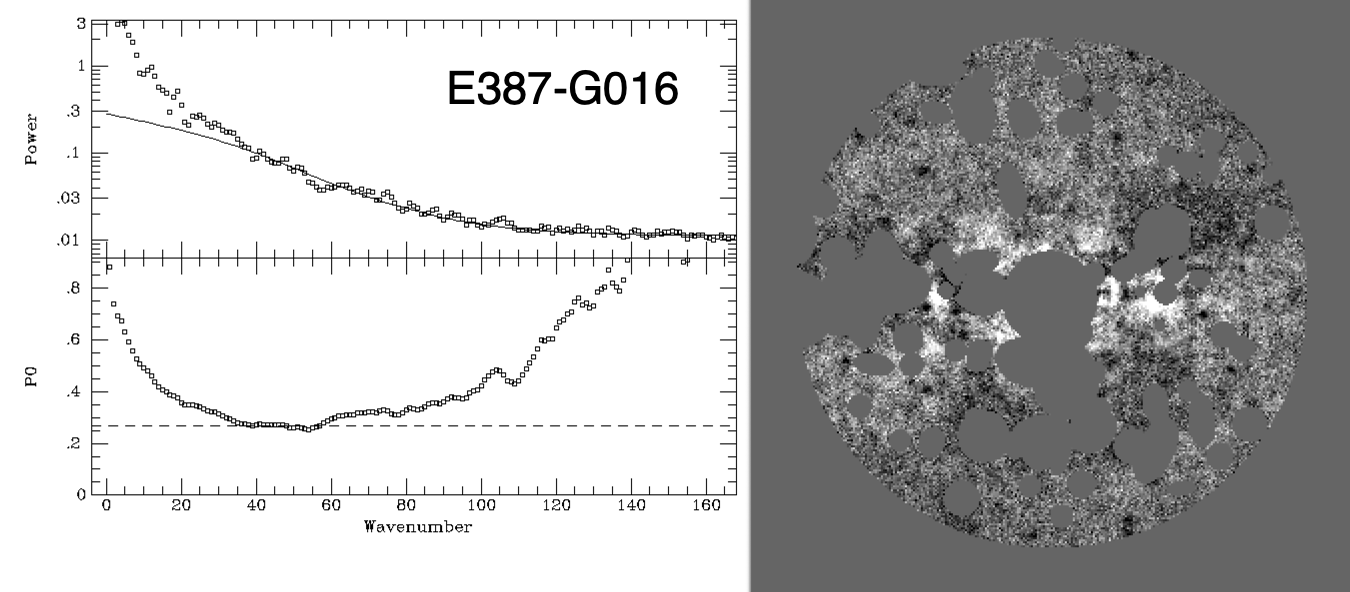}

\caption{SBF sample:35-46}

\label{fig:sample_35-46}

\vspace{-60pt} 
\end{figure*}

\begin{figure*}
\centering

\includegraphics[scale = 0.185]
 {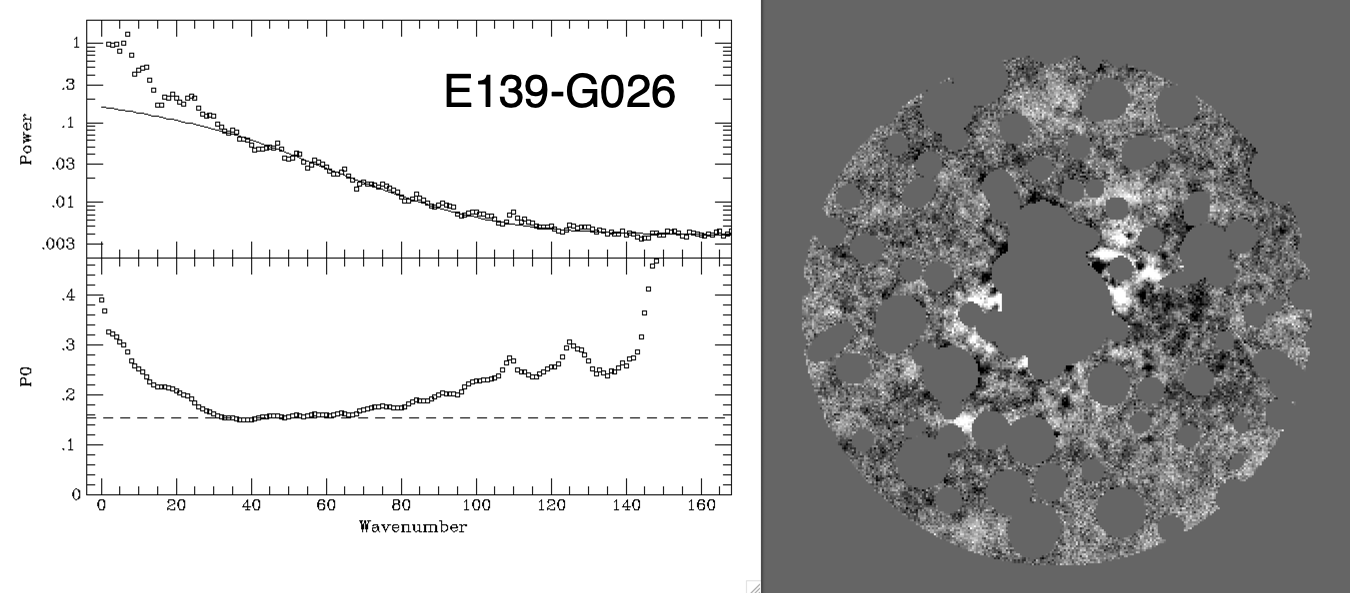}
\includegraphics[scale = 0.185]
 {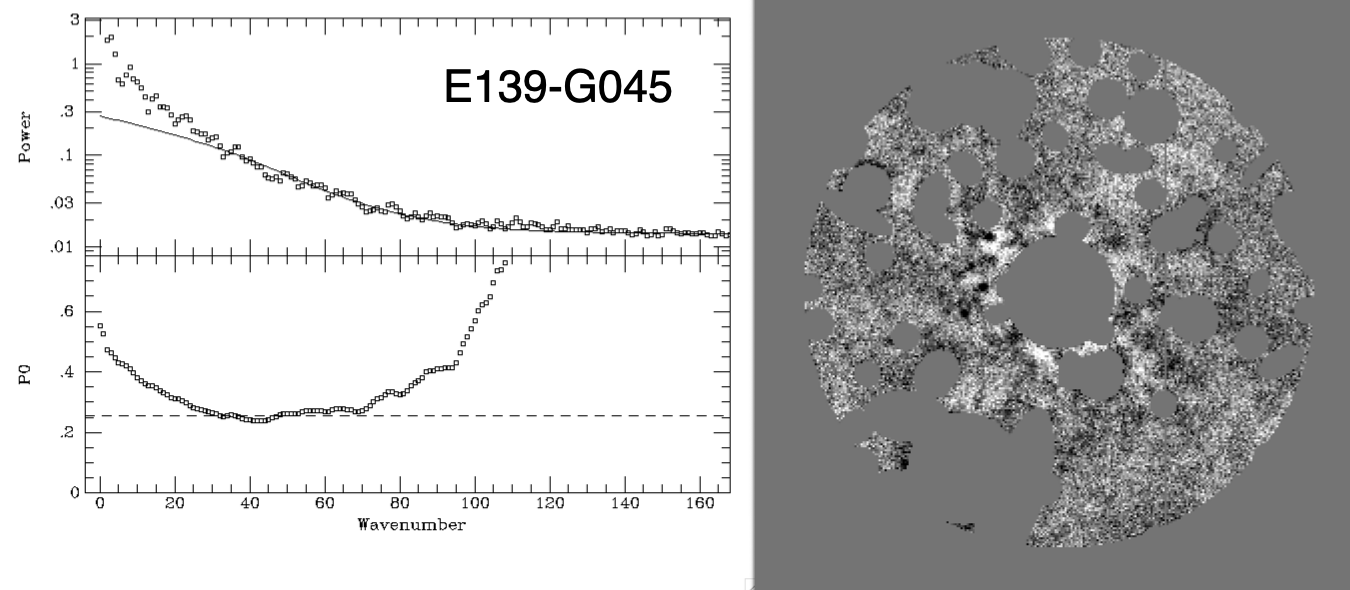}
\includegraphics[scale = 0.185]
 {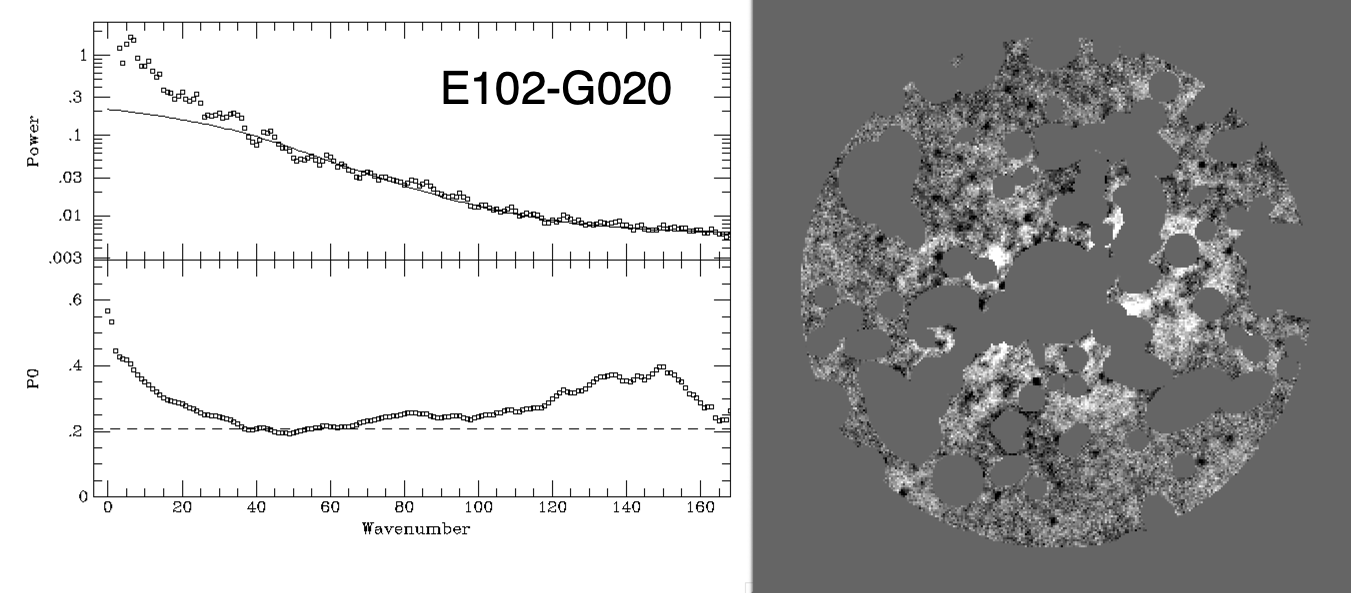}
\includegraphics[scale = 0.185]
 {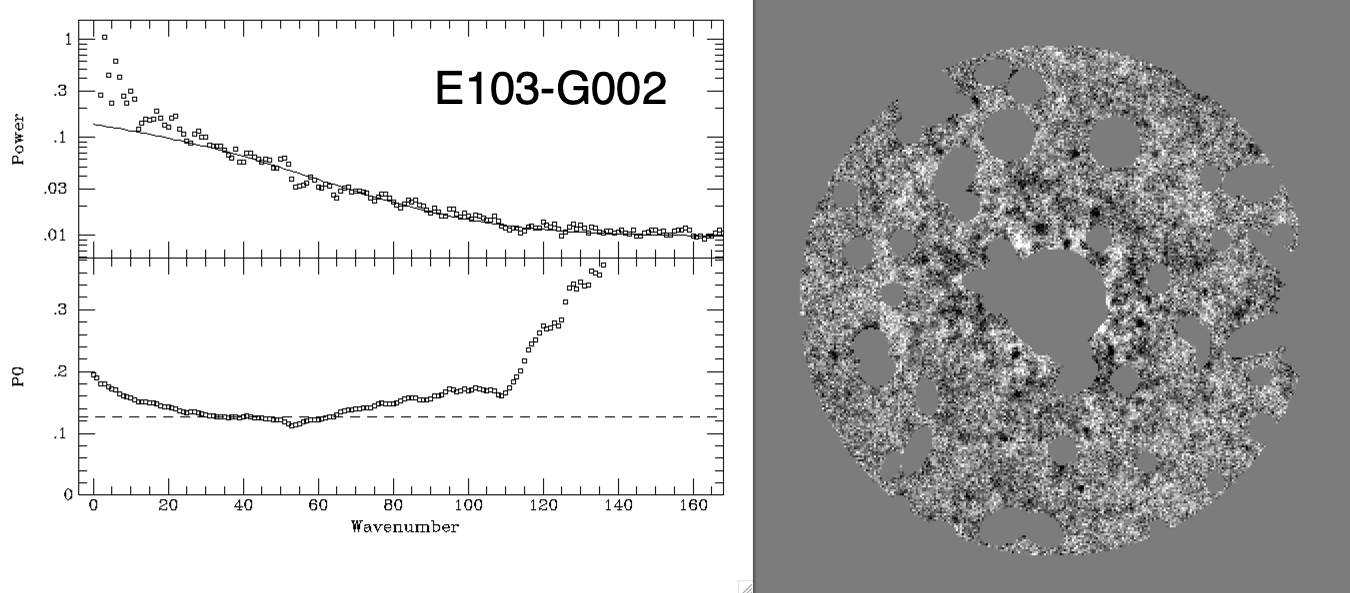}
\includegraphics[scale = 0.185]
 {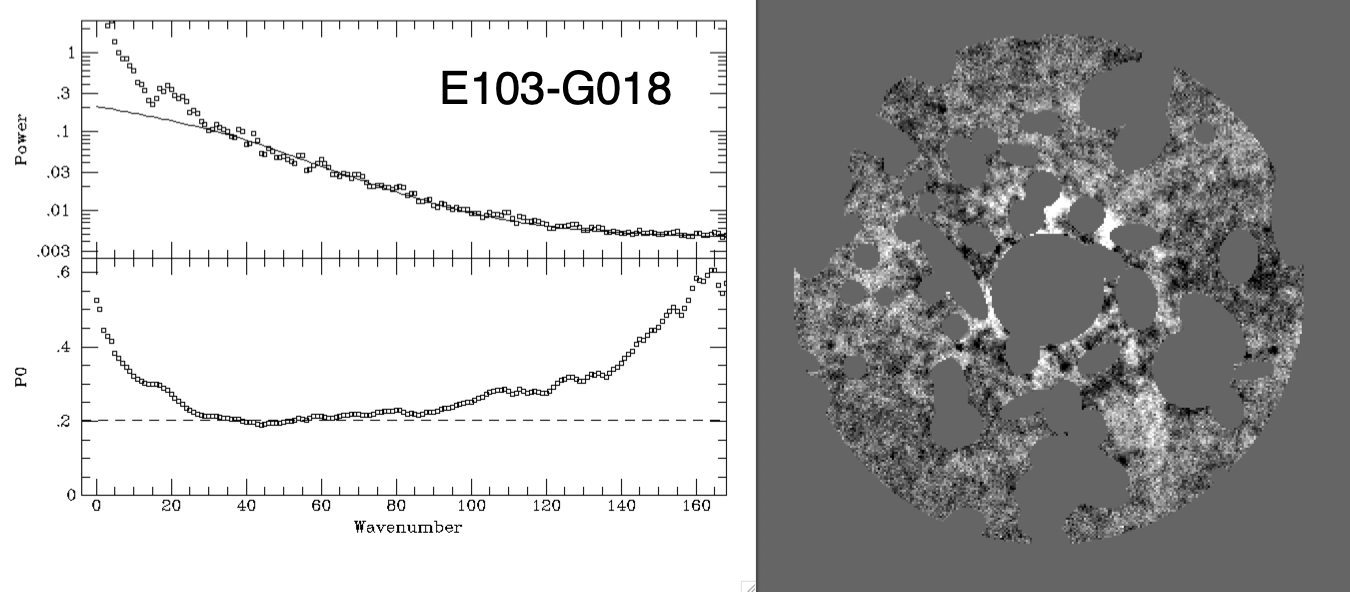}
\includegraphics[scale = 0.185]
 {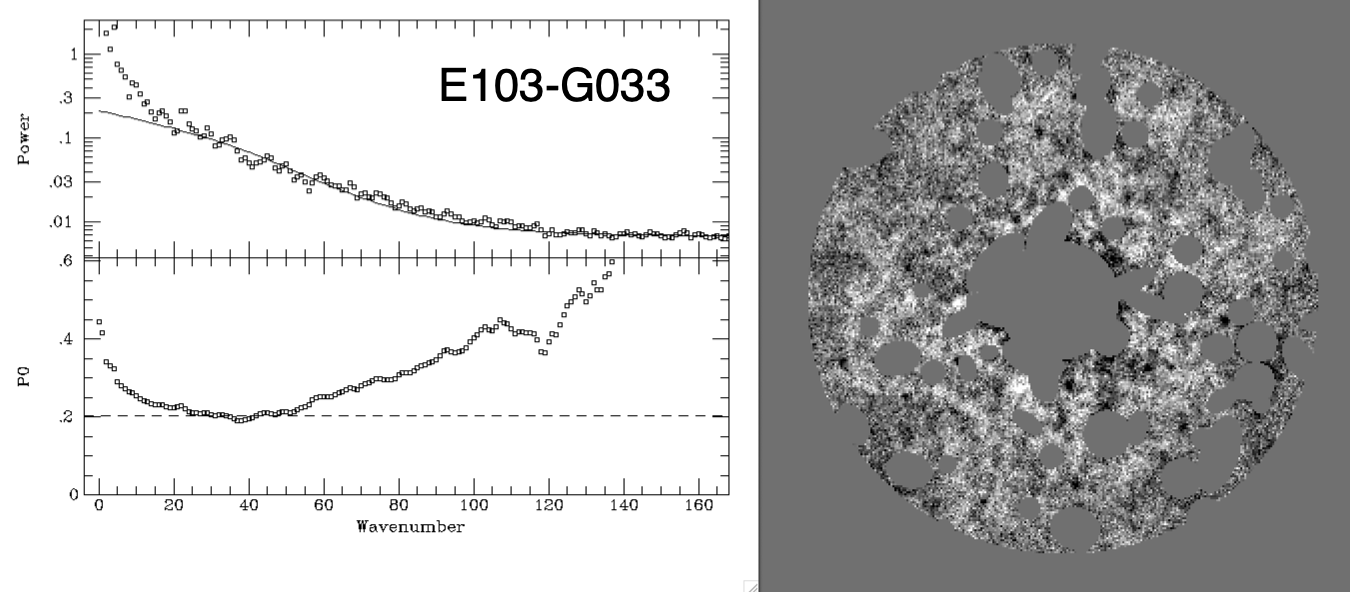}
\includegraphics[scale = 0.185]
 {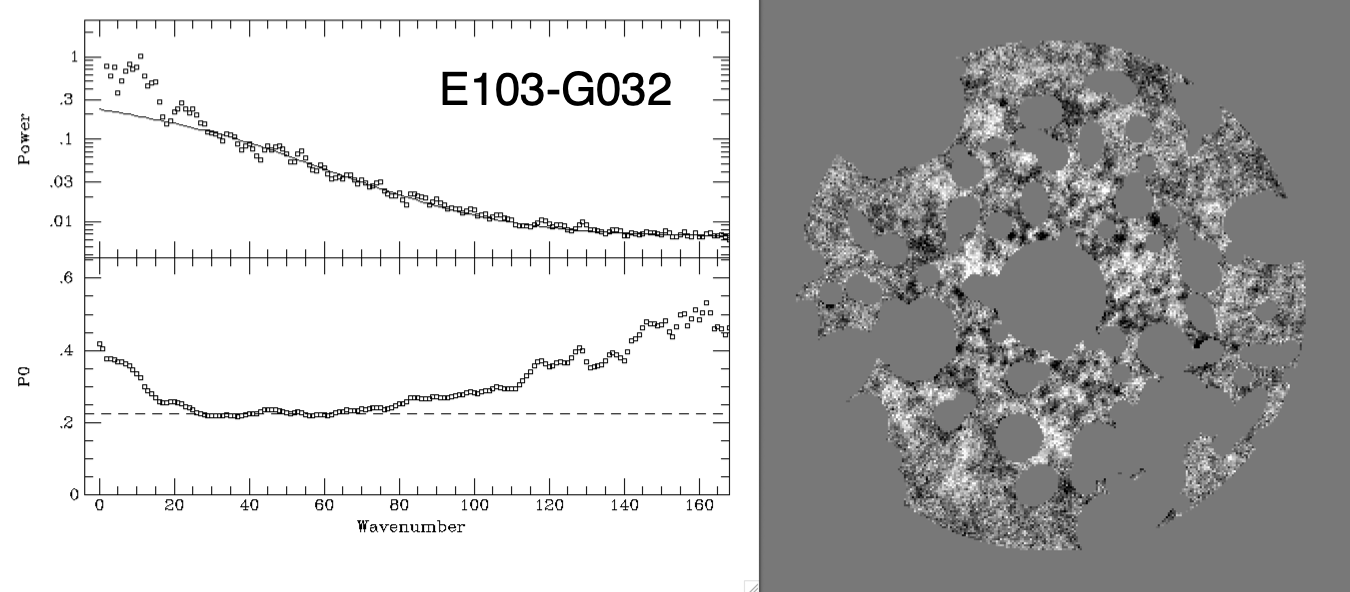}
\includegraphics[scale = 0.185]
 {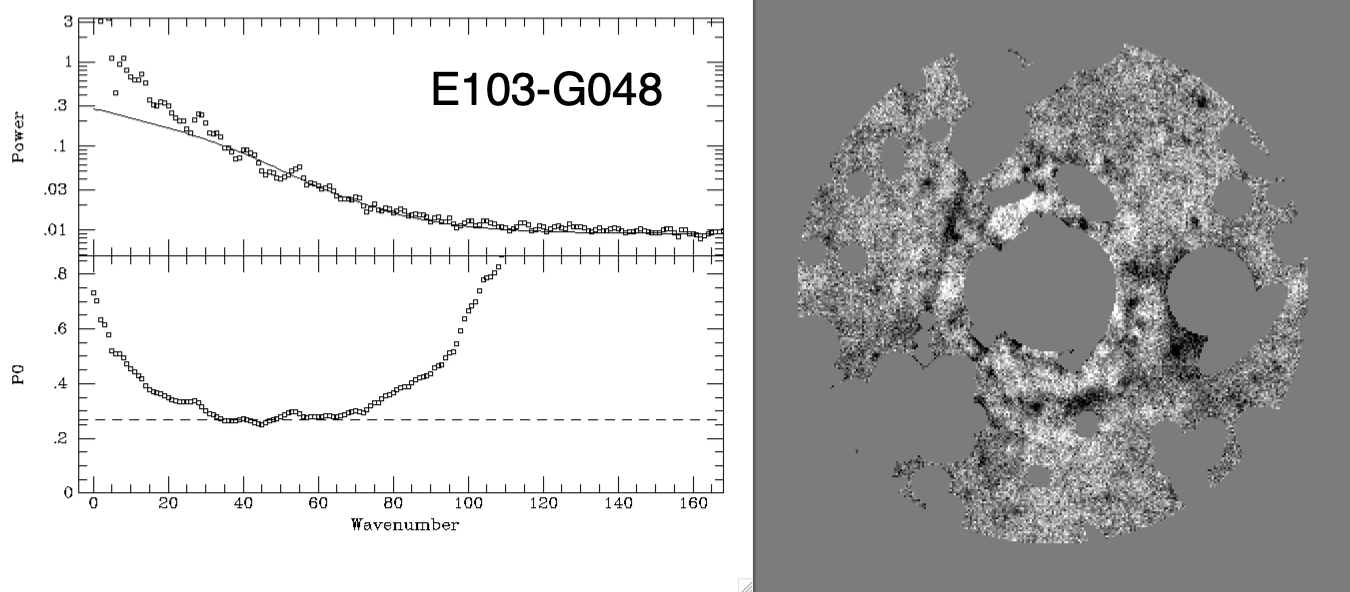}
\includegraphics[scale = 0.185]
 {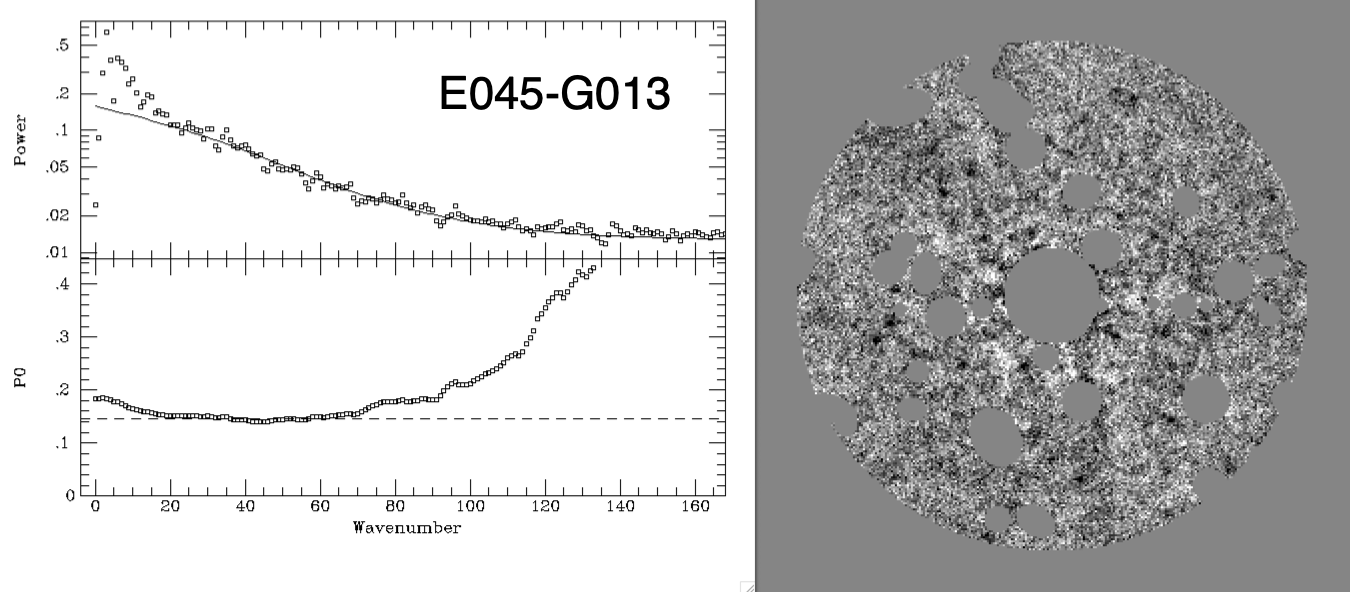}
\includegraphics[scale = 0.185]
 {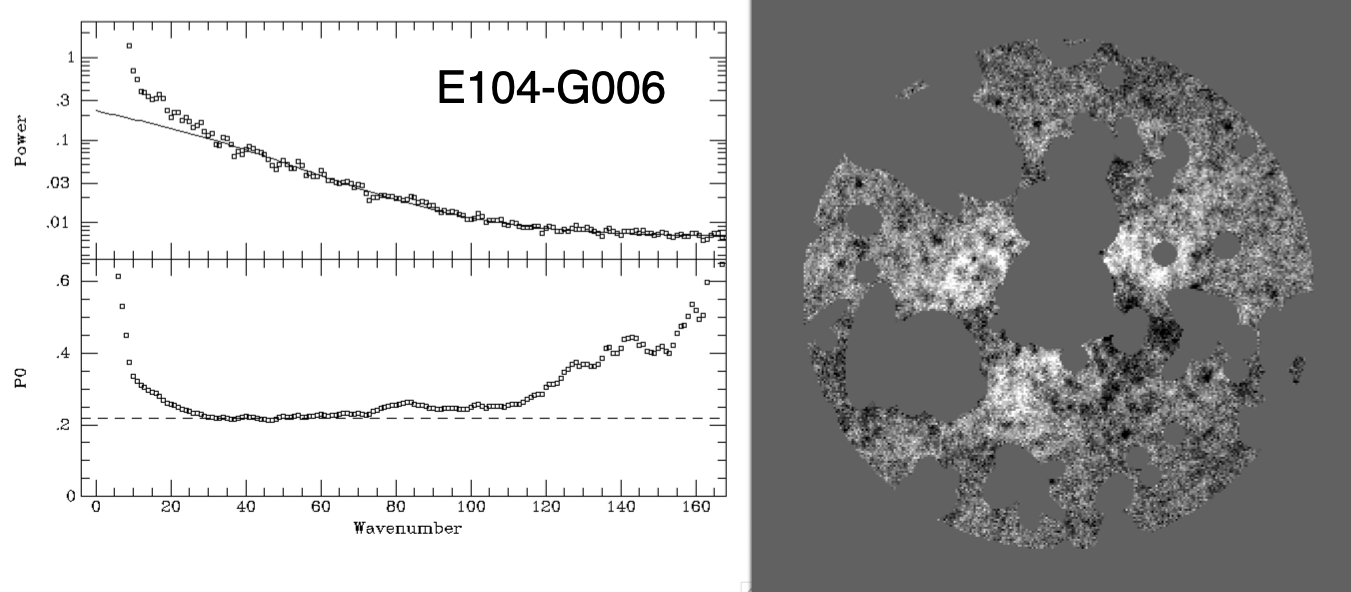}
\includegraphics[scale = 0.185]
 {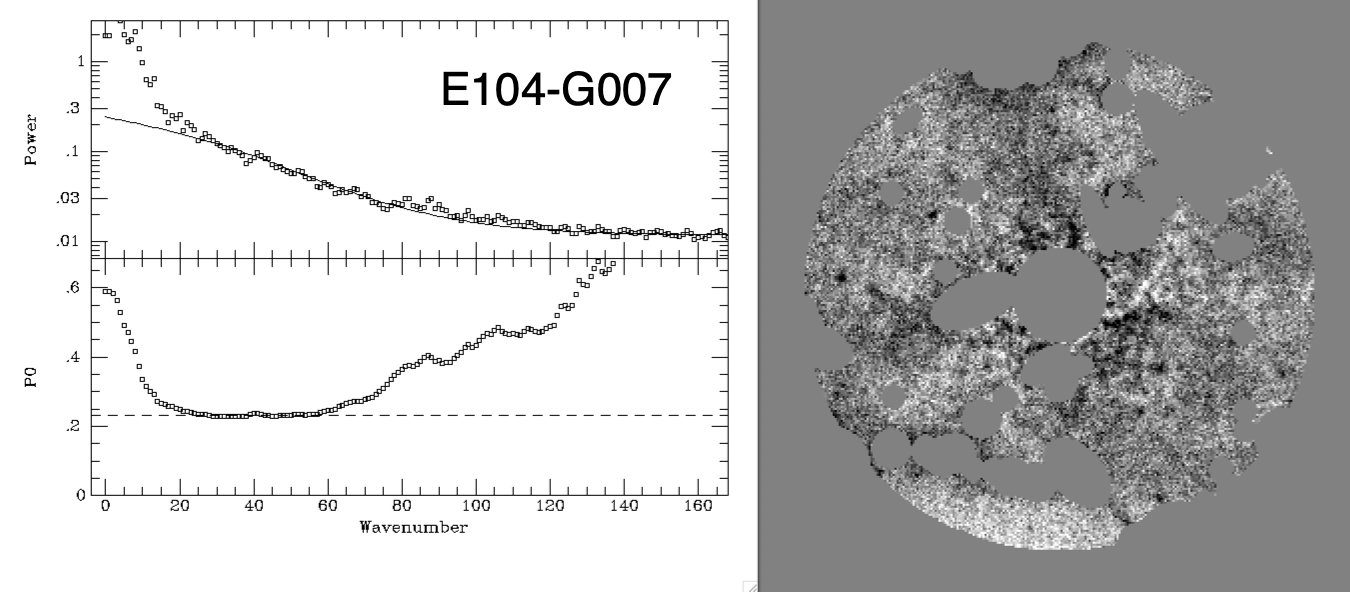}
\includegraphics[scale = 0.185]
 {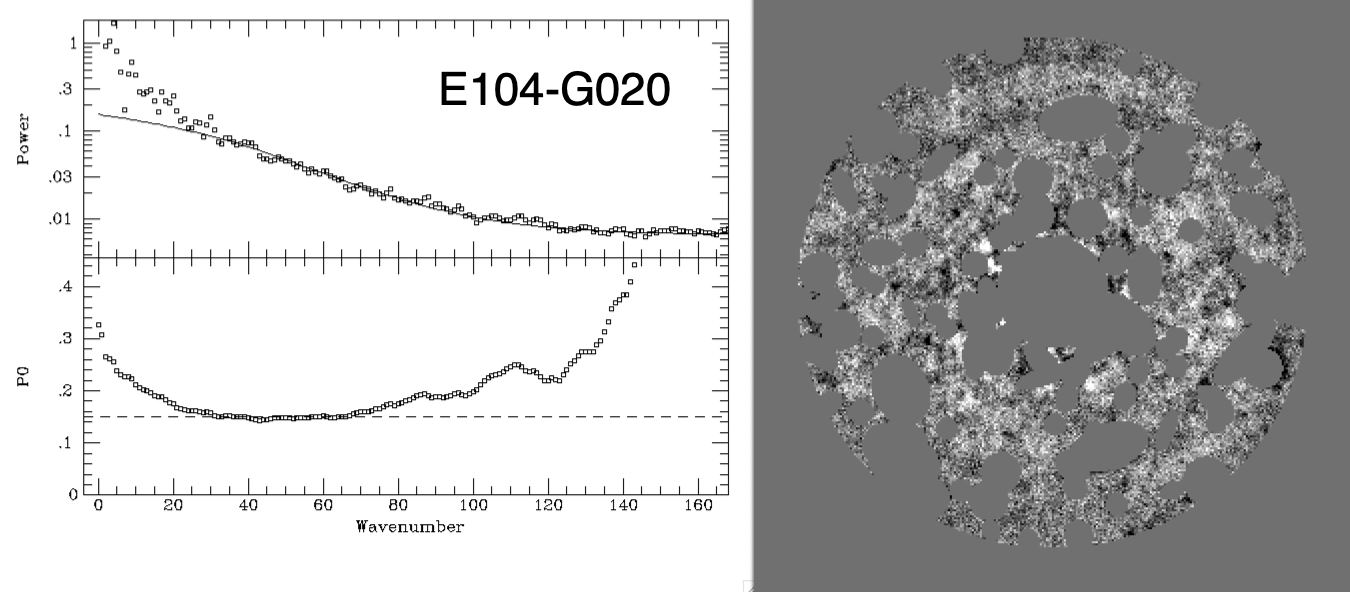}

\caption{SBF sample:47-58}

\label{fig:SBF_sample_47-58}

\vspace{-60pt} 
\end{figure*}

\begin{figure*}
\centering
\includegraphics[scale = 0.185]
 {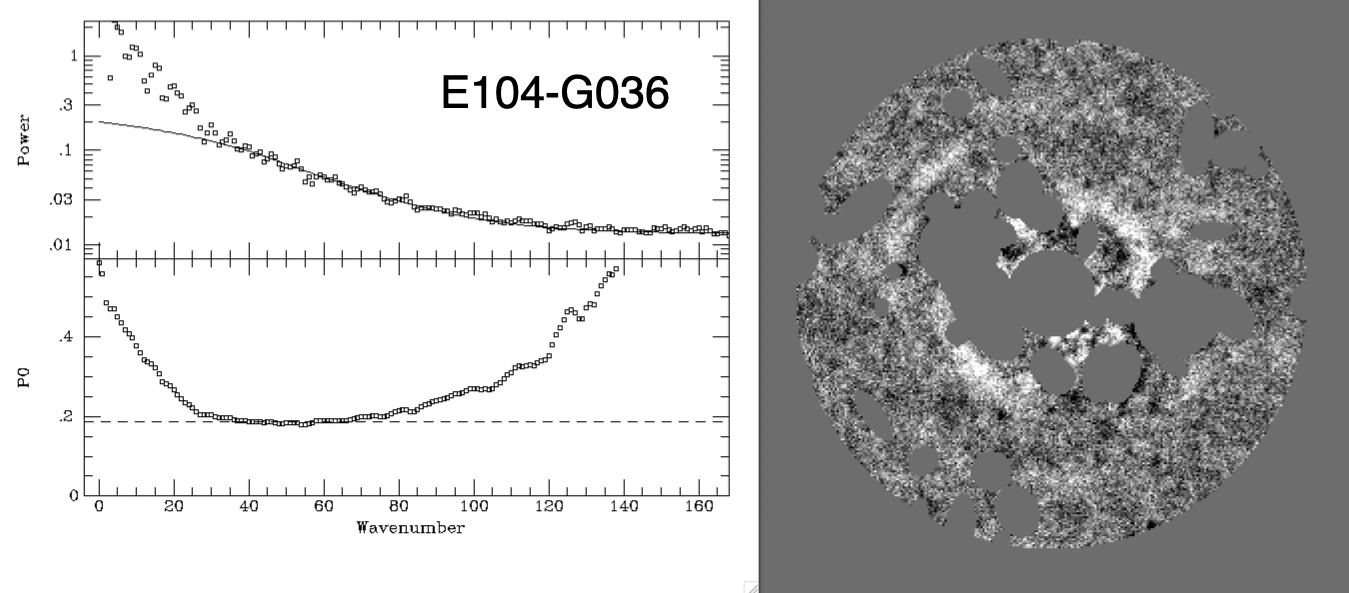}
\includegraphics[scale = 0.185]
 {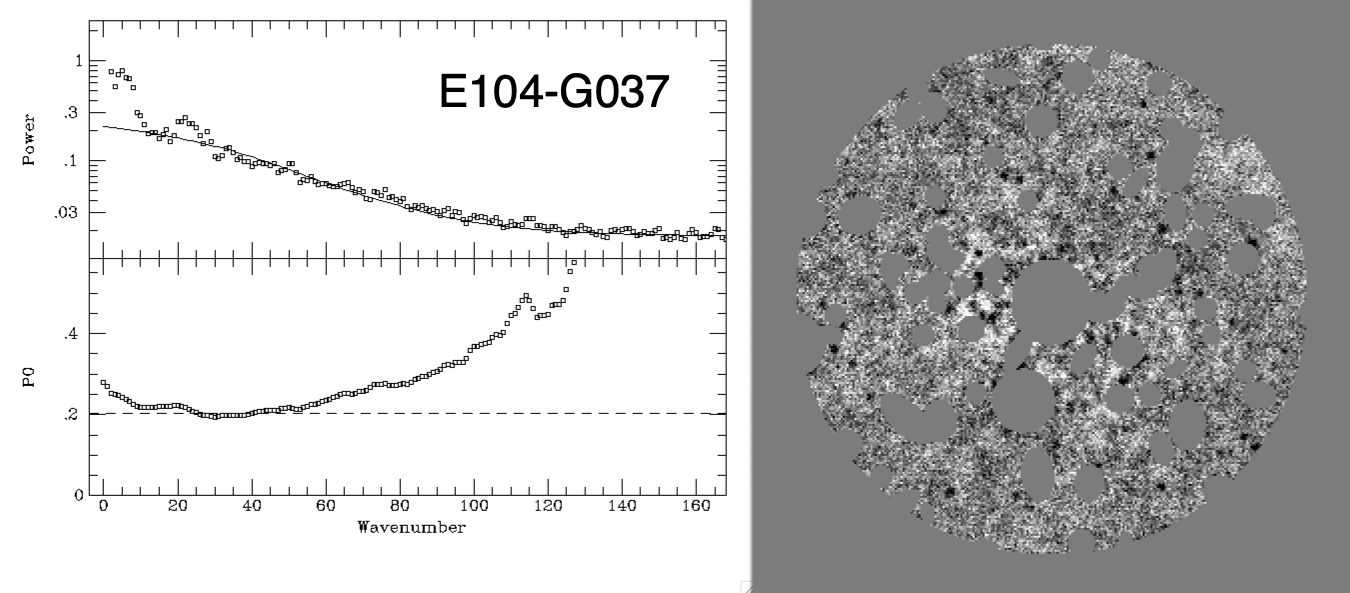}
\includegraphics[scale = 0.185]
 {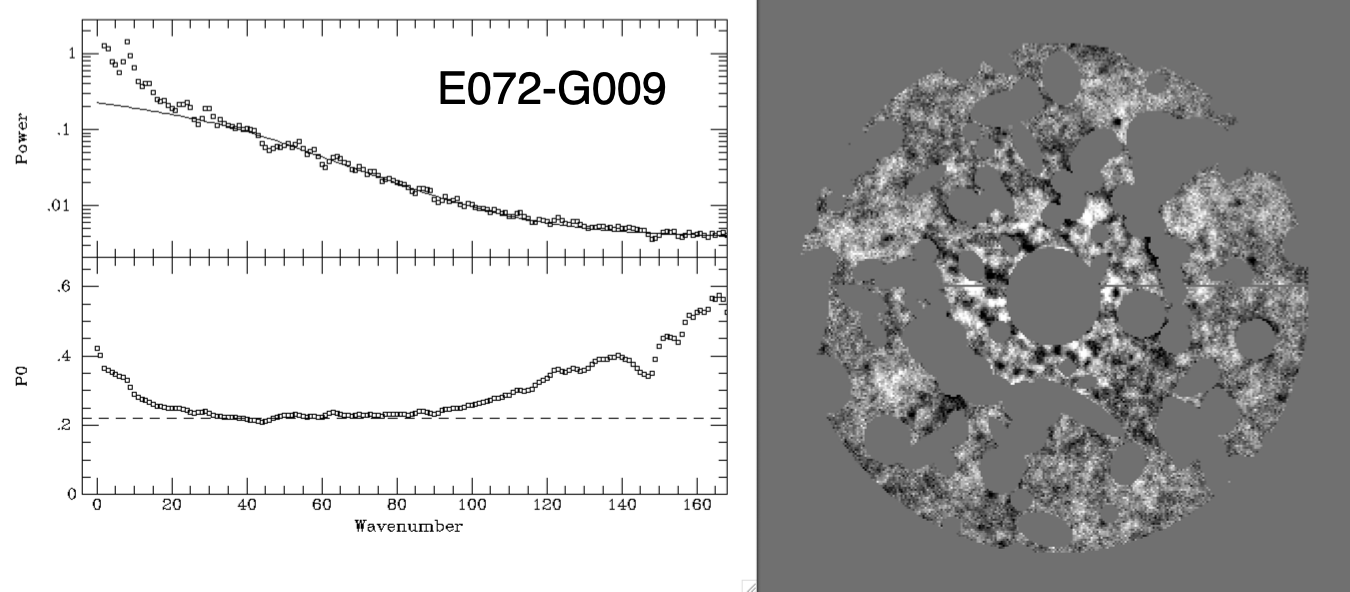}
\includegraphics[scale = 0.185]
 {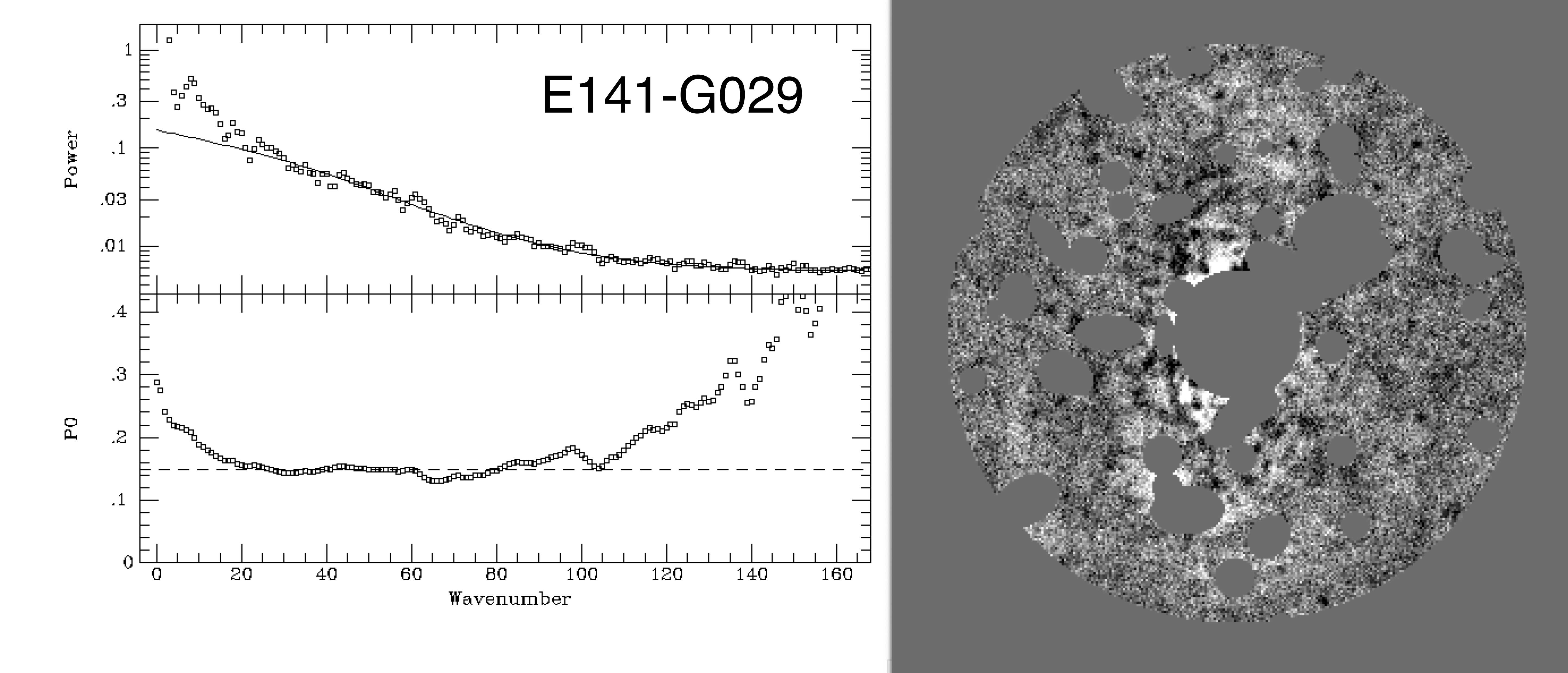}
\includegraphics[scale = 0.185]
 {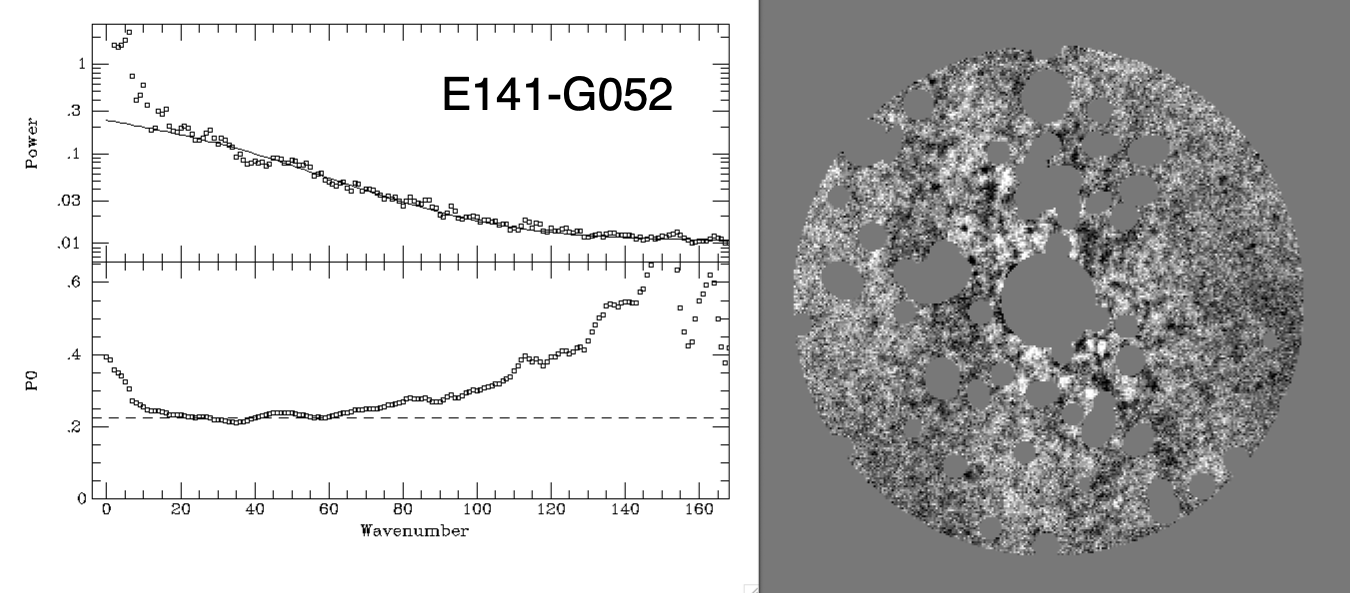}
\includegraphics[scale = 0.185]
 {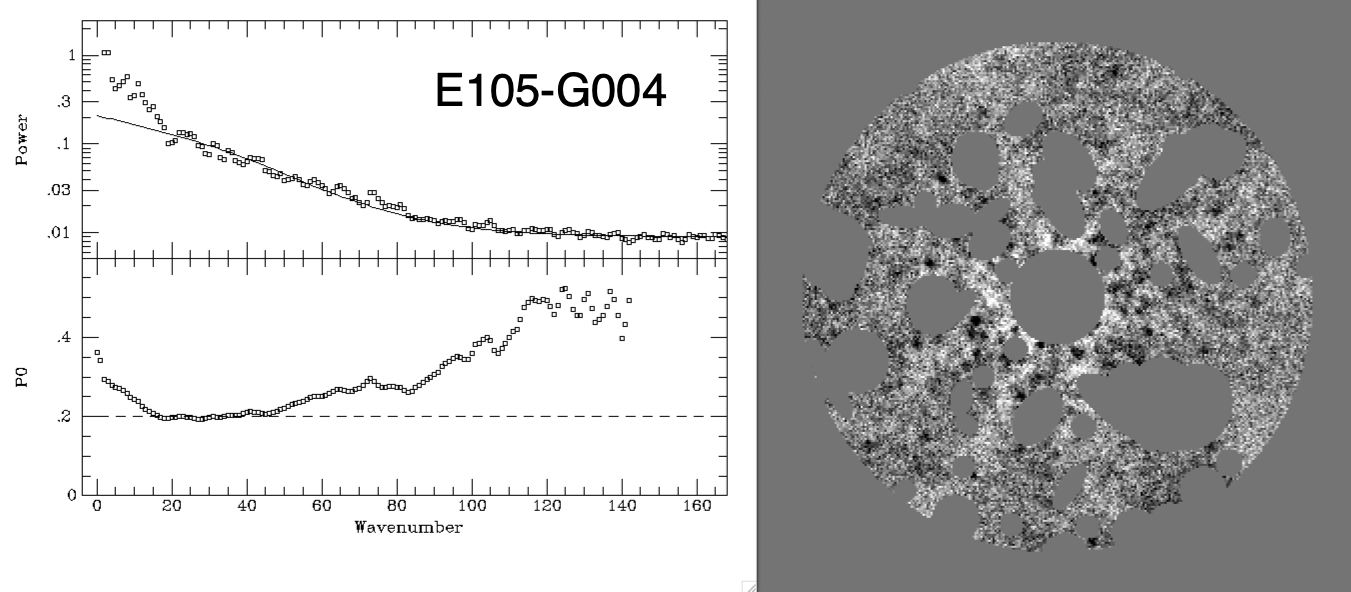}
\includegraphics[scale = 0.185]
 {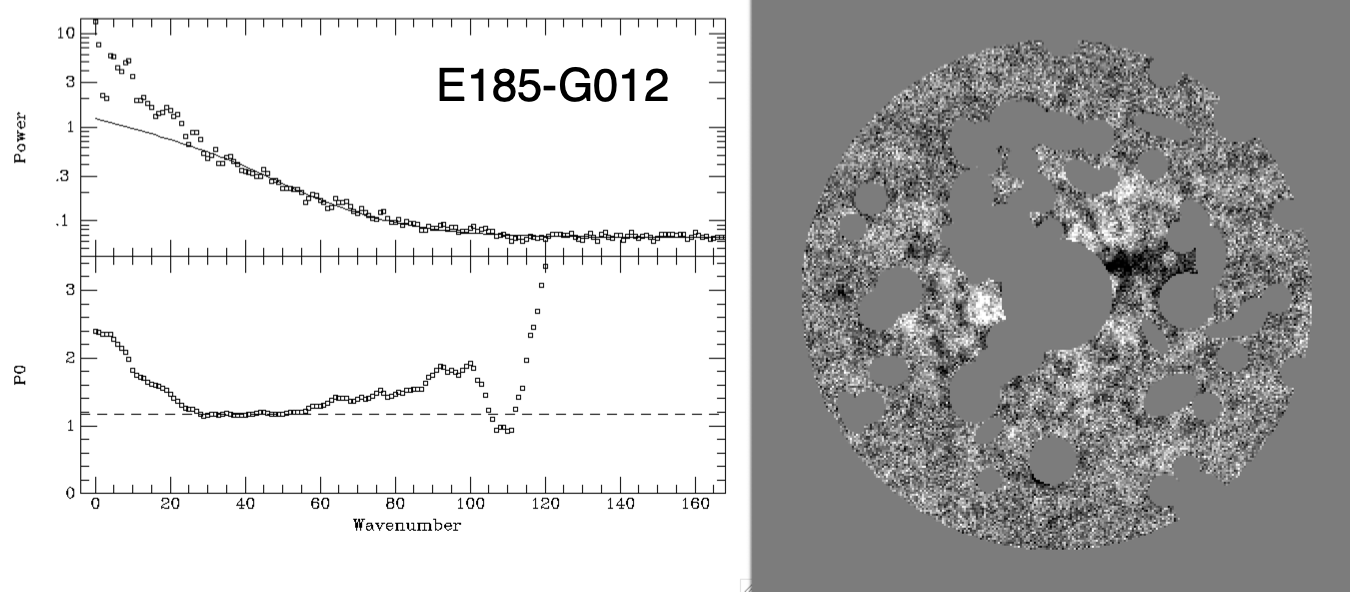}
\includegraphics[scale = 0.185]
 {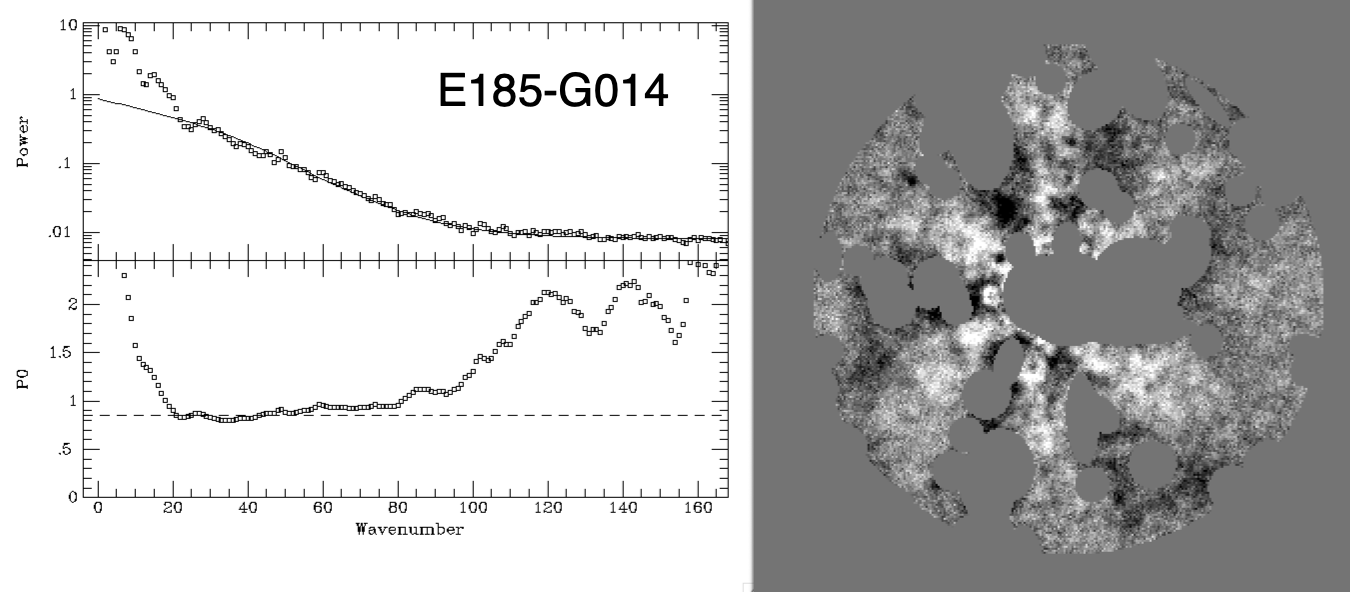}

\caption{SBF sample:59-66}

\label{fig:SBF_sample_59-66}
\end{figure*}

\clearpage

\bibliography{SBF}{}
\bibliographystyle{aasjournal}










\end{document}